\pdfoutput=1

\pdfoutput=1

\documentclass[11pt]{scrartcl}
\usepackage[a4paper, total={16cm, 24cm}]{geometry}
\usepackage{microtype} 
\usepackage{authblk}
\title{Understanding the Impact of Proportionality \\ in Approval-Based Multiwinner Elections}
\date{\vspace{-1.5cm}}

\author[1]{Niclas Boehmer}
\author[1]{Lara Glessen}
\author[2]{Jannik Peters}
\affil[1]{Hasso Plattner Institute, University of Potsdam, Germany} 
\affil[2]{National University of Singapore}
\affil[ ]{\texttt{niclas.boehmer@hpi.de, lara.glessen@hpi.de, peters@nus.edu.sg}} 

\usepackage{xcolor}
\usepackage{graphicx}
\usepackage{pgfplots}
\pgfplotsset{compat=1.17}

\usepackage{amsmath, amssymb, amsthm, mathtools, nicefrac, siunitx}

\usepackage{thmtools}
\usepackage{thm-restate}
\newtheorem{theorem}{Theorem}[section]

\newtheorem{observation}[theorem]{Observation}

\newtheorem{lemma}[theorem]{Lemma}
\newtheorem{corollary}[theorem]{Corollary}
\theoremstyle{definition}
\newtheorem{definition}[theorem]{Definition}

\usepackage[pagebackref]{hyperref}
\hypersetup{
	pdfencoding=auto,
	psdextra,
	colorlinks=true,
	citecolor=green!50!black,
	linkcolor=red!60!black,
}

\usepackage[nameinlink]{cleveref}

\usepackage{natbib}
\usepackage{algorithm}
\usepackage{algpseudocode}

\usepackage[inline]{enumitem}
\usepackage{booktabs}
\usepackage{subcaption}

\usepackage[textsize=tiny]{todonotes}

\newcommand{\prooflink}[1]{\marginline{\vspace{0.5cm}\footnotesize \hyperlink{restated#1}{\hypertarget{original#1}{[Proof]}}}}
\newcommand{\restatehere}[1]{%
	\marginline{\vspace{0.6cm}\footnotesize \hyperlink{original#1}{\hypertarget{restated#1}{[Main]}}}%
	\csname #1\endcsname*%
}

\usepackage{tabularx}
\usepackage{thm-restate}
\usepackage{xspace}
\usepackage[T1]{fontenc}
\usepackage{lmodern}
\usepackage{adjustbox}
\usepackage{multirow}

\DeclareMathOperator{\jr}{JR}

\DeclareMathOperator{\ejrp}{EJR+}

\newcommand{\problemdefX}[3]{
  \begin{center}
    \begin{minipage}{0.95\columnwidth}
      \noindent
      \textsc{#1}
      
      \vspace{2pt}
      \setlength{\tabcolsep}{3pt}
      \begin{tabularx}{\columnwidth}{@{}lX@{}}
        \textbf{Input:} 		& #2 \\
        \textbf{Question:} 	& #3
      \end{tabularx}
    \end{minipage}
  \end{center}
}

\newcommand{\diffcomm}{\textsc{Diff-Committees}\xspace}
\newtheorem{exmp}{Example}[section]

\newcommand{\mis}{\textsc{Multicolor Independent Set}\xspace}
\newcommand{\restrsat}{\textsc{(3,B2)-Satisfiability}\xspace}
\newcommand{\jrntwojr}{\textsc{JR-not-2JR}\xspace}
\newcommand{\jrnejr}{\textsc{JR-not-EJR+}\xspace}

\begin{document}

\maketitle

\bigskip
{\footnotesize\tableofcontents}

\newpage

\begin{abstract}
	\begin{center}
		\textbf{\textsf{Abstract}} \smallskip
	\end{center}
	Despite extensive theoretical research on proportionality in approval-based multiwinner voting, its impact on which committees and candidates can be selected in practice remains poorly understood. We address this gap by (i) analyzing the computational complexity of several natural problems related to the behavior of proportionality axioms, and (ii) conducting an extensive experimental study on both real-world and synthetic elections. Our findings reveal substantial variation in the restrictiveness of proportionality across instances, including previously unobserved high levels of restrictiveness in some real-world cases. We also introduce and evaluate new measures for quantifying a candidate’s importance for achieving proportional outcomes, which differ clearly from assessing candidate strength by approval score.
	
	\end{abstract}

\section{Introduction}
Proportionality in multiwinner voting is one of the most actively studied topics in computational social choice. Given potentially conflicting preferences of voters over candidates, the goal is to select a fixed-size subset of candidates that proportionally represents the preferences of the voters. Applications of proportional multiwinner voting arise in contexts such as participatory budgeting \citep{PPS21a}, blockchain systems \citep{BBC+24a}, or civic participation platforms \citep{RML25a}. One of the building blocks of proportional multiwinner voting are so-called proportionality axioms: properties that certify the proportionality of an outcome. Within the study of proportionality, \emph{approval-based} multiwinner voting, in which each voter votes for a subset of the candidates they approve of, has emerged as a simple, yet expressive model.
Accordingly, researchers have put significant effort into the development of axioms capturing various notions of proportionality, their theoretical analysis, and the design of voting rules that satisfy them \citep[e.g.,][]{ABC+16a, PeSk20a, BrPe23a, HKP+25a, ALS+23a, KSS25a}.

Despite the extensive theoretical coverage of proportionality axioms, research into their practical impact remains limited. We are only aware of two papers whose experiments explicitly focus on how proportionality axioms constrain the set of outcomes in approval-based multiwinner elections: \citet{BFNK19a} and \citet{BrPe23a} (see \Cref{sec:RW} for further related work). 
Both of these papers measure the strength of proportionality axioms through their restrictiveness, i.e., they compute the fraction of committees satisfying an axiom in elections sampled from various synthetic models.  The conclusions \citeauthor{BFNK19a} and \citeauthor{BrPe23a} draw can be viewed as quite negative. \citeauthor{BFNK19a} remark that even for extended justified representation (EJR)  \citep{ABC+16a},  the strongest axiom they study: ``satisfying EJR is quite easy'' \citep[page~114]{BFNK19a}, as in the majority of their elections, a majority of \emph{all} committees satisfy EJR. Similarly, although the EJR+ axiom introduced by \citeauthor{BrPe23a} is stronger than EJR, their findings suggest that committees satisfying EJR+ remain quite common in most of their synthetic sampling models.

This leaves us in an unsatisfying state. One could interpret the prior work as evidence that proportionality axioms impose only extremely mild constraints, raising doubts about their importance and practical relevance: If proportionality axioms were rarely restrictive, imposing them would not change the set of selectable committees. Thus, appealing to proportionality as a justification for a rule, as commonly done, would be of very limited practical significance. 
 At the same time, both studies focus exclusively on synthetically generated elections, raising questions about the extent to which their conclusions translate to realistic instances and whether proportionality axioms might be even less powerful on real-world preference data.
This motivates our \emph{first research question}: to what extent do proportionality axioms constrain which committees can be selected in practice?

\begin{figure}[t]
    \centering
\begin{tikzpicture}[scale=0.5]

\definecolor{darkgray176}{RGB}{176,176,176}
\definecolor{green}{RGB}{0,128,0}
\input{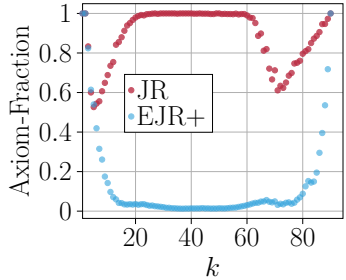}

\begin{axis}[
tick align=outside,
tick pos=left,
x grid style={darkgray176},
xlabel={$k$},
xmajorgrids,
xmin=1, xmax=94.45,
xtick style={color=black},
y grid style={darkgray176},
ylabel={Axiom-Fraction},
ymajorgrids,
ymin=-0.0367335464389048, ymax=1.04936826411614,
ytick style={color=black},
legend style={at={(0.5,0.4)}, anchor=south east, legend columns=1, draw=black, font=\huge},
legend cell align={left},
label style={font =\huge},
ticklabel style={color=black, font=\huge},
]
\addplot [draw=blue, fill=blue, mark=*, only marks, opacity=0.8]
table{%
x  y
1 1
2 1
3 0.833333333333333
4 0.600240096038415
5 0.526315789473684
6 0.543773790103317
7 0.555555555555556
8 0.606428138265615
9 0.648929266709929
10 0.688231245698555
11 0.735294117647059
12 0.754147812971342
13 0.841750841750842
14 0.855431993156544
15 0.879507475813544
16 0.912408759124088
17 0.938967136150235
18 0.951474785918173
19 0.967117988394584
20 0.979431929480901
21 0.987166831194472
22 0.992063492063492
23 0.99009900990099
24 0.99601593625498
25 0.997008973080758
26 0.999000999000999
27 1
28 0.998003992015968
29 1
30 1
31 1
32 1
33 1
34 1
35 0.999000999000999
36 1
37 1
38 1
39 1
40 1
41 0.999000999000999
42 0.999000999000999
43 1
44 0.999000999000999
45 0.999000999000999
46 0.999000999000999
47 0.997008973080758
48 0.997008973080758
49 1
50 1
51 0.997008973080758
52 0.998003992015968
53 0.997008973080758
54 0.997008973080758
55 0.99601593625498
56 0.999000999000999
57 0.995024875621891
58 0.997008973080758
59 0.997008973080758
60 0.99009900990099
61 0.979431929480901
62 0.9765625
63 0.951474785918173
64 0.886524822695035
65 0.898472596585804
66 0.811688311688312
67 0.819672131147541
68 0.720980533525595
69 0.666222518321119
70 0.712250712250712
71 0.610873549175321
72 0.634115409004439
73 0.623441396508728
74 0.650618087182824
75 0.685400959561343
76 0.698812019566737
77 0.746825989544436
78 0.750750750750751
79 0.760456273764259
80 0.796178343949045
81 0.809061488673139
82 0.819000819000819
83 0.850340136054422
84 0.87260034904014
85 0.886524822695035
86 0.905797101449275
87 0.946073793755913
88 0.946969696969697
89 0.972762645914397
90 1
};
\addlegendentry{JR}

\addplot [draw=green, fill=green, mark=*, only marks, opacity=0.6]
table{%
x  y
1 1
2 1
3 0.823723228995058
4 0.614250614250614
5 0.539665407447383
6 0.419111483654652
7 0.315258511979823
8 0.260960334029228
9 0.175008750437522
10 0.142348754448399
11 0.0971817298347911
12 0.0709521782318717
13 0.0594459636190703
14 0.0416666666666667
15 0.0407265618636475
16 0.0341635065423115
17 0.033213763783712
18 0.0351506204084502
19 0.0332679064506471
20 0.0358770135973882
21 0.0323258445126879
22 0.0349283967865875
23 0.0352249110570996
24 0.0311575011684063
25 0.0275603571822291
26 0.0284179715251925
27 0.0248997783919723
28 0.0237287331229386
29 0.0219992960225273
30 0.0202843871072436
31 0.0171738682420828
32 0.0171974960445759
33 0.0147412179194245
34 0.0145011600928074
35 0.0135261257118124
36 0.0130458038172022
37 0.0126347176772335
38 0.0133840141335189
39 0.0130729207519544
40 0.0140175780428658
41 0.0136597093213856
42 0.0138908181691902
43 0.0138153986433279
44 0.0153031555106663
45 0.0136537411250683
46 0.014100593634992
47 0.0131495897328003
48 0.0135650239422673
49 0.0132247143461701
50 0.0145810854159984
51 0.0139070453091536
52 0.0137296629367749
53 0.0148933635172167
54 0.0167425663005626
55 0.0176812772954718
56 0.0184189199145362
57 0.0215331610680448
58 0.0240373058987549
59 0.0267436884895165
60 0.0302910974464605
61 0.0337575532525403
62 0.0383788762665029
63 0.0421123557651815
64 0.0424899086466964
65 0.0484637006881845
66 0.0488376636061731
67 0.0561009817671809
68 0.0474811262523147
69 0.0442145288941946
70 0.0492101766645342
71 0.0316876861651562
72 0.0374335554390956
73 0.0343159122885282
74 0.042369290738073
75 0.0444483954129256
76 0.037575620937136
77 0.0428357249946455
78 0.0581327752586909
79 0.0696281854894861
80 0.0880281690140845
81 0.125328988595062
82 0.152299725860493
83 0.14446691707599
84 0.149543891132047
85 0.194969779684149
86 0.295159386068477
87 0.395256916996047
88 0.535331905781585
89 0.717875089734386
90 1
};
\addlegendentry{EJR+}

\end{axis}

\end{tikzpicture}

    \caption{Percentage of committees satisfying JR and EJR+ depending on the committee size $k$, in the Warsaw Praga-Północ PB election 2022.}
    \label{fig:praga_axioms}
\end{figure}

To illustrate the significant, yet nuanced, practical impact proportionality may have, and to motivate further investigation, we  present \Cref{fig:praga_axioms}.
This plot is based on a participatory budgeting election conducted in the Praga-Północ district of Warsaw in 2022. Treating this election's votes and candidates as an approval-based multiwinner election, we approximate, for each possible committee size $k$, the percentage of committees satisfying the weak proportionality axiom JR (red) and the stronger axiom EJR+ (blue).\footnote{Note that the depicted values do contain small sampling errors, e.g., resulting in some slightly higher reported percentages for JR than EJR+.}
We observe that for certain values of $k$, only a small fraction of committees satisfy EJR+, and that the satisfaction rate for JR can fluctuate in seemingly unpredictable ways.
This example also highlights a potential limitation of prior experimental work on proportionality, which typically evaluates restrictiveness only for a single, hand-picked value~of~$k$.

In addition to studying the impact of proportionality on the space of feasible committees, as our \emph{second research question}, we investigate the impact of proportionality on the merit of different candidates.  
In elections based on the principle of individual excellence, in which candidates with the most approvals are selected, a candidate’s merit is naturally quantified by their approval score.
In contrast, when aiming to select a committee that satisfies a proportionality axiom, assessing a candidate’s strength becomes more complex, as the proportionality of a committee containing the candidate always depends on which other candidates are present.
Good measures for a candidate’s strength under proportionality would contribute to improving the transparency and our understanding of proportional multiwinner elections.

\subsection{Our Contributions}
Our investigation consists of two parts: an algorithmic analysis (\Cref{sec:AA}) and an experimental analysis (\Cref{sec:EA}), in both of which we focus on the axioms JR and EJR+, which can be verified and satisfied in polynomial time  (see \Cref{sub:prelims} for definitions). Proofs and additional experimental results can be found in the appendix. 

In the algorithmic part, we study the complexity of a range of computational problems aimed at understanding the impact of proportionality in a given instance.  
Our selection of questions serves two purposes. First, we study the problems that we would like to compute in our experimental analysis. Second, we provide new results on existing computational problems and introduce new problems related to the behavior of proportionality axioms, which might be of independent interest and help to understand the boundaries of tractability when dealing with (easy-to-verify) proportionality axioms.
These problems include
\begin{enumerate*}[label=(\roman*)]
  \item deciding whether two proportionality axioms are equivalent in a given instance,
  \item finding a pair of distant proportional committees, 
  \item  finding a proportional committee containing a given subset of candidates, 
  \item and counting committees (or those containing a given candidate) that satisfy a particular axiom.
\end{enumerate*}
We show that all of our problems are NP-hard (resp.\ \#P-hard), highlighting that even for easy-to-verify and easy-to-construct axioms, slightly more nuanced computational questions immediately yield computational hardness.
Complementing these results, we develop ILP- and sampling-based algorithms for our experiments and investigate our problem's parameterized complexity.  While simple brute-force algorithms yield fixed-parameter tractability (FPT) with respect to the number of candidates $m$ for all problems, we develop slightly more involved algorithms showing FPT with respect to the number of voters~$n$, a parameterization of interest in case only a few opinions need to be aggregated, e.g., in expert or criteria aggregation.

In our experimental analysis, we investigate the impact of proportionality on the space of feasible committees and on candidates' strength.
Due to the general lack of real-world data on approval-based multiwinner elections \citep{BFJ+24a}, we base our analysis on voting data from participatory budgeting instances available in pabulib \citep{FFP+23a}, complemented by synthetic data generated using the resampling and Euclidean models.
Our experimental findings include the following observations:
\begin{enumerate*}[label=(\roman*)]
  \item The restrictiveness of proportionality axioms can vary drastically and in unexpected ways with small changes in~$k$, highlighting that the choice of $k$ deserves more careful consideration than it typically receives in experimental studies.
  \item While JR and EJR+ often do not impose strong restrictions, there are some real-world instances where they can significantly constrain the outcome space. Even in such cases, the set of proportional committees tends to be highly diverse.
  \item We introduce two measures for candidates' importance for proportionality, one of which is inspired by the Banzhaf power index in cooperative game theory. Both measures turn out to be highly correlated. We demonstrate that using candidates' approval scores to assess their proportionality merit can be highly misleading. 
  \item Proportional multiwinner voting rules tend to favor candidates with a high approval score over those important for proportionality. 
\end{enumerate*}

\subsection{Related Work}\label{sec:RW}

For an overview of experimental work in approval-based multiwinner voting (and beyond), we refer to the recent survey-style “guide” by \citet{BFJ+24a}.

The work closest to ours is the aforementioned paper by \citet{BFNK19a}, which was the first to experimentally investigate the impact of proportionality axioms. They considered elections with 100 candidates, 100 voters, and a committee size of 10, sampled from different variants of the Impartial Culture and Euclidean model.
In their experiments, they evaluated how many committees satisfy JR, PJR, and EJR, analyzed the size of the smallest candidate set satisfying JR, and studied the satisfaction and coverage scores achievable by JR committees.\footnote{In a similar setup, satisfaction and coverage scores of JR committees were studied by \citet{EFI+24a}, while the size of JR-satisfying candidate sets was examined  by \citet{EFI+22b}.}
\citeauthor{BFNK19a} also introduced and studied related algorithmic questions, such as the complexity of counting JR, PJR, or EJR committees.
Our work differs from theirs in that we focus our experiments on real-world voting data and explore new aspects of the impact of proportionality, such as its dependency on the committee size and its relation to candidate importance, also resulting in the study of different computational problems.

In their work on verifiable proportionality axioms, \citet{BrPe23a} also conducted experimental studies on how often random committees satisfy different proportionality axioms, using synthetic elections from four models (not including the Euclidean model).  While they observed that their new axiom EJR+ is significantly harder to satisfy than existing ones, such as EJR, for a large part of their parameter space, still over $50\%$ of the committees satisfy it.

\citet{FLSS23a} and \citet{BBC+24a} examined how frequently committees selected by various approval-based multiwinner voting rules satisfy proportionality axioms. Their main finding is that
common proportionality axioms are frequently satisfied by a large array of rules. 
\citet{FLSS23a} studied this question using elections sampled from various synthetic models \citep{SFJ+22a}, along with some instances from pabulib, whereas \citet{BBC+24a} analyzed large real-world elections from the proof-of-stake Polkadot blockchain. 

Finally, a recent study by \citet{BBMP25a} empirically examined proportionality and proportional rules in the setting of ordinal multiwinner voting, using a dataset from Scottish local government elections. Among other findings, they observed that the classic PSC axiom \citep{Dumm84a} is satisfied by many committees on their elections.

\section{Preliminaries} \label{sub:prelims}
\paragraph{Model and Notation.} For $t \in \mathbb{N}$, we let $[t] \coloneqq \{1, \dots, t\}$. Let $N = [n]$ be a set of \emph{voters} and $C = \{c_1, \dots, c_m\}$ be a set of \emph{candidates}. Each voter $i \in N$ possesses an \emph{approval ballot} $A_i \subseteq C$ indicating the candidates approved by this voter. For a given candidate $c \in C$, we denote by $N_c \coloneqq \{i \in N \colon c \in A_i\}$ the set of approvers of $c$ and refer to $|N_c|$ as the \emph{approval score} of $c$.  Together, $A = (A_i)_{i \in N}$ forms an approval profile. We let $k \le m$ denote the committee size and refer to any set $W \subseteq C$ of candidates of size $\lvert W\rvert = k$ as a \emph{committee}. Together, the tuple $\mathcal{I} = (N, C, A, k)$ forms an \emph{approval-based multiwinner election}.

\paragraph{Proportionality Axioms.} The central objects we study in our work are \emph{proportionality axioms}. We focus on two easy-to-verify axioms: \emph{justified representation (JR)} \citep{ABC+16a} and \emph{extended justified representation plus (EJR+)} \citep{BrPe23a}.

\begin{definition}[EJR+]
    Given an election $\mathcal{I} = (N, C, A, k)$ and parameter $t \in [k]$, a subset $W\subseteq C$ satisfies \emph{$t$-extended justified representation plus ($t$-EJR+)} if for any candidate $c \notin W$, $\ell \in [t]$, and set $N' \subseteq N_c$ of voters with $\lvert N'\rvert \ge \ell\frac{n}{k}$, there exists an $i \in N'$ with $\lvert A_i \cap W\rvert \ge \ell$. 
\end{definition}
A committee satisfies justified representation (JR) if it satisfies $1$-EJR+, and EJR+ if it satisfies $k$-EJR+. By definition, any committee satisfying EJR+ also satisfies JR. EJR+ and JR committees are guaranteed to exist and can be computed in polynomial time \citep{BrPe23a}.
For some election $\mathcal{I}$, we denote by $\ejrp(\mathcal{I})$, resp.\ $\jr(\mathcal{I})$,  the set of committees satisfying EJR+, resp.\ JR, in $\mathcal{I}$. 
Moreover, we refer to $\nicefrac{\lvert\ejrp(\mathcal{I})\rvert}{{m \choose k}}$, resp.\ $\nicefrac{\lvert\jr(\mathcal{I})\rvert}{{m \choose k}}$, as the election's \emph{EJR+-fraction}, resp.\ \emph{JR-fraction}.
Further, for some candidate $c\in C$, we let $\ejrp(\mathcal{I}, c)\subseteq \ejrp(\mathcal{I})$, resp.\ $\jr(\mathcal{I}, c)\subseteq \jr(\mathcal{I})$,  be the set of committees satisfying EJR+, resp.\ JR, that contain $c$. 

\newcommand{\clrstr}{20} 
\newcommand{\strstr}{40} 
\begin{exmp}
    \begin{figure}[h!]
    \centering
    \begin{tikzpicture}
    [yscale=0.4,xscale=0.8,
    voter/.style={anchor=south}]
    
        \foreach \i in {1,...,10}
    		\node[voter] at (\i-0.5, -1) {$\i$};
        
       \draw[fill=teal!\clrstr] (0, 0) rectangle (4, 1);
        \draw[fill=teal!\clrstr] (0, 1) rectangle (4, 2);
        \draw[fill=teal!\clrstr] (0, 2) rectangle (2, 3);
        \draw[fill=teal!\clrstr] (2,  2) rectangle (4, 3);
        \draw[fill=magenta!\clrstr] (4,  0) rectangle (7, 1);
        \draw[fill=magenta!\clrstr] (5,  1) rectangle (10, 2);
        \draw[fill=magenta!\clrstr] (8,  2) rectangle (9, 3);
        \draw[fill=magenta!\clrstr] (8,  0) rectangle (10, 1);
        \draw[fill=magenta!\clrstr] (9,  2) rectangle (10, 3);
        \node at ( 2, 0.5) {$c_{1}$};
        \node at ( 2, 1.5) {$c_{2}$};
        \node at ( 1, 2.5) {$c_{3}$};
        \node at ( 3, 2.5) {$c_{4}$};
        \node at ( 5.5, 0.5) {$c_{5}$};
        \node at ( 7.5, 1.5) {$c_{6}$};
        \node at ( 9, 0.5) {$c_{7}$};
        \node at ( 8.5, 2.5) {$c_{8}$};
        \node at ( 9.5, 2.5) {$c_{9}$};
    \end{tikzpicture}
    \caption{Example instance to illustrate JR and EJR+.}
    \label{fig:jrex}
\end{figure}
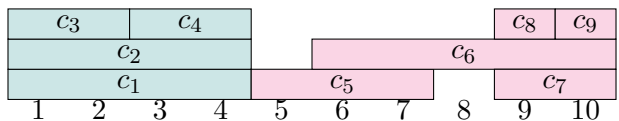
Consider the instance depicted in \Cref{fig:jrex}: voters correspond to integers on the $x$-axis, while candidates correspond to boxes, with every voter approving the candidates above them, e.g., voter $1$ approves $c_1, c_2$, and $c_3$. For $k=5$, the committee $\{c_1, c_3, c_4, c_5, c_7\}$  satisfies JR, as the only ``uncovered'' voter $8$ is on its own not enough to constitute a JR violation. It, however, does not satisfy EJR+ as witnessed by candidate $c_6$. This candidate is approved by five voters, but no voter approves two candidates in the committee. However, each one of them approves at most one candidate,  leading to an EJR+ violation (as $\nicefrac{n}{k} = 2$). The committee $\{c_1, c_2, c_3, c_4, c_6\}$ satisfies~EJR+.
\end{exmp}

\paragraph{Parameterized Algorithms.}
In our algorithmic analysis, we make use of the framework of parameterized algorithms \citep{cygan2015parameterized}, focusing on the complexity classes FPT and W[1]. A parameterized problem $X$ consists of an instance $\mathcal{I}$ together with a parameter $\ell \in \mathbb{N}$.  A fixed-parameter tractable (FPT) algorithm for $(\mathcal{I},\ell)$ is an algorithm running in time $\mathcal{O}(f(\ell) \cdot \lvert \mathcal{I}\rvert^{\mathcal{O}(1)})$ for some computable function $f$. Under standard complexity theoretical
assumptions, if $(\mathcal{I},\ell)$ is W[1]-hard, then it does not admit an FPT algorithm.

\paragraph{Measures of Candidate Importance for Proportionality.} 
We introduce two ways in which we can quantify a candidate's importance for proportional representation. 
As our first measure, we define the \emph{JR-prevalence}, resp.,   \emph{EJR+-prevalence}, of some candidate $c\in C$ as the fraction of committees fulfilling the axiom that contain $c$, i.e., $\nicefrac{\lvert\jr(\mathcal{I}, c)\rvert}{\lvert\jr(\mathcal{I})\rvert}$, resp.\ 
$\nicefrac{\lvert\ejrp(\mathcal{I}, c)\rvert}{\lvert\ejrp(\mathcal{I})\rvert}$. For our second measure, we draw inspiration from the raw Banzhaf power index \citep{banzhaf1964weighted} from cooperative game theory, which counts how often an agent is ``pivotal'' for a coalition to be winning. 
Translating this idea, we count the number of proportional committees
for which proportionality is violated after we remove the candidate: We let the \emph{JR-power-index} (resp.\ \emph{EJR+-power-index}) of candidate $c\in C$ be $|\{W\in \jr(\mathcal{I}, c)\colon  W\setminus \{c\}\text{ does not satisfy JR} \}|$ (resp.\ $|\{W\in \ejrp(\mathcal{I}, c)\colon  W\setminus \{c\}\text{ does not satisfy EJR+} \}|$).

\section{Algorithmic Analysis}\label{sec:AA}
In this section, we present an  algorithmic analysis of combinatorial problems we would like to solve to better understand the impact of proportionality axioms in practice.

\subsection{Differentiating Proportionality Axioms}

We start with a simple problem: can we determine whether, for a given instance, EJR+ is a strictly stronger property than JR, i.e., is there a committee satisfying JR, but not~EJR+? 
\problemdefX{JR-not-EJR+}
	{Election $\mathcal{I} = (N, C ,A, k)$.}
	{Is there a committee~$W \subseteq C$ satisfying JR but not EJR+?}
Despite the significant conceptual differences between EJR+ and JR and the fact that both can be verified in polynomial time, it turns out that checking whether JR and EJR+ coincide on a given instance is NP-complete.

\prooflink{jrnotejr}
\begin{restatable}{theorem}{jrnotejr}
\label{th:jrnotejr}
    \jrnejr is NP-complete, and W[1]-hard when parameterized by the committee size $k$.
    \label{thm:jrnejrfpt}
\end{restatable}
In fact, our reduction from a variant of \textsc{Multicolored Independent Set} even shows that it is hard to check the existence of a committee satisfying JR, but not $2$-EJR+.

We complement this negative result in two ways. Firstly, we present a simple polynomial-size ILP for \jrnejr in \Cref{app:ilps}. Secondly, we show that \jrnejr is in FPT when parameterized by the number of voters~$n$ by introducing equivalence classes of candidates: We say that two candidates are equivalent if they are approved by the exact same set of voters. Since there are at most 
 $2^n$ such classes, we can formulate an ILP where the number of variables is a function in $n$ by replacing candidates by their equivalence class. Using Lenstra's algorithm, which solves ILPs in FPT time with respect to the number of variables \citep{Lens83a},  we get an FPT algorithm for our problem. 

\prooflink{fptnjrnejr}
\begin{restatable}{theorem}{fptnjrnejr}\label{fptnjrnejr}
\label{th:fptnjrnejr}
    \jrnejr is in FPT when parameterized by the number of voters $n$.
\end{restatable}

\subsection{Distances Between Proportional Committees}
We further want to understand the space of proportional committees. To measure its \emph{richness} and \emph{diversity}, we compute how different proportional committees can be. For two committees $W_1$ and $W_2$, we define their distance as $d(W_1, W_2) := \lvert W_1 \setminus W_2\rvert$.
\citet{DFPS25a} prove that JR committees can be widely different and even isolated: there exists an instance with a JR-committee $W_1$ such that the closest other JR-committee $W_2$ has  distance~$d(W_1, W_2) = k-1.$
This motivates us to study the following problem to better understand the set of JR/EJR+-committees: 

\problemdefX{X-Diff-Committees}
	{Election $\mathcal{I} = (N, C ,A, k)$ and~$k^\prime \leq k.$}
	{Are there two committees~$W_1$ and~$W_2$, both satisfying axiom X, with~$ d(W_1 ,W_2) \geq k^\prime $?}

\citet{BFNK19a} showed that for any election, every candidate that is approved by at least one voter belongs to at least one JR-committee. Thus, we conclude that any election with~$m > k$ has two JR-committees~$W_1$ and~$W_2$ with~$d(W_1, W_2) \geq 1$. We show that already for $k' \geq 2$ the problem becomes NP-complete.

\prooflink{jrdiffhard}
\begin{restatable}{theorem}{jrdiffhard}
\label{th:jrdiffhard}
For each $k^\prime \geq 2$,  \textsc{JR-}\diffcomm and \textsc{EJR+-}\diffcomm are NP-complete and W[1]-hard  when parameterized by the committee size $k$.
\end{restatable}

We again complement the negative result by giving a polynomial-size ILP formulation in \Cref{app:ilps}.
Using a similar approach as for \Cref{fptnjrnejr}, we additionally show that both \textsc{JR-} and \textsc{EJR+-Diff-Committees}  are in FPT when parameterized by $n$. For \textsc{JR-Diff-Committees} we give an explicit combinatorial algorithm, while for \textsc{EJR+-Diff-Committees} we again employ Lenstra's algorithm.

\prooflink{jrdiffilp}
\begin{restatable}{theorem}{jrdiffilp}
\label{th:jrdiffilp}
    \textsc{JR-Diff-Committees} and \textsc{EJR+-Diff-Committees} are in FPT when parameterized by $n$.
\end{restatable}
\subsection{Candidate Containment}
To measure the importance of (subsets of)  candidates for proportionality, we aim to solve the following  problem:

\problemdefX{$p$-Candidates-X}
	{Election $\mathcal{I} = (N, C ,A, k)$ and set of candidates $C^\prime \subseteq C$ with $|C^\prime| = p$.}
	{Is there a committee~$W \subseteq C$ with $C^\prime \subseteq W$, satisfying the proportionality axiom X?}
Indeed, as mentioned previously, for any single candidate $c$, there always exists a committee satisfying JR containing $c$, i.e., any instance of $1$\textsc{-Candidates-JR} is a 'Yes-Instance'. For EJR+ it follows from \citet[Theorem~4.14]{DFPS25a} that $1$\textsc{-Candidates-EJR+} is also always a 'Yes-Instance'. Neither of these results transfers to pairs of candidates, for which the problem becomes computationally intractable.

\prooflink{twocanhard}
\begin{restatable}{theorem}{twocanhard}
\label{th:twocanhard}
$2$-\textsc{Candidates-JR} is NP-complete, and W[1]-hard when parameterized by the committee size~$k$.
\end{restatable}
Finally, we again prove membership in FPT, when parameterized by the number of voters. For JR, this again follows from a combinatorial algorithm, while for EJR+ we again employ Lenstra's algorithm.

\prooflink{canjrfpt}
\begin{restatable}{theorem}{canjrfpt}
\label{th:canjrfpt}
    $p$-\textsc{Candidates-JR} and $p$-\textsc{Candidates-EJR+}  are in FPT when~parameterized~by~$n$.
\end{restatable}

\subsection{Counting Committees}
To measure the restrictiveness of proportionality axioms, we will count the number of proportional committees. \citet{BFNK19a} have shown that counting the number of JR committees is $\#P$-hard, and $\#W[1]$-hard when parameterized by $k$.  We complement this result by showing that one can count the number of JR committees in FPT time in $n$. We achieve this via a dynamic programming approach, again by introducing equivalence classes of candidates.

\prooflink{countingjr}
\begin{restatable}{theorem}{countingjr}
\label{th:countingjr}
    For a given election $\mathcal{I}$, the number of JR committees $\lvert\jr(\mathcal{I})\rvert$ can be computed in FPT time in $n$.
\end{restatable}
It remains an open problem whether these results generalize to EJR+. 
Although most of the measures relevant to our experiments are computationally intractable, we still aim to compute them. For counting problems, we turn to a sampling-based approach. We present an exemplary sampling-based approximation algorithm for the counting variant of the \textsc{$1$-Candidates-JR} problem; observe that the \#P-hardness of \textsc{$1$-Candidates-JR} follows from the \#P-hardness of counting JR committees. In particular, we give a sampling bound on the number of samples needed to accurately approximate the JR-prevalence of a given candidate. The sampling complexity of this problem is polynomial in the instance size, error, and in $\nicefrac{{m \choose k}}{\lvert\jr(\mathcal{I})\rvert}$, i.e., the inverse of the fraction of JR committees in the given election.

\prooflink{proplearn}
\begin{restatable}{proposition}{proplearn}
\label{pr:proplearn}
There is an algorithm that, given an election $\mathcal{I}$, candidate $c$, and two positive rational numbers $\varepsilon$ and $\delta$, outputs a value~$\tilde{\alpha}$ such that the JR-prevalence of $c$ lies, with probability at least~$1 - \delta$, in the interval~$[\tilde{\alpha} - \varepsilon, \tilde{\alpha} + \varepsilon]$. This algorithm has an expected running time polynomial in~$|C|, |V|, \nicefrac{1}{\varepsilon}, \ln(\nicefrac{1}{\delta})$ and~$\nicefrac{{m \choose k}}{\lvert\jr(\mathcal{I})\rvert}$.
\end{restatable}
To compute an instance's JR/EJR+-fraction, a simple Monte Carlo estimation with Hoeffding's inequality leads to a similar $\varepsilon$-approximation with probability 
 at least $1-\delta$ in time polynomial in~$|C|$, $|V|$, $\nicefrac{1}{\varepsilon}$, and $\ln(\nicefrac{1}{\delta})$.

\section{Experiments} \label{sec:EA}
We present our experimental investigation. After introducing our dataset, we examine the restrictiveness of different proportionality axioms in \Cref{sec:restr} and the importance of individual candidates for proportionality in \Cref{sec:cI}.

\paragraph{Dataset.} 
We present our results on a real-world dataset in the main body, and relegate results on two synthetic datasets based on the resampling and Euclidean models to \Cref{app:exp}.
Given the general scarcity of real-world data from approval-based multiwinner elections, we rely on pabulib, the primary source of participatory budgeting elections, as our dataset \citep{FFP+23a}. Participatory budgeting generalizes multiwinner voting by assigning costs to candidates and selecting a subset that respects a total budget constraint. We include all pabulib instances with approval ballots and an average ballot size of at least four as of May 2nd, 2025. We do not apply any preprocessing, resulting in 369 instances with $n \in [69, 95899]$ and $m \in [7, 138]$. While these instances include voters' approval ballots over candidates, they do not specify a value for $k$ (since the budget constraint takes the role of $k$ in participatory budgeting). As we discuss in \Cref{sec:restr}, we set $k=\lfloor\frac{m}{2}\rfloor$.\footnote{A similar approach is taken by \citet{SFJ+22a} and \citet{FLSS23a}. The only large-scale, real-world approval-based datasets we are aware of are from two blockchains \citep{BBC+24a}, but their size (up to $10000$ candidates and $40000$ voters) is prohibitive for our experiments. Our combination of participatory budgeting data and synthetic approval-based multiwinner voting data represents the most informative and widely used approach in the literature \citep{BFJ+24a}. While the costs of a project in a participatory budgeting exercise certainly influences its approval score, the connection in Pabulib is much weaker and more heterogeneous than one might expect (see \citep{FFP+23a} Fig. B.6). Further, based on measured distances between instances, \citet{SFJ+22a} argue that Pabulib instances without costs yield realistic approval elections.}

\subsection{How Restrictive Are Proportionality Axioms?}\label{sec:restr}
We analyze the restrictiveness of JR and EJR+. To shed light on the effect of the choice of~$k$, and select a value for~$k$, we  analyze how the JR/EJR+-fraction changes with changing $k$. Subsequently, we take a closer look at the JR- and EJR+-fractions for $k = \lfloor\frac{m}{2}\rfloor$, and the average and maximum distances between JR, resp., EJR+-committees.

\paragraph{Dependency on $k$.}
We start by analyzing the JR- and EJR+-fraction for varying values of $k$ to understand $k$'s influence on the impact of proportionality and to provide guidance on how to choose values of $k$ that yield instances which are interesting from the perspective of proportionality.\footnote{For each instance and each $k \in [m]$, we estimate the JR-fraction (resp.\ EJR+-fraction) by sampling until we found $1000$ JR (resp.\ $1000$ EJR+) committees. We apply a timeout of $15$ minutes per $k$ and instance, resulting in the exclusion of $9$ instances that did not finish for all $k$ and for which all computed EJR+-fractions~exceeded~$0.95$.}

\begin{figure*}[t]
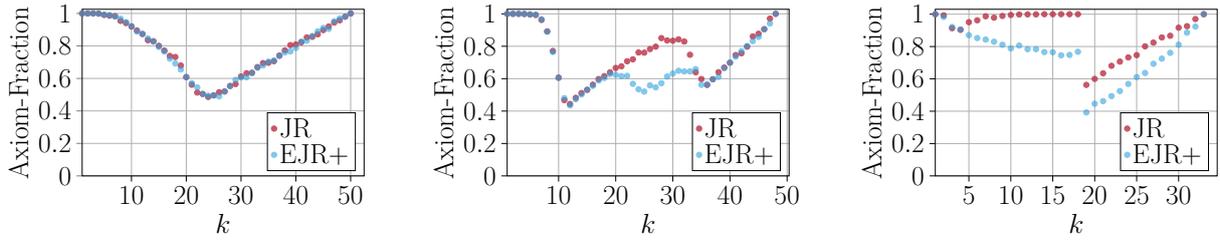

	\centering
	\begin{subfigure}[t]{0.3\textwidth}
	\centering
\begin{tikzpicture} [scale=0.5]

\definecolor{darkgray176}{RGB}{176,176,176}
\definecolor{green}{RGB}{0,128,0}
\input{tikz_figures/color-def}

\begin{axis}[
tick align=outside,
tick pos=left,
x grid style={darkgray176},
xlabel={$k$},
xmajorgrids,
xmin=1, xmax=52.45,
xtick style={color=black, font=\huge},
y grid style={darkgray176},
ylabel={Axiom-Fraction},
ymajorgrids,
ymin=0, ymax=1.02569275644142,
legend style={at={(0.97,0.03)}, anchor=south east, legend columns=1, draw=black, font=\huge},
legend cell align={left},
label style={font =\huge},
ticklabel style={color=black, font=\huge},
height =6cm,
width=9cm,
]
\addplot [draw=blue, fill=blue, mark=*, only marks, opacity=0.8]
table{%
x  y
1 1
2 1
3 1
4 0.999000999000999
5 0.991080277502478
6 0.986193293885602
7 0.983284169124877
8 0.954198473282443
9 0.940733772342427
10 0.919117647058823
11 0.895255147717099
12 0.873362445414847
13 0.835421888053467
14 0.829187396351575
15 0.800640512409928
16 0.771604938271605
17 0.738552437223043
18 0.732064421669107
19 0.679347826086957
20 0.60790273556231
21 0.561167227833894
22 0.513083632632119
23 0.504540867810293
24 0.486144871171609
25 0.496277915632754
26 0.515198351365276
27 0.51975051975052
28 0.553709856035437
29 0.565291124929339
30 0.611995104039168
31 0.632111251580278
32 0.634920634920635
33 0.669792364367046
34 0.694444444444444
35 0.696378830083565
36 0.706214689265537
37 0.740740740740741
38 0.773395204949729
39 0.805152979066023
40 0.809061488673139
41 0.822368421052632
42 0.852514919011083
43 0.856164383561644
44 0.871839581517001
45 0.897666068222621
46 0.919963201471941
47 0.943396226415094
48 0.957854406130268
49 0.980392156862745
50 1
};
\addlegendentry{JR}

\addplot [draw=green, fill=green, mark=*, only marks, opacity=0.6]
table{%
x  y
1 1
2 1
3 1
4 0.999000999000999
5 0.992063492063492
6 0.99009900990099
7 0.98135426889107
8 0.968054211035818
9 0.946969696969697
10 0.922509225092251
11 0.88809946714032
12 0.874125874125874
13 0.846023688663283
14 0.826446280991736
15 0.79428117553614
16 0.765696784073507
17 0.721500721500722
18 0.691562932226833
19 0.654878847413229
20 0.608642726719416
21 0.571102227298686
22 0.544662309368192
23 0.503778337531486
24 0.493583415597236
25 0.49480455220188
26 0.489715964740451
27 0.51975051975052
28 0.553097345132743
29 0.591016548463357
30 0.606428138265615
31 0.606796116504854
32 0.632911392405063
33 0.66577896138482
34 0.679809653297077
35 0.703729767769177
36 0.710227272727273
37 0.738007380073801
38 0.759878419452888
39 0.766283524904215
40 0.788022064617809
41 0.827129859387924
42 0.832639467110741
43 0.861326442721792
44 0.889679715302491
45 0.910746812386157
46 0.904977375565611
47 0.952380952380952
48 0.968992248062015
49 0.979431929480901
50 1
};
\addlegendentry{EJR+}
\end{axis}

\end{tikzpicture} 
 \caption{Warszawa -- Obszar~3 (2019)}
        \label{fig:k-dependency-1}
\end{subfigure}
\hfill
\begin{subfigure}[t]{0.3\textwidth}
	\centering
\begin{tikzpicture} [scale=0.5]

\definecolor{darkgray176}{RGB}{176,176,176}
\definecolor{green}{RGB}{0,128,0}
\input{tikz_figures/color-def}

\begin{axis}[
tick align=outside,
tick pos=left,
x grid style={darkgray176},
xlabel={$k$},
xmajorgrids,
xmin=1, xmax=50.35,
xtick style={color=black},
y grid style={darkgray176},
ylabel={Axiom-Fraction},
ymajorgrids,
ymin=0, ymax=1.02825141365811,
ytick style={color=black},
legend style={at={(0.97,0.03)}, anchor=south east, legend columns=1, draw=black, font=\huge},
legend cell align={left},
label style={font =\huge},
ticklabel style={color=black, font=\huge},
height =6cm,
width=9cm,
]
\addplot [draw=blue, fill=blue, mark=*, only marks, opacity=0.8]
table{%
x  y
1 1
2 1
3 1
4 0.999000999000999
5 0.99601593625498
6 0.99304865938431
7 0.962463907603465
8 0.892060660124888
9 0.771010023130301
10 0.606796116504854
11 0.467289719626168
12 0.443852640923213
13 0.481927710843373
14 0.511247443762781
15 0.530785562632696
16 0.563063063063063
17 0.597728631201435
18 0.61576354679803
19 0.638977635782748
20 0.666222518321119
21 0.676132521974307
22 0.707213578500707
23 0.719424460431655
24 0.761614623000762
25 0.760456273764259
26 0.782472613458529
27 0.797448165869219
28 0.850340136054422
29 0.836120401337793
30 0.834028356964137
31 0.8424599831508
32 0.829187396351575
33 0.747384155455904
34 0.639386189258312
35 0.597728631201435
36 0.561482313307131
37 0.596302921884317
38 0.639795265515035
39 0.668449197860963
40 0.698324022346369
41 0.745712155108128
42 0.75642965204236
43 0.797448165869219
44 0.860585197934596
45 0.87719298245614
46 0.905797101449275
47 0.970873786407767
48 1
};
\addlegendentry{JR}

\addplot [draw=green, fill=green, mark=*, only marks, opacity=0.6]
table{%
x  y
1 1
2 1
3 1
4 1
5 0.997008973080758
6 0.995024875621891
7 0.967117988394584
8 0.889679715302491
9 0.764525993883792
10 0.603864734299517
11 0.480076812289966
12 0.434971726837756
13 0.473260766682442
14 0.50251256281407
15 0.53276505061268
16 0.555864369093941
17 0.588581518540318
18 0.604229607250755
19 0.628140703517588
20 0.624219725343321
21 0.616142945163278
22 0.617283950617284
23 0.568181818181818
24 0.534188034188034
25 0.520291363163371
26 0.558659217877095
27 0.546448087431694
28 0.5720823798627
29 0.613496932515337
30 0.631711939355654
31 0.648929266709929
32 0.644329896907217
33 0.644745325596389
34 0.658761528326746
35 0.562113546936481
36 0.564334085778781
37 0.596302921884317
38 0.611246943765281
39 0.665335994677312
40 0.699300699300699
41 0.735835172921266
42 0.769822940723634
43 0.801282051282051
44 0.831946755407654
45 0.857632933104631
46 0.904159132007233
47 0.940733772342427
48 1
};
\addlegendentry{EJR+}

\end{axis}

\end{tikzpicture}
 \caption{Ursynów Wysoki Południowy (2018)}
        \label{fig:k-dependency-3}
	\end{subfigure}
	\hfill
	\begin{subfigure}[t]{0.3\textwidth}
	\centering
\begin{tikzpicture}[scale=0.5]

\definecolor{darkgray176}{RGB}{176,176,176}
\definecolor{green}{RGB}{0,128,0}
\input{tikz_figures/color-def}

\begin{axis}[
tick align=outside,
tick pos=left,
x grid style={darkgray176},
xlabel={$k$},
xmajorgrids,
xmin=1, xmax=34.6,
xtick style={color=black},
y grid style={darkgray176},
ylabel={Axiom-Fraction},
ymajorgrids,
ymin=0, ymax=1.03037676609105,
ytick style={color=black},
legend style={at={(0.97,0.03)}, anchor=south east, legend columns=1, draw=black, font = \huge},
legend cell align={left},
label style={font =\huge},
ticklabel style={color=black, font=\huge},
height =6cm,
width=9cm,
]
\addplot [draw=blue, fill=blue, mark=*, only marks, opacity=0.8]
table{%
x  y
1 1
2 0.99304865938431
3 0.91324200913242
4 0.903342366757001
5 0.950570342205323
6 0.962463907603465
7 0.987166831194472
8 0.978473581213307
9 0.989119683481701
10 0.995024875621891
11 0.998003992015968
12 0.999000999000999
13 0.999000999000999
14 1
15 1
16 1
17 1
18 1
19 0.56274620146314
20 0.599880023995201
21 0.634517766497462
22 0.68073519400953
23 0.706713780918728
24 0.732600732600733
25 0.746825989544436
26 0.801282051282051
27 0.824402308326463
28 0.856164383561644
29 0.866551126516464
30 0.916590284142988
31 0.925925925925926
32 0.969932104752667
33 1
};
\addlegendentry{JR}

\addplot [draw=green, fill=green, mark=*, only marks, opacity=0.6]
table{%
x  y
1 1
2 0.983284169124877
3 0.920810313075506
4 0.904977375565611
5 0.870322019147084
6 0.852514919011083
7 0.843170320404722
8 0.829875518672199
9 0.8110300081103
10 0.788643533123028
11 0.8058017727639
12 0.783699059561129
13 0.783699059561129
14 0.765696784073507
15 0.764525993883792
16 0.745712155108128
17 0.74794315632012
18 0.767459708365311
19 0.392464678178964
20 0.446229361892013
21 0.462320850670365
22 0.493583415597236
23 0.524658971668416
24 0.568828213879408
25 0.610873549175321
26 0.63653723742839
27 0.693481276005548
28 0.725163161711385
29 0.76103500761035
30 0.8110300081103
31 0.885739592559787
32 0.923361034164358
33 1
};
\addlegendentry{EJR+}

\end{axis}

\end{tikzpicture}
 \caption{Praga-Północ (2017)}
        \label{fig:k-dependency-4}
	\end{subfigure} 
    \caption{JR-fraction (red) and EJR+-fraction (blue) for different committee sizes $k\in [m]$ across selected pabulib instances.} 
    \label{fig:jr-ejr-dependency} 
\end{figure*}

We begin by discussing some intuition. At first glance, increasing $k$ has two opposing effects on the constraints imposed by JR (and analogously by EJR+).
First, more candidates surpass the approval threshold $\nicefrac{n}{k}$ and may thus induce JR violations.
Second, in a randomly selected committee, more voters approve at least one candidate and can therefore not be part of a JR violation.
These two effects explain the very high JR-fraction at extreme $k$: at very low $k$, no candidate meets the approval threshold $\nicefrac{n}{k}$; at very high $k$, nearly every voter approves someone in the committee.
However, there is a third, slightly more subtle effect: a candidate $c$ with approval score above $\nicefrac{n}{k}$ can still impose additional constraints as $k$ increases, since every $\nicefrac{n}{k}$-subset of $N_c$ must be ``satisfied'', and their number grows with $k$.
Our results demonstrate that the interplay of these three effects can be quite~intricate.

We present results for four  instances in \Cref{fig:jr-ejr-dependency,fig:praga_axioms}.
\Cref{fig:k-dependency-1,fig:praga_axioms} represent typical behaviors observed in our dataset, while \Cref{fig:k-dependency-3,fig:k-dependency-4} illustrate more exceptional cases.
The pattern in \Cref{fig:k-dependency-1} is relatively simple and aligns with basic intuition. \Cref{fig:praga_axioms} already exhibits more complex behavior: JR and EJR+ begin to diverge. JR becomes easy to satisfy between $k = 20$ and $k = 60$, but is restrictive before and after, including sharp increases and drops.
\Cref{fig:k-dependency-3} contains even more local extrema, and \Cref{fig:k-dependency-4} reveals a striking non-continuity, where changing $k$ by just one dramatically affects the behavior of the axiom-fractions.

One explanation for the sudden drops in the JR-fraction in \Cref{fig:praga_axioms,fig:k-dependency-3,fig:k-dependency-4} could be the first effect: new candidates surpass the $\nicefrac{n}{k}$ threshold as $k$ increases.
This explanation is at best partial. In \Cref{fig:k-dependency-4}, indeed, a new candidate passes the threshold at the jump at $k=19$.
In contrast, in \Cref{fig:k-dependency-3}, all candidates have an approval score of at least $\nicefrac{n}{23}$, i.e., all are ``relevant'' for $k \geq 23$, and yet we still observe rapid changes and a non-monotonous behavior~afterwards.

\begin{figure}[t]
	\centering
	\begin{tikzpicture}[scale=0.68]

		\definecolor{darkgray176}{RGB}{176,176,176}
		\definecolor{green}{RGB}{0,128,0}
		\definecolor{lightgray204}{RGB}{204,204,204}
		\input{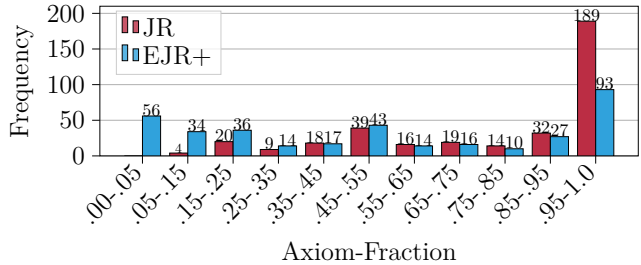}
        
		\begin{axis}[
legend cell align={left},
legend style={
  fill opacity=0.8,
  draw opacity=1,
  text opacity=1,
  at={(0.03,0.97)},
  anchor=north west,
  draw=lightgray204,
  font=\Large
},
height =4.5cm,
width=12cm,
tick align=outside,
tick pos=left,
x grid style={darkgray176},
xlabel={Axiom-Fraction},
xmin=-0.94, xmax=10.94,
xtick style={color=black},
xtick={0,1,2,3,4,5,6,7,8,9,10},,
xticklabel style={rotate=45.0,anchor=east},
xticklabels={
  .00-.05,
  .05-.15,
  .15-.25,
  .25-.35,
  .35-.45,
  .45-.55,
  .55-.65,
  .65-.75,
  .75-.85,
  .85-.95,
  .95-1.0
},
y grid style={darkgray176},
ylabel={Frequency},
ymajorgrids,
ymin=0, ymax=210,
ytick style={color=black},
tick label style={color=black, font=\Large},
label style={font =\Large},
]
\draw[draw=black,fill=blue] (axis cs:-0.4,0) rectangle (axis cs:0,0);
\addlegendimage{ybar,ybar legend,draw=black,fill=blue}
\addlegendentry{JR}

\draw[draw=black,fill=blue] (axis cs:0.6,0) rectangle (axis cs:1,4);
\draw[draw=black,fill=blue] (axis cs:1.6,0) rectangle (axis cs:2,20);
\draw[draw=black,fill=blue] (axis cs:2.6,0) rectangle (axis cs:3,9);
\draw[draw=black,fill=blue] (axis cs:3.6,0) rectangle (axis cs:4,18);
\draw[draw=black,fill=blue] (axis cs:4.6,0) rectangle (axis cs:5,39);
\draw[draw=black,fill=blue] (axis cs:5.6,0) rectangle (axis cs:6,16);
\draw[draw=black,fill=blue] (axis cs:6.6,0) rectangle (axis cs:7,19);
\draw[draw=black,fill=blue] (axis cs:7.6,0) rectangle (axis cs:8,14);
\draw[draw=black,fill=blue] (axis cs:8.6,0) rectangle (axis cs:9,32);
\draw[draw=black,fill=blue] (axis cs:9.6,0) rectangle (axis cs:10,189);
\draw[draw=black,fill=green] (axis cs:-2.77555756156289e-17,0) rectangle (axis cs:0.4,56);
\addlegendimage{ybar,ybar legend,draw=black,fill=green}
\addlegendentry{EJR+}

\draw[draw=black,fill=green] (axis cs:1,0) rectangle (axis cs:1.4,34);
\draw[draw=black,fill=green] (axis cs:2,0) rectangle (axis cs:2.4,36);
\draw[draw=black,fill=green] (axis cs:3,0) rectangle (axis cs:3.4,14);
\draw[draw=black,fill=green] (axis cs:4,0) rectangle (axis cs:4.4,17);
\draw[draw=black,fill=green] (axis cs:5,0) rectangle (axis cs:5.4,43);
\draw[draw=black,fill=green] (axis cs:6,0) rectangle (axis cs:6.4,14);
\draw[draw=black,fill=green] (axis cs:7,0) rectangle (axis cs:7.4,16);
\draw[draw=black,fill=green] (axis cs:8,0) rectangle (axis cs:8.4,10);
\draw[draw=black,fill=green] (axis cs:9,0) rectangle (axis cs:9.4,27);
\draw[draw=black,fill=green] (axis cs:10,0) rectangle (axis cs:10.4,93);
\draw (axis cs:0.2,56.5) node[
  scale=0.9,
  anchor=base,
  text=black,
  rotate=0.0
]{56};
\draw (axis cs:0.8,4.5) node[
  scale=0.7,
  anchor=base,
  text=black,
  rotate=0.0
]{4};
\draw (axis cs:1.2,34.5) node[
  scale=0.9,
  anchor=base,
  text=black,
  rotate=0.0
]{34};
\draw (axis cs:1.8,20.5) node[
  scale=0.9,
  anchor=base,
  text=black,
  rotate=0.0
]{20};
\draw (axis cs:2.2,36.5) node[
  scale=0.9,
  anchor=base,
  text=black,
  rotate=0.0
]{36};
\draw (axis cs:2.8,9.5) node[
  scale=0.9,
  anchor=base,
  text=black,
  rotate=0.0
]{9};
\draw (axis cs:3.2,14.5) node[
  scale=0.9,
  anchor=base,
  text=black,
  rotate=0.0
]{14};
\draw (axis cs:3.8,18.5) node[
  scale=0.9,
  anchor=base,
  text=black,
  rotate=0.0
]{18};
\draw (axis cs:4.2,17.5) node[
  scale=0.9,
  anchor=base,
  text=black,
  rotate=0.0
]{17};
\draw (axis cs:4.8,39.5) node[
  scale=0.9,
  anchor=base,
  text=black,
  rotate=0.0
]{39};
\draw (axis cs:5.2,43.5) node[
  scale=0.9,
  anchor=base,
  text=black,
  rotate=0.0
]{43};
\draw (axis cs:5.8,16.5) node[
  scale=0.9,
  anchor=base,
  text=black,
  rotate=0.0
]{16};
\draw (axis cs:6.2,14.5) node[
  scale=0.9,
  anchor=base,
  text=black,
  rotate=0.0
]{14};
\draw (axis cs:6.8,19.5) node[
  scale=0.9,
  anchor=base,
  text=black,
  rotate=0.0
]{19};
\draw (axis cs:7.2,16.5) node[
  scale=0.9,
  anchor=base,
  text=black,
  rotate=0.0
]{16};
\draw (axis cs:7.8,14.5) node[
  scale=0.9,
  anchor=base,
  text=black,
  rotate=0.0
]{14};
\draw (axis cs:8.2,10.5) node[
  scale=0.9,
  anchor=base,
  text=black,
  rotate=0.0
]{10};
\draw (axis cs:8.8,32.5) node[
  scale=0.9,
  anchor=base,
  text=black,
  rotate=0.0
]{32};
\draw (axis cs:9.2,27.5) node[
  scale=0.9,
  anchor=base,
  text=black,
  rotate=0.0
]{27};
\draw (axis cs:9.8,189.5) node[
  scale=0.9,
  anchor=base,
  text=black,
  rotate=0.0
]{189};
\draw (axis cs:10.2,93.5) node[
  scale=0.9,
  anchor=base,
  text=black,
  rotate=0.0
]{93};
\end{axis}
\end{tikzpicture}
        \caption{Number of pabulib instances (in total 360) with given JR-fractions (red) and EJR+-fractions (blue). }\label{freq-dist} 
\end{figure}

The key takeaway from this analysis is that the dependency of JR/EJR+-fractions on $k$ is more complex and less predictable than one might expect.
We observe sudden jumps, multiple local extrema, and differing behaviors between JR and EJR+.
This calls for caution when conducting experiments: the choice of $k$, often treated as a generic hyperparameter, can substantially influence an instance's behavior from the perspective of proportionality.

In our remaining experiments, we aim to select a value of $k$ that maximizes the number of instances that are non-trivial from the perspective of proportionality, i.e., have an EJR+-fraction below $95\%$. 
We pick $k = \lfloor \frac{m}{2} \rfloor$ , as it maximizes the number of non-trivial instances among all $k = \lfloor \frac{m}{c} \rfloor$ for~$c \in \{2, \ldots, 10\}$; see \Cref{fig:best-k-histo} in \Cref{app:exp}.
While we have seen above that the chosen $k$ can have a strong impact on the instance level, we verify the robustness of dataset-level trends. For this, we compared the distribution of JR- resp.\ EJR+-fractions for $k = \lfloor \frac{m}{2} \rfloor$ to the distributions for $k = \lfloor \frac{m}{c} \rfloor$ for~$c \in \{2, \ldots, 10\}$ and for $k$ equal to the total budget of the instance divided by the average-project cost (as an estimation for the average number of affordable projects). 
General trends turn out to be stable in both cases.

\paragraph{JR/EJR+-Fractions.}
In \Cref{freq-dist}, we show a frequency histogram for the JR- (red) and EJR+-fractions (blue) of the pabulib instances with $k = \lfloor\frac{m}{2}\rfloor$.\footnote{For every instance, we sampled committees until we reach $5000$ JR committees and return $5000$ divided by the total number of samples as the JR-fraction. We parallelized instance-wise on $128$ kernels  with a total timeout of five days, resulting in $360$/$369$ solved instances. We proceed analogously for EJR+. \label{fot}}
We observe a diverse picture: While many instances exhibit high fractions and thus confirm previous observations by \citet{BFNK19a}, and \citet{BrPe23a} that JR and EJR+ impose only extremely mild constraints on many instances, there are also numerous instances where only a small fraction of committees satisfy these axioms. This is especially pronounced for EJR+, with $56$ instances exhibiting an EJR+-fraction of at most $5\%$. For JR, this effect is slightly less pronounced but still present, as  $24$ instances have a JR-fraction of at most $25\%$. These findings suggest that proportionality notions can matter in real-world voting instances, as they can significantly reduce the space of feasible committees.

Prior work suggests that different proportionality axioms tend to coincide in practice. For instance, \citet{BBC+24a} note that ``proportionality axioms seem to lose their discriminative power in practice''. However, \Cref{freq-dist} shows that this is not generally the case, as the EJR+-fraction is often substantially lower than the JR-fraction for our chosen value of $k$.
We shed further light on this question with the help of the \jrnejr problem:\footnote{We used Gurobi with the ILP implementation for \jrnejr provided in \Cref{app:ilpjrnejr} with a timeout of $30$ minutes per instance, resulting in $24$ unsolved instances.} For $275$ pabulib instances, there exists a committee that satisfies JR but not EJR+, whereas in $70$ instances, every JR committee also satisfies EJR+, rendering the two axioms effectively equivalent.
These results draw a mixed picture: while there are cases where the distinction between proportionality axioms becomes irrelevant, there are also many instances where the ``strength'' of the considered axiom continues to matter.

\begin{figure}[t]
    \centering
    \input{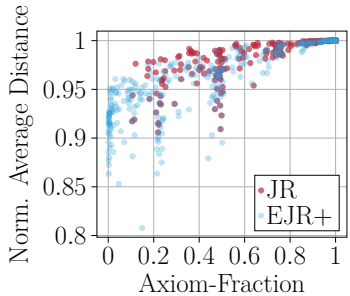}
    \caption{Each point represents one instance and one axiom (JR in red, EJR+ in blue). We plot the correlation between the axiom fraction and the normalized average distance among committees fulfilling the axiom.} 
    \label{fig:ratio-vs-distance}
\end{figure}

\paragraph{Distances Between Committees.}
After establishing that EJR+ and even JR can impose quite strong restrictions on real-world instances, the next natural question is: 
Does each of these axioms lead to a solution space 
consisting of only similar committees?
We answer this question in the negative.
For this, we start by measuring the average distance between two JR, resp.\ EJR+, committees,\footnote{We approximate these values by calculating the average distance between pairs of committees 
from the set of 5000 sampled JR, resp.\ EJR+, committees from Footnote \ref{fot}.} normalized by the expected distance between two randomly drawn committees of size~$k$.
In \Cref{fig:ratio-vs-distance}, each red (resp.\ blue) dot represents an instance, with the $x$-coordinate indicating the instance's JR-fraction (resp.\ EJR+-fraction) and the $y$-coordinate the instance's normalized average distance between two JR (resp.\ EJR+) committees.
We observe that the normalized average distance---even for EJR+---is typically above $0.9$, suggesting that proportional committees are widely distributed over the space of all committees.
Moreover, while there is some correlation between instances' normalized distance and their JR/EJR+-fraction, the general differences in terms of average distance across instances are small, further confirming the heterogeneity of proportional committees, even in cases where only a few committees fulfill proportionality.

Examining the maximum, instead of the average distance, between proportional committees paints a similar picture.\footnote{We compute the maximum distance by solving the ILP formulation for \textsc{Diff-Committees} from \Cref{ilp-dimm-comm} using Gurobi. We set a time limit of $30$ minutes for each instance, resulting in all instances being solved for JR, and $269$ instances for EJR+.} For JR,  we find that $329$ out of $369$ instances admit disjoint JR committees, while for all other instances we can find two JR committees with an overlap strictly smaller than four. For EJR+, we found disjoint committees for $189$ out of $269$ solved instances, and a maximum overlap of eight among all solved instances. Given our choice of $k = \lfloor\frac{m}{2}\rfloor$, the existence of disjoint proportional committees is quite remarkable, as it implies that one can partition the candidates into two halves in a way that both are representative. 

\begin{figure}[t]
    \centering
\begin{tikzpicture}[scale=0.7]

\definecolor{darkgray176}{RGB}{176,176,176}
\definecolor{green}{RGB}{0,128,0}
\input{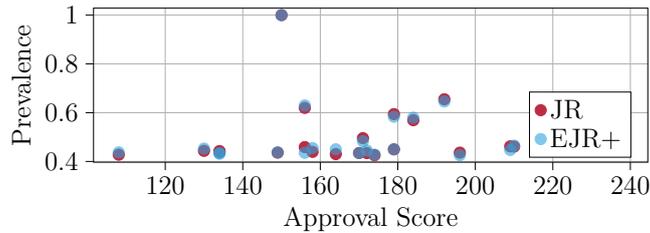}

\begin{axis}[
tick align=outside,
tick pos=left,
width =12cm,
height=4.5cm,
x grid style={darkgray176},
xlabel={Approval Score},
xtick distance =20,
xmajorgrids,
xmin=101.5, xmax=244.5,
xtick style={color=black},
y grid style={darkgray176},
ylabel={Prevalence},
ymajorgrids,
ymin=0.39688, ymax=1.02872,
ytick style={color=black},
legend style={at={(0.97,0.03)}, anchor=south east, legend columns=1, draw=black, font =\Large},
legend cell align={left},
label style={font =\Large},
ticklabel style={color=black, font=\Large},
]
\addplot [draw=blue, fill=blue, mark=*, only marks, opacity=1, mark size=3 ]
table{%
x  y
179 0.4496
170 0.4346
134 0.4426
171 0.4956
158 0.4394
130 0.4436
179 0.5942
184 0.5692
210 0.461
150 1
149 0.4362
134 0.4392
209 0.462
164 0.4302
174 0.4256
192 0.655
156 0.459
156 0.6196
172 0.4344
196 0.4362
238 0.445
108 0.4278
};
\addlegendentry{JR}

\addplot [draw=green, fill=green, mark=*, only marks, opacity=0.6, mark size=3]
table{%
x  y
179 0.4502
170 0.4342
134 0.4318
171 0.483
158 0.4548
130 0.4528
179 0.5834
184 0.5796
210 0.4646
150 1
149 0.4374
134 0.434
209 0.447
164 0.4498
174 0.428
192 0.646
156 0.4356
156 0.6296
172 0.4452
196 0.426
238 0.4498
108 0.4372
};
\addlegendentry{EJR+}

\end{axis}

\end{tikzpicture}
    \caption{Each point corresponds to one candidate in the Warszawa, Wysokie-Okoecie (2017) election and one axiom (JR in red, EJR+ in blue). We plot the candidate's approval score against its axiomatic prevalence.}
    \label{fig:approval_vs_axiom_fraction}
\end{figure}

\subsection{What Makes a Candidate Important for Proportionality?}\label{sec:cI}

We analyze three candidate importance measures: the prevalence and power index, which aim to capture importance for proportionality, and the traditional approval score (see \Cref{sub:prelims} for definitions).\footnote{To compute the prevalence and power index of a candidate, we sampled $5000$ JR committees and $5000$ EJR+ committees using rejection sampling.  For the prevalence, we take the fraction of sampled committees containing the candidate. For the power index, we count how many sampled committees containing the candidate stop to satisfy the respective axiom when the candidate is removed. 
} In this section, we only consider instances with an EJR+-fraction $\leq 95\%$, resulting in $267$ instances. We do this because when nearly all committees are proportional, there are no substantial differences in candidates' importance for proportionality, skewing the results.

\paragraph{Correlation Between Measures.}

We are interested in finding out 
\begin{enumerate*}[label=(\roman*)]
  \item how closely our two new ways to measure candidates' importance for proportionality are related, and
  \item whether they differ from the established approach of assessing candidates' merit by approval score. 
\end{enumerate*}
To this end, we compute the Pearson correlation coefficient (PCC) between each pair of measures across all candidates within one instance. 
Regarding question (a), we find a very strong correlation for EJR+, with an average PCC value of $0.99$. For JR, the average PCC is with $0.72$ lower; however, this is also partly due to the fact that, in contrast as for EJR+, our dataset contains instances with JR-ratio very close to 1: for these cases, the prevalences of candidates do not vary significantly, and therefore, small sampling errors can have quite a large impact on the PCC.
Regarding question (b), we find that the approval score is notably less correlated with our two proportionality-based measures: the average PCC between approval score and prevalence is only $0.22$ for JR and $0.65$ for EJR+.
To make this observation more tangible, we show in \Cref{fig:approval_vs_axiom_fraction} the election from Warszawa, Wysokie-Okoecie (2017), which plots candidates' approval score ($x$-axis) against their prevalence ($y$-axis).
We observe virtually no connection between approval scores and prevalence: candidates with the lowest and highest approval scores have similar prevalence, while a candidate with a medium approval score appears in all JR and EJR+ committees.
While it is intuitive that such behavior can appear in the worst case, it is noteworthy that it also occurs in real-world instances.
Practically, a candidate's strength and merit to be selected is often assessed by their approval score. However, we conclude that from the perspective of proportionality, this approach falls short and can be actively misleading.

\paragraph{Measures and Proportional Voting Rules.}
Lastly, we investigate whether proportional voting rules select candidates that are important for proportionality, i.e., have a high prevalence or power index. We focus on the popular voting rule Method of Equal Shares (MES) with seq-Phragmén completion, which always returns an EJR+ committee \citep{PeSk20a} and defer results for seq-PAV and seq-Phragmén, which exhibit similar behavior, to the appendix.

\begin{figure}[t]
	\centering
\begin{tikzpicture}[scale =0.7]

\definecolor{darkgray176}{RGB}{176,176,176}
\definecolor{green}{RGB}{0,128,0}
\definecolor{lightgray204}{RGB}{204,204,204}
\definecolor{orange}{RGB}{255,165,0}
\input{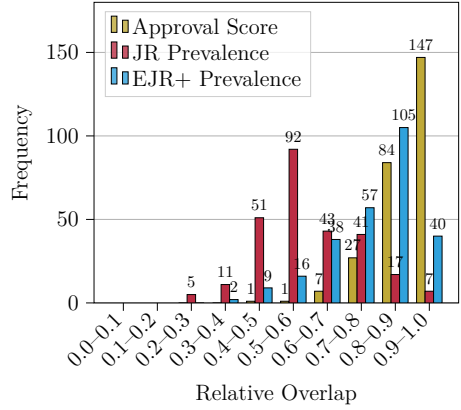}

\begin{axis}[
legend cell align={left},
legend style={
  fill opacity=0.8,
  draw opacity=1,
  text opacity=1,
  at={(0.03,0.97)},
  anchor=north west,
  draw=lightgray204,
  font = \large
},
tick align=outside,
tick pos=left,
x grid style={darkgray176},
xlabel={Relative Overlap},
xmin=-0.8625, xmax=9.8625,
xtick style={color=black},
xtick={0,1,2,3,4,5,6,7,8,9},
xticklabel style={rotate=45.0,anchor=east},
xticklabels={0.0–0.1,0.1–0.2,0.2–0.3,0.3–0.4,0.4–0.5,0.5–0.6,0.6–0.7,0.7–0.8,0.8–0.9,0.9–1.0},
y grid style={darkgray176},
ylabel={Frequency},
ymajorgrids,
ymin=0, ymax=180,
ytick style={color=black},
tick label style={color=black, font=\large},
label style={font =\large},
]
\addlegendimage{ybar,ybar legend,draw=black,fill=orange}
\addlegendentry{Approval Score}

\draw[draw=black,fill=orange] (axis cs:1.625,0) rectangle (axis cs:1.875,0);
\draw[draw=black,fill=orange] (axis cs:2.625,0) rectangle (axis cs:2.875,0);
\draw[draw=black,fill=orange] (axis cs:3.625,0) rectangle (axis cs:3.875,1);
\draw[draw=black,fill=orange] (axis cs:4.625,0) rectangle (axis cs:4.875,1);
\draw[draw=black,fill=orange] (axis cs:5.625,0) rectangle (axis cs:5.875,7);
\draw[draw=black,fill=orange] (axis cs:6.625,0) rectangle (axis cs:6.875,27);
\draw[draw=black,fill=orange] (axis cs:7.625,0) rectangle (axis cs:7.875,84);
\draw[draw=black,fill=orange] (axis cs:8.625,0) rectangle (axis cs:8.875,147);

\addlegendimage{ybar,ybar legend,draw=black,fill=blue}
\addlegendentry{JR Prevalence}

\draw[draw=black,fill=blue] (axis cs:1.875,0) rectangle (axis cs:2.125,5);
\draw[draw=black,fill=blue] (axis cs:2.875,0) rectangle (axis cs:3.125,11);
\draw[draw=black,fill=blue] (axis cs:3.875,0) rectangle (axis cs:4.125,51);
\draw[draw=black,fill=blue] (axis cs:4.875,0) rectangle (axis cs:5.125,92);
\draw[draw=black,fill=blue] (axis cs:5.875,0) rectangle (axis cs:6.125,43);
\draw[draw=black,fill=blue] (axis cs:6.875,0) rectangle (axis cs:7.125,41);
\draw[draw=black,fill=blue] (axis cs:7.875,0) rectangle (axis cs:8.125,17);
\draw[draw=black,fill=blue] (axis cs:8.875,0) rectangle (axis cs:9.125,7);

\addlegendimage{ybar,ybar legend,draw=black,fill=green}
\addlegendentry{EJR+ Prevalence}

\draw[draw=black,fill=green] (axis cs:2.125,0) rectangle (axis cs:2.375,0);
\draw[draw=black,fill=green] (axis cs:3.125,0) rectangle (axis cs:3.375,2);
\draw[draw=black,fill=green] (axis cs:4.125,0) rectangle (axis cs:4.375,9);
\draw[draw=black,fill=green] (axis cs:5.125,0) rectangle (axis cs:5.375,16);
\draw[draw=black,fill=green] (axis cs:6.125,0) rectangle (axis cs:6.375,38);
\draw[draw=black,fill=green] (axis cs:7.125,0) rectangle (axis cs:7.375,57);
\draw[draw=black,fill=green] (axis cs:8.125,0) rectangle (axis cs:8.375,105);
\draw[draw=black,fill=green] (axis cs:9.125,0) rectangle (axis cs:9.375,40);

\draw (axis cs:3.75,2) node[
  scale=0.8,
  anchor=south,
  text=black,
  rotate=0.0
]{1};
\draw (axis cs:4.75,2) node[
  scale=0.8,
  anchor=south,
  text=black,
  rotate=0.0
]{1};
\draw (axis cs:5.75,8) node[
  scale=0.8,
  anchor=south,
  text=black,
  rotate=0.0
]{7};
\draw (axis cs:6.75,28) node[
  scale=0.8,
  anchor=south,
  text=black,
  rotate=0.0
]{27};
\draw (axis cs:7.75,85) node[
  scale=0.8,
  anchor=south,
  text=black,
  rotate=0.0
]{84};
\draw (axis cs:8.75,148) node[
  scale=0.8,
  anchor=south,
  text=black,
  rotate=0.0
]{147};
\draw (axis cs:2,6) node[
  scale=0.8,
  anchor=south,
  text=black,
  rotate=0.0
]{5};
\draw (axis cs:3,12) node[
  scale=0.8,
  anchor=south,
  text=black,
  rotate=0.0
]{11};
\draw (axis cs:4,52) node[
  scale=0.8,
  anchor=south,
  text=black,
  rotate=0.0
]{51};
\draw (axis cs:5,93) node[
  scale=0.8,
  anchor=south,
  text=black,
  rotate=0.0
]{92};
\draw (axis cs:6,44) node[
  scale=0.8,
  anchor=south,
  text=black,
  rotate=0.0
]{43};
\draw (axis cs:7,42) node[
  scale=0.8,
  anchor=south,
  text=black,
  rotate=0.0
]{41};
\draw (axis cs:8,18) node[
  scale=0.8,
  anchor=south,
  text=black,
  rotate=0.0
]{17};
\draw (axis cs:9,8) node[
  scale=0.8,
  anchor=south,
  text=black,
  rotate=0.0
]{7};
\draw (axis cs:3.25,3) node[
  scale=0.8,
  anchor=south,
  text=black,
  rotate=0.0
]{2};
\draw (axis cs:4.25,10) node[
  scale=0.8,
  anchor=south,
  text=black,
  rotate=0.0
]{9};
\draw (axis cs:5.25,17) node[
  scale=0.8,
  anchor=south,
  text=black,
  rotate=0.0
]{16};
\draw (axis cs:6.25,39) node[
  scale=0.8,
  anchor=south,
  text=black,
  rotate=0.0
]{38};
\draw (axis cs:7.25,58) node[
  scale=0.8,
  anchor=south,
  text=black,
  rotate=0.0
]{57};
\draw (axis cs:8.25,106) node[
  scale=0.8,
  anchor=south,
  text=black,
  rotate=0.0
]{105};
\draw (axis cs:9.25,41) node[
  scale=0.8,
  anchor=south,
  text=black,
  rotate=0.0
]{40};
\end{axis}

\end{tikzpicture}
 \caption{Number of pabulib instances in which the committee returned by MES has a given normalized overlap with the set of $k$ candidates with the highest approval score (olive),  JR-prevalence (red), and EJR+-prevalence (blue).}
        \label{fig:histogram-mes}
\end{figure}

To answer this question, we measure the overlap between the $k$ candidates selected by MES and the $k$ candidates with the highest approval score, JR-prevalence, and EJR+-prevalence. Notably, the committee of the $k$ candidates with the highest JR-prevalence (resp.\ EJR+-prevalence) satisfies JR (resp.\ EJR+) in all our instances.
\Cref{fig:histogram-mes} presents a histogram showing how often different normalized overlaps occur across instances. We find that MES frequently selects over $90\%$ of the $k$ candidates with the highest approval score. In contrast, the overlap with the $k$ candidates with the highest prevalence is notably lower, particularly for JR-prevalence. While this does not question the proportionality of the outcome computed by MES, it highlights that the rule tends to select a particular type of proportional committee: committees composed of generally popular candidates.

\section{Conclusion}
We conducted an algorithmic and experimental investigation into the impact of proportionality on feasible committees and candidate importance in approval-based multiwinner elections. We believe our findings can inform future discussions around proportionality in several ways:
\begin{enumerate*}[label=(\roman*)]
\item Our observation that proportionality axioms can significantly restrict the outcome space on real-world voting data highlights their practical relevance and can help guide discussions on the adoption of proportional voting rules.
\item Our experiments highlighting the influence of the committee size~$k$ can inform the design and interpretation of future experiments, emphasizing the need for greater care when selecting $k$.
\item Our prevalence and power index measures offer a new perspective on candidate importance under proportionality and can contribute to the transparency of election outcomes.
\end{enumerate*}
A promising direction for future research is to move beyond fixed voting rules and explore rule-free approaches that optimize various objectives subject to proportionality constraints. This could be relevant for certain applications and contribute to a more nuanced understanding of the space of proportional outcomes.

\section*{Acknowledgements}
This work was partially supported by the Singapore Ministry of Education under grant number MOE-T2EP20221-0001. This research was (partially) funded by the HPI Research School on Foundations of AI (FAI).
\newpage

\bibliography{bibfile, abb, algo}
\newpage

\setcounter{secnumdepth}{2}
\renewcommand{\thesubsection}{\thesection.\arabic{subsection}}
\clearpage
\appendix

\section{Integer Linear Programming Formulations}
\label{app:ilps}
\subsection{ILP for \jrnejr}
\label{app:ilpjrnejr}
Recall that in this problem we ask whether there is a committee of size~$k$ that satisfies JR but not EJR+. \citet{BFNK19a} already introduced a set of constraints that are sufficient for satisfying JR, and it is easy to find a set of constraints that are sufficient for violating EJR+. We combine these two sets of constraints, yielding an almost straightforward ILP formulation for our problem: For each candidate $c_j \in C$ we add a binary variable $x_j$ indicating whether $c_j$ is on the committee. Further, for each voter $i \in N$ we add a variable $y_i$ to track the approval-score of voter $i$. In addition to these two types of variables, we introduce an integer variable~$\ell$ indicating the size of the selected cohesive group violating EJR+. Further, we add for each voter~$i$ a binary variable~$v_i$ with the intended meaning that~$v_i =1$ if voter~$i$ is in the group that violates EJR+; and for each candidate~$c_j$ a binary variable~$u_j$, with the intended meaning that~$u_j=1$ if candidate $c_j$ witnesses the EJR+ violation. To ensure that our variables comply to their intended meanings, we add the following constraints:
\begin{align}
    \sum_{c_j \in C} x_j &= k \\
    y_i &\geq x_j & \quad \forall i \in N, c_j \in A_i \\
    y_i &\leq \sum_{c_j \in A_i} x_j & \quad \forall i \in N \\
    \sum_{i \in N: c_j \in A_i} (1 - y_i) & < \nicefrac{n}{k} & \quad \forall c_j \in C \\
    \sum_{i \in N} v_i & \geq \nicefrac{\ell n}{k}\\
    \sum_{c_j \in C} u_j & = 1\\
    u_j + x_j &\le 1 &\quad \forall c_j \in C\\
    u_j &\leq [c_j \in A_i] + (1 - v_i) & \quad \forall i \in N, c_j \in C \\
    \sum_{c_j \in C} [c_j \in A_i] \cdot x_j & < \ell + m(1 - v_i) & \quad \forall i \in N \\
    \ell  & \in \mathbb{N}_{\geq 2} \\
    y_i, v_i, x_j, u_j &\in \{0, 1\}  &\quad \forall i \in N, c_j \in C
\end{align}

We give an informal explanation why the above constraints yield indeed an ILP formulation for the problem \textsc{JR-not-EJR+}: The first constraint ensures that the selected committee is of size~$k$. The next three types of constraints ensure that the committee satisfies JR: On the one hand the constraints of type (2) make sure that~$y_i =1$ holds if voter~$i$ approves some candidate in the committee, and on the other hand the constraints of type (3) make sure that~$y_i=0$ holds if voter~$i$ does not approve any candidate in the committee. Constraints of type (4) ensure that for all candidates~$c_j$ there are strictly less than~$\nicefrac{n}{k}$ voters approving~$c_j$ who do not approve any candidate in the committee, that is, JR is satisfied. 

However, the next four types of constraints make sure EJR+ is violated: Constraints of type (5) and (6) ensure that we select at least~$\nicefrac{\ell n}{k}$ voters as well as a single candidate (to form the violating cohesive group). Constraints of type (7) give us that the ``witness'' candidate $c_j$ with $u_j = 1$ cannot be selected by the committee.
Constraints of type (8) ensure that if a candidate~$c_j$ is selected (i.e.~$u_j=1$), then it is approved by all selected voters, i.e.,~by each voter~$i$ with~$v_i = 1$. Note that if~$v_i = 0$ or~$u_j=0$, then the corresponding constraint is trivially satisfied. To ensure that the selected voters and candidates really form  group violating EJR+, we introduce the last type of constraints: For a selected voter,  i.e., for whom $v_i =1$ holds, the constraint ensures that it approves strictly less than~$\ell$ candidates in the committee. We finish by observing that for a non-selected voter this constraint is again trivially satisfied.
\subsection{ILP for \textsc{Diff-Committees}} \label{ilp-dimm-comm}
In this problem we want to maximize the distance that two JR-committees~$W_1$ and~$W_2$ can have in a given instance, i.e.,\ we search for two committees that share a minimum number of candidates. To this end, we introduce for each candidate~$c_j$ two binary variables~$x_j$ and~$a_j$ with the intended meanings that~$x_j=1$ if~$c_j \in W_1$, and~$a_j=1$ if~$c_j \in W_2$. Analogously, we introduce two binary variables~$v_i$ and~$b_i$ for each voter~$i$ with the intended meaning that~$y_i=1$ if voter~$i$ approves at least one candidate in~$W_1$, and~$b_i=1$ if voter~$i$ approves at least one candidate in~$W_2$. In addition, we introduce for each candidate~$c_j$ a new binary variable~$z_j$ with the intended meaning that~$z_j =1$ if $c_j \in W_1 \setminus W_2$.

\begin{align}
	\text{maximize } \sum_{c_j \in C} z_j \\
    \sum_{c_j \in C} x_j &= k \\
    \sum_{c_j \in C} a_j &= k \\
    y_i &\geq x_j & \quad \forall i \in N, c_j \in A_i \\
    b_i &\geq a_j & \quad \forall i \in N, c_j \in A_i \\
    y_i &\leq \sum_{c_j \in A_i} x_j & \quad \forall i \in N \\
    b_i &\leq \sum_{c_j \in A_i} a_j & \quad \forall i \in N \\
    \sum_{i \in N: c_j \in A_i} (1 - y_i) &< \nicefrac{n}{k} & \quad \forall c_j \in C \\
    \sum_{i \in N: c_j \in A_i} (1 - b_i) &< \nicefrac{n}{k} & \quad \forall c_j \in C \\
    z_j &\leq x_j & \quad \forall c_j \in C \\
    z_j &\leq 1 - a_j & \quad \forall c_j \in C \\
    y_i, b_i, x_j, a_j, z_j &\in \{0, 1\}  &\quad \forall i \in N, c_j \in C   
\end{align}

With the explanations given in~\Cref{app:ilpjrnejr}, we observe that the constraints of type (18)-(25) make sure that both selected committees are of size~$k$ and satisfy JR. Constraints (26) and (27) ensure that~$z_j=1$ only if~$x_j = 1$ and~$a_j =0$ hold. By maximizing the number of~$z_j$s with~$z_j=1$, we maximize the distance two JR-committees can have and thus, we obtain an ILP formulation for our problem.

We can also adapt the ILP for JR to work for EJR+. Now, instead of using the variables $y_i$ and $b_i$ to indicate whether voter $i$ is represented or not, we introduce $k$ variables $y_{i, 1}, \dots, y_{i,k}$ with variable $y_{i, \ell}$ indicating that voter $i$ approves less than $\ell$ candidates in the first committee. 
\begin{align}
	\text{maximize } \sum_{c_j \in C} z_j \\
    \sum_{c_j \in C} x_j &= k \\
    \sum_{c_j \in C} a_j &= k \\
    \ell \cdot y_{i, \ell} &\leq \sum_{c_j \in A_i} x_j & \quad \forall i \in N, \ell \in [k] \\
    \ell \cdot b_{i, \ell} &\leq \sum_{c_j \in A_i} a_j & \quad \forall i \in N, \ell \in [k] \\
    \sum_{i \in N: c_j \in A_i} (1 - y_{i, \ell})  &< \ell \cdot \nicefrac{n}{k} + n\cdot x_j & \quad \forall c_j \in C, \ell \in [k] \\
    \sum_{i \in N: c_j \in A_i} (1 - b_{i, \ell}) &< \ell \cdot \nicefrac{n}{k} + n\cdot a_j & \quad \forall c_j \in C, \ell \in [k] \\
    z_j &\leq x_j & \quad \forall c_j \in C \\
    z_j &\leq 1 - a_j & \quad \forall c_j \in C \\
    y_{i, \ell}, b_{i, \ell}, x_j, a_j, z_j &\in \{0, 1\}  &\quad \forall i \in N, c_j \in C, \ell \in [k]   
\end{align}
To see that the ILP is correct, we argue why both selected committees satisfy EJR+: Assume that, without loss of generality, the committee $W = \{c_j \in C \colon x_j = 1\}$ does not satisfy EJR+. Thus, there exists some $c_{j'} \notin W$ together with $N' \subseteq N_{j'}$ and $\ell \in [k]$ such that $\lvert N'\rvert \ge \ell \frac{n}{k}$ and $\lvert A_i \cap W\rvert < \ell$ for all $i \in N'$. As $\sum_{c_j \in A_i} x_j < \ell$ for all $i \in N'$ we know that $y_{i, \ell} = 0$. Hence, for candidate $c_{j'}$ inequality (29) is violated, as the left-hand side is at least $\ell \frac{n}{k}$. On the other hand, it is easy to see any pair of EJR+ committees corresponds to a feasible variable assignment for the above ILP.

\subsection{ILP for \textsc{$p$-Candidates-JR}}
In this problem we are given in addition to the election instance a subset~$C_p \subseteq C$ and ask whether there is a JR-committee of size~$k$ containing~$C_p$. Our ILP formulation below consists of the same type of variables and constraints as we used in the previous section to satisfy JR. The only additional constraints, i.e.,\ fixing the candidate-variables~$x_j =1$ for~$c_j \in C_p$ in (38), result by the requirement~$C_p \subseteq C$. 

\begin{align}
    \sum_{c_j \in C} x_j &= k \\
    y_i &\geq x_j & \quad \forall i \in N, c_j \in A_i \\
    y_i & \leq  \sum_{c_j \in A_i} x_j & \quad \forall i \in N \\
    \sum_{i \in N : c_j \in A_i} (1 - y_i) & < \nicefrac{n}{k} & \quad \forall c_j \in C \\
    x_1, \ldots, x_p &= 1 & \quad \forall c_j \in C_p \\
    y_i, x_j &\in \{0, 1\} & \quad \forall i \in N, c_j \in C
\end{align}

\subsection{ILP for \textsc{$p$-Candidates-EJR+}}

Finally, to create the ILP for EJR+, we again add additional binary variables $y_{i, \ell}$ for each $i \in N$ and $\ell \in [k]$ with $y_{i, \ell} = 1$ only if voter $i$ approves at least $\ell$ candidates in the outcome. 

\begin{align}
    \sum_{c_j \in C} x_j &= k \\
    \ell \cdot y_{i, \ell} & \le  \sum_{c_j \in A_i} x_j & \quad \forall i \in N, \ell \in [k] \\
    \sum_{i \in N : c_j \in A_i} (1 - y_{i, \ell}) & < \ell \nicefrac{n}{k} + n\cdot x_j & \quad \forall c_j \in C, \ell \in [k] \\
    x_1, \ldots, x_p &= 1 & \quad \forall c_j \in C_p \\
    y_{i, \ell}, x_j &\in \{0, 1\} & \quad \forall i \in N, c_j \in C, \ell \in [k]
\end{align}

\section{Missing Proofs}
\subsection{Missing Proofs for the \jrnejr Problem}
\restatehere{jrnotejr}

We prove our theorem in two steps: First, we sketch a hardness proof to show that \mis\ remains NP-hard on 2-regular graphs. Second, we reduce \mis on 2-regular graphs to \jrnejr.

\problemdefX{Multicolor Independent Set}
	{A graph~$G=(V,E)$ with a vertex partition~$V= V_1 \cup \cdots \cup V_h$.}
	{Is there an independent set~$V^\prime \subseteq V$ such that~$|V^\prime \cap V_i| = 1$ for all~$i \in [h]$?}

We say that a vertex~$v$ has color~$i$ if~$v \in V_i$, and that a set~$V^\prime\subseteq V$ is~$h$-colorful if~$|V^\prime \cap V_i| \geq 1$ for all~$i \in [h]$.

\begin{lemma}
\mis\ is NP-hard, also when restricted to 2-regular graphs with an even number of vertices and each color appearing exactly three times.
\end{lemma}

We can show the lemma by reducing from a restricted version of~\textsc{3-SAT} which is NP-hard by \cite{Berman}. 

\problemdefX{(3,B2)-Satisfiability}
	{A formula~$\phi = C_1 \land \ldots \land C_m$ in CNF where each clause~$C_i$ contains exactly three literals and each literal appears exactly twice.}
	{Is there a satisfying assignment for~$\phi$?}
	
\begin{proof}
We construct an instance~$G$ of \mis\ with~$h:=m$ colors. For every clause~$C_i$, we insert~$|C_i|$ vertices of color~$i$ (corresponding to its literals). We insert an edge between two vertices~$v$ and~$v^\prime$, corresponding to literals~$\ell$ and~$\ell^\prime$ in the given~\restrsat\ instance, if~$\ell = \neg \ell^\prime$. 
One observes that a satisfying assignment for~$\phi$ corresponds to an $h$-colorful independent set in~$G$ by selecting for each clause the vertex corresponding to a literal which ensures that the clause is satisfied. For the reverse direction, assume we are given an~$h$-colorful independent set~$V^\prime$, we assign truth values to the variables such that the literals in~$\phi$, corresponding to the vertices in~$V^\prime$, evaluate to `true'. 
Due to the assumption that every literal appears exactly twice, the number of vertices is even and, by construction, is 2-regular. Thus, the lemma follows. 
\end{proof}

\begin{proof}[Proof of Theorem]
Given a 2-regular graph~$G=(V,E)$ together with a vertex partition~$V= V_1 \cup \cdots \cup V_h$, with~$|V_i|=3$ for all~$i \in [h]$ and~$|V|$ even, we construct an instance of \jrntwojr\ as follows:

\begin{itemize}
\item For every color~$i \in [h]$,  we add $ \frac{|E|}{2}$ \textit{color-$i$-voters} \footnote{Since~$G$ is 2-regular, we know that~$|V| = |E|$ holds and thus, the number of edges must also be even.}  and for every vertex~$u \in V_i$, we add a \textit{vertex-candidate}~$c_u$ which is approved by all color-$i$-voters.
\item For each edge~$e=\{u, u^\prime\} \in E$, we add an \textit{edge-candidate}~$c_e$ and an \textit{edge-voter}~$v_e$ who approves~$c_e$. We let edge-voter~$v_e$ also approve the two vertex-candidates~$c_u$ and~$c_{u^\prime}$. 
\item We add two \textit{dummy-candidates} $d_1$ and~$d_2$ which are approved by all edge-voters.
\item We add $ \frac{|E|^2}{2} - |E| \cdot h - |E|$ \textit{filling-voters} as well as \textit{filling-candidates}. Each filling-candidate is approved by exactly one filling-voter. 
\end{itemize}

Note that~$|E| = 3h$ holds, yielding
\begin{align*}
  \frac{|E|^2}{2} - |E| \cdot h - |E| 
 =  |E| \cdot \Big(\frac{|E|}{2} - h -1\Big) 
 =  |E| \cdot \Big(\frac{3h}{2} - h -1 \Big) \geq 0.
\end{align*}
Thus, the above number of filling-voters and filling-candidates is well-defined.  Last, let~$n$ be the total number of voters and set~$k := |E| - h $. To identify the cohesive groups in our election, we observe that
\begin{align*}
\frac{n}{k}  = \frac{h \cdot \frac{|E|}{2} + |E| + \frac{|E| ^2}{2} - h \cdot |E| - |E|}{|E| - h }  = \frac{\frac{|E|^2}{2} - \frac{|E| \cdot h}{2}}{|E| -h} = \frac{|E|}{2}.
\end{align*}

We find cohesive groups of four types in our constructed instance:
\begingroup
\renewcommand{\theenumi}{\arabic{enumi}}
\renewcommand{\labelenumi}{\theenumi.}
\begin{enumerate}
    \item Each color~$i \in [h]$ induces one 1-cohesive group of color-$i$-voters who approve all vertex-candidates which derive from vertices in~$V_i$.
    \item Each vertex-candidate~$c_u$, derived from a vertex~$u \in V_i$, induces a 1-cohesive group among the color-$i$-voters and the two edge-voters approving~$c_u$. Clearly, also every subset of these voters of size at least~$\frac{|E|}{2}$ forms a 1-cohesive group.
    \item Every~$\frac{|E|}{2}$-large subset of edge-voters induces a 1-cohesive group (they all approve the two dummy-candidates).
    \item All edge-voters together induce one 2-cohesive group.
\end{enumerate}
\endgroup

$(\Rightarrow)$ To show correctness of our reduction, we assume first that there is an $h$-colorful independent set~$V^\prime =\{v_1, \ldots , v_h\}$ in~$G$ and construct a JR-committee~$W$ which does not satisfy 2-JR. To this end, we add the~$h$ vertex-candidates, corresponding to the vertices in~$V^\prime$, to~$W$ and afterwards~$k-h = |E| -2h $ edge-candidates approved by edge-voters who do not approve any selected vertex-candidate (note that since~$G$ is 2-regular, at most~$2h$ edge-voters can approve~$h$ vertex-candidates).

We claim that~$W$ fulfills JR, but violates 2-JR. Clearly, for each~$i \in [h]$ all color-$i$-voters approve one candidate in the committee since~$V^\prime$ is an~$h$-colorful set. Note that this also implies that no cohesive group of the second type (in our enumeration above) violates JR. Further, all edge-voters approve exactly one candidate: since~$V^\prime$ is an independent set, $2h$ distinct edge-voters approve vertex-candidates in~$W$, and by construction, the other~$|E| - 2h$ edge-voters approve an edge-candidate. As there are no further 1-cohesive groups, the committee~$W$ satisfies JR. However, all edge-voters together induce a 2-cohesive group and no edge-voter approves two candidates in~$W$, implying that~$W$ violates 2-JR.
\bigskip

$(\Leftarrow)$ For the reverse direction assume there exists a committee~$W$ satisfying JR, but not 2-JR. To give the 1-cohesive groups within color-$i$-voters enough representation for every~$i \in [h]$, there has to be at least one vertex-candidate~$c_u$ in~$W$ with~$u \in V_i$. Since there is only one 2-cohesive group, namely the one which consists of all edge-voters, and~$W$ violates 2-JR---but satisfies JR---we conclude that no edge-voter is allowed to approve two candidates in~$W$. In particular, no edge-voter approves two vertex-candidates in~$W$. Therefore, no two vertices $u, u^\prime$ with~$c_u, c_{u^\prime} \in W$ share an edge in~$G$. Thus, the vertices~$\{u \in V \vert c_u \in W\}$ form an~$h$-colorful independent set in~$G$ (theoretically, the independent set could be strictly larger than~$h$ but this does not cause any problems).

Since \mis\ is W[1]-hard when parameterized by the number of colors~\citep{Pie03}, and by construction we have~$k = 2h$, that is, the committee-size is a function only depending on the number of colors in our reduction, we may even conclude W[1]-hardness with respect to~$k$.

\end{proof}

\restatehere{fptnjrnejr}
\begin{proof}

We claim that the given ILP is a formulation for \jrnejr. Note that in comparison to the ILP formulation given in \Cref{app:ilps}, we do not introduce variables for each candidate, but for each equivalence class, allowing us to bound the number of variables by a function in~$n$, as argued above. Accordingly, we define by~$\mathcal{Q}_i$ the set of equivalence classes approved by voter~$i$.

To argue why the given formulation is indeed correct, let us first explain the intended meanings of the constraints that ensure that a JR-committee is induced by them, that are, constraints of type (45)-(49), and the therein appearing variables: For each equivalence class~$[c_j] \in \mathcal{Q}$, we introduce a binary variable~$x_j$ and an integer variable~$z_j$ with the intended meaning that~$x_j =1$ if in the committee there is at least one candidate of the equivalence class~$[c_j]$, and~$x_j =0$ otherwise; and~$z_j$ denoting how many candidates of the equivalence class are in the committee. Note that the inequalities of type (48) ensure that~$z_j =0$ if $x_j =0$, and take a value in~$\{1, \ldots, |[c_j]|\}$, otherwise. 
Further, we adapt the equality that ensures that we select exactly~$k$ candidates for the committee by summing over the variables~$z_j$, instead of~$x_j$, for~$[c_j] \in \mathcal{Q}$, resulting in equality (45).
Observe that, as before, we have a variable~$y_i$ for each voter~$i \in [n]$, indicating whether voter~$i$ approves at least one candidate in the committee. Moreover, observe that the remaining constraints, that are, constraints involving the variables~$y_i$, are taken over from the regular ILP, with the small modification that we introduce constraints for each equivalence class, instead of for each candidate, and accordingly sum over equivalence classes, instead of candidates, if applicable.  

Second, let us consider the constraints that ensure EJR+ is violated, namely constraints of type (49)-(54), and the therein involved variables: As before, we have a variable~$v_i$ for each voter~$i \in [n]$, indicating whether a voter is part of the violating cohesive group, or not. Further, we have again the integer variable~$\ell$, reflecting the size of the cohesive group. 
Next, we have, similar to above, a binary variable~$t_j$ and an integer variable~$u_j$ for each equivalence class~$[c_j] \in \mathcal{Q}$: The variable~$u_j$ denotes whether the equivalence class~$[c_j]$ is present in the violating cohesive group, and variable~$t_j$ denotes how many representatives of that equivalence class are selected. Similar to above, we have constraints of type (36) to ensure~$t_j =0$, if~$u_j=0$; and~$t_j \in [1, \ldots, |[c_j]|]$, otherwise.  These constraints only slightly  modified from the corresponding constraints in the standard ILP, in the way that we sum over, and introduce constraints for, equivalence classes, instead of candidates. In addition, we have to exchange the variables~$x_j$ by~$z_j$ to count the total number of candidates a voter approves.

\begin{align}
    \sum_{[c_j] \in \mathcal{Q}} z_j &= k \\
    y_i &\geq x_j & \quad \forall i \in N, [c_j] \in \mathcal{Q}_i \\
    y_i &\leq \sum_{[c_j] \in \mathcal{Q}_i} x_j & \quad \forall i \in N \\
    x_j &\leq z_j \leq |[c_j]| \cdot x_j & \quad \forall [c_j] \in \mathcal{Q}\\
    \sum_{i \in N: [c_j] \in \mathcal{Q}_i} (1 - y_i) & < \nicefrac{n}{k} & \quad \forall [c_j] \in \mathcal{Q} \\
    \sum_{i \in N} v_i & \geq \nicefrac{\ell n}{k}\\
    \sum_{[c_j] \in \mathcal{Q}} t_j &\geq 1\\
    u_j &\leq t_j \leq |[c_j]| \cdot u_j & \quad \forall [c_j] \in \mathcal{Q}\\
    u_j &\leq [[c_j] \in \mathcal{Q}_i] + (1 - v_i) & \quad \forall i \in N, [c_j] \in \mathcal{Q} \\
    \sum_{c_j \in C} [[c_j] \in \mathcal{Q}_i] \cdot z_j & < \ell + m(1 - v_i) & \quad \forall i \in N \\
    \ell  & \in \mathbb{N}_{\geq 2} \\
    z_j, t_j &\in \mathbb{N} &\quad \forall  [c_j] \in \mathcal{Q}\\
    y_i, v_i, x_j, u_j &\in \{0, 1\}  &\quad \forall i \in N,  [c_j] \in \mathcal{Q}
\end{align}

\end{proof}
\subsection{Missing proofs for the \textsc{X-Diff-Committees} Problem}

\restatehere{jrdiffhard}
\begin{proof}
To show our theorem, we reduce from \textsc{X3C} which is known to be NP-hard by~\cite{Karp2010}.

\problemdefX{X3C}
	{A set~$X=\{x_1, x_2, \ldots, x_{3q}\}$ and a collection~$\cal{S}$ of 3-element subsets of~$X$.}
	{Does~$\cal{S}$ contain an exact cover of~$X$, i.e., a subcollection~$\cal{S}^\prime \subseteq \cal{S}$ such that every element of~$X$ occurs in exactly one set of~$\cal{S}^\prime$?}

Given an instance~$(X, \cal{S})$ of \textsc{X3C}, we construct an instance of \textsc{Diff-Comm} with~$n:=q^4$ voters and set~$k:=q$. For each~$p \in [q-1]$ we add~$q^3 +3q -1$ voters, and a candidate~$c_p$ who is approved by all of these voters. We call the voters \textit{block-$p$-voters}, and the corresponding candidate \textit{block-$p$-candidate}. Next, we add for each~$S=\{x_i, x_j, x_\ell\} \in \cal{S}$ a candidate~$c_S$ (we call them \textit{set-candidates}) who is, for all~$p \in [q-1]$, approved by the~$i^{th}, j^{th}$ and~$\ell^{th}$ block-$p$-voter. Last, we add another candidate~$c_q$ and~$q^3-3q^2+4q-1$ voters who jointly approve~$c_q$.

To identify the cohesive groups in our election, let us first verify that we have indeed~$q^4$ voters:
\begin{align*}
n &= (q-1) \cdot (q^3 + 3q -1) + q^3-3q^2+4q-1 \\
&= q^4 +3q^2 -q -q^3 -3q +1 + q^3-3q^2+4q-1 = q^4
\end{align*}

Hence, a 1-cohesive group consists of at least~$\nicefrac{n}{k} =q^3$ voters, yielding that for all~$p \in [q-1]$ the candidate~$c_p$ induces a 1-cohesive group of size~$q^3 + 3q -1$, but that no other candidate is approved by at least~$q^3$ voters. It follows that the~$q-1$ block-candidates form together with any of the remaining ones a JR-committee.

We claim that if $(X, \cal{S})$ has an exact cover~$\cal{S}^\prime$, then our created election instance has two JR-committees~$W_1, W_2$ with~$d(W_1, W_2) \geq 2$; and otherwise each pair~$(W_1, W_2)$ of JR-committees has distance at most one. 

$(\Rightarrow)$ To show the correctness of our reduction, we transform an exact cover~$\cal{S}^\prime$ of~$(X, \cal{S})$ into a JR-committee~$W_2$ with~$\{c_1,\ldots,c_{q-1}\} \cap W_2 = \emptyset$. To this end, we put all candidates corresponding to sets in~$\cal{S}^\prime$ into~$W_2$ and show that~$W_2$ is indeed a JR-committee: Since~$\cal{S}^\prime$ is an exact cover of~$X$ and all sets in~$\cal{S}$ are of size three, we conclude~$|\mathcal{S}^\prime| = q$, and hence,~$|W_2|=q$ holds, as well. As the subcollection~$\cal{S}^\prime$ covers all~$3q$ elements of~$X$, also~$3q$ voters of each 1-cohesive group approve a candidate in~$W_2$. That is, in each induced 1-cohesive group only~$q^3-1$ voters do not approve any candidate in the committee. Hence, we have shown that our committee~$W_2$ satisfies JR.

$(\Leftarrow)$ For the reverse direction we assume there are two JR-committees~$W_1, W_2$ with~$d(W_1, W_2) \geq 2$. We claim that in this case one of the two committees, say~$W_2$, has to consist of~$q$ set-candidates. For now, we assume the correctness of the previous statement, and postpone its proof to the end. Remember that if a committee~$W_2$ in our constructed election satisfies JR, then at least~$3q$ voters of each 1-cohesive group have to approve a candidate in~$W_2$. Due to symmetry of the instance, it is sufficient to consider only the possibilities of selecting~$q$ set-candidates such that~$3q$ block-1-voters approve someone in~$W_2$: Each set-candidate is approved by exactly three block-1-voters, that means, the~$q$ set-candidates in~$W_2$ need to be approved by disjoint sets of block-1-voters. By construction, these candidates yield a 1-to-1 correspondence to a partition of~$X$ into~$q$ disjoint subsets.

It remains to show that if two JR-committees $W_1$ and~$W_2$ exist with~$d(W_1, W_2) \geq 2$, then one of them, say~$W_2$, has to consist of~$q$ set-candidates. To this end we observe that if there exist two JR-committees~$W_1$ and~$W_2$ with distance at least two, then not both of them can contain all~$q-1$ block-candidates. Due to symmetry, we may assume, without loss of generality,~$c_1 \notin W_2$. But the only way we can build a committee not containing $c_1$ such that at least~$3q$ block-1-voters approve someone in the committee is by selecting `enough' set-candidates: Since each set-candidate is approved by exactly three block-1-voters,~$W_2$ has to contain~$q$ set-candidates. As a committee consists of exactly~$q$ candidates, the claim follows. 
\end{proof}

\restatehere{jrdiffilp}
\begin{proof}
We claim that~\Cref{alg2} decides correctly whether a given instance has two JR-committees with distance at least~$k^\prime$, and its running time lies in~$\mathcal{O}(2^{2^n} \cdot nm )$.

\begin{algorithm}
\caption{}\label{alg2}
\textbf{Input:} An instance~$(N, C, k, k^\prime)$ of \diffcomm
\begin{algorithmic}[1]
\State Construct the quotient~$\mathcal{Q}= C / \sim$
\State{$d_{max}, d \gets 0$}
\State{$L \gets \emptyset$}
\For{$X \in 2^{Q}$ with~$|X| \leq k$}
	\If{$X$ satisfies JR}
		\State{add~$X$ to~$L$}
	\EndIf
\EndFor
\For{$\{X_1, X_2\} \in \binom{L}{2}$}
	\State{$k_c \gets  |X_1 \cap X_2|$}
	\For{$[c] \in X_1 \cap X_2$}
		\If{$|[c]| > 1$}
			 \State {$k_c \gets k_c-1$}
		\EndIf
	\EndFor
	\If{$m - |X_1 \cup X_2| \geq 2k - |X_1| - |X_2|$}
			\State{$d \gets k - k_c$}
		\Else 
			\State{$d \gets 2k-m$}
	\EndIf
	\If{$d > d_{max}$}
		\State{$d_{max} \gets d$}
	\EndIf
\EndFor
\If{$k^\prime \leq d_{max}$}
	\State{\Return Yes}
\Else
	\State{\Return No}
\EndIf
	
\end{algorithmic}
\end{algorithm}

Let us first describe the algorithm: By iterating over all subsets of~$\mathcal{Q}$ in the first for-loop, we can interpret~$L$ as a list of all justifying groups, up to equivalence, that do not contain two equivalent candidates. Subsequently, we consider the sets in~$L$ as subsets of~$C$ by selecting a representative of each equivalence class to show explicitly how to construct the corresponding committees.

In the second for-loop, we iterate over all pairs~$\{X_1, X_2\} \in \binom{L}{2}$ and look at their shared candidates. If~$c \in X_1 \cap X_2$ is equivalent to some candidate~$c^\prime$, then we replace~$c$ in~$X_2$ by~$c^\prime$, allowing us to reduce~$k_c$, namely the number of shared candidates, by one. To construct two JR-committees containing the justifying group~$X_1$ resp.~$X_2$, while maximizing their distance, we would like to complement the sets~$X_1$ and~$X_2$ by disjoint sets of candidates (that are also disjoint with~$X_1 \cup X_2$, of course). To this end, observe that if~$m - |X_1 \cup X_2| \geq 2k - |X_1| - |X_2|$ holds, then, we may choose arbitrary~$2k - |X_1| - |X_2|$ candidates from~$ C \setminus (X_1 \cup X_2)$, add~$k - |X_1|$ of them to~$X_1$, and the remaining ones to~$X_2$. Hence, we have~$d(X_1, X_2) = k-k_c$.

Otherwise, there are not enough candidates to complement~$X_1$ and~$X_2$ with by disjoint sets, that is,
\begin{align*}
 m - |X_1 \cup X_2| & < 2k - |X_1| - |X_2| \\
\Leftrightarrow  m + k_c & < 2k.
\end{align*}

In this case, we first fill the remaining seats with candidates from~$C \setminus (X_1 \cup X_2)$ until none non-chosen candidate is leftover. Afterwards, we fill the remaining seats in~$X_1$ with candidates from~$X_2$ and vice versa. Thus,~$d(X_1, X_2) = 2k-m$. 

In the last step, we check whether the distance between~$X_1$ and~$X_2$ is greater than the current best one, and if so, update~$d_{max}$.

Let us argue now why our algorithm (implicitly) finds two JR-committees of maximum distance. Therefore, assume by contradiction, there are two JR-committees~$W_1^\prime$ and~$W_2^\prime$ with~$d(W_1^\prime, W_2^\prime) > d_{max}$. We delete candidates from~$W_1^\prime$ and~$W_2^\prime$ such that each equivalence class is represented by at most one candidate, and call the thereby resulting sets~$W_1$ resp.~$W_2$.
Note that at some point in the iteration process we considered a pair~$X_1, X_2$ such that~$X_1 \sim W_1$ and~$X_2 \sim W_2$ hold, and that, after having replaced equivalent candidates that appear in~$X_1$ and~$X_2$, we have~$|X_1 \cap X_2| \leq  |W_1 \cap W_2|$. In the next step of the algorithm we checked whether there are enough candidates left to fill the remaining spots, i.e., candidates that neither appear in~$X_1$, nor in~$X_2$, such that~$X_1$ and~$X_2$ can be complemented by disjoint sets. If this applied, then we complemented~$X_1$ and~$X_2$ in this way, resulting in, say,~$X_1^\prime$ resp.~$X_2^\prime$. However, putting everything together, we obtain
\begin{align*}
d_{max} \geq d(X_1^\prime, X_2^\prime) = k - k_c = k - |X_1 \cap X_2| \geq  k - |W_1 \cap W_2| \geq d(W_1^\prime, W_2^\prime),
\end{align*}
contracting our assumption.  In the other case, i.e., we do not have enough candidates to proceed as above, we constructed two JR-committees, say again~$X_1^\prime$ and~$X_2^\prime$, with distance~$2k-m$. However, observe that there is no way an instance with~$m < 2k$ candidates can produce two JR-committees with strictly greater distance. Therefore, $d_{max} \geq d(X_1^\prime, X_2^\prime) \geq d(W_1^\prime, W_2^\prime)$ holds, yielding again a contradiction to our assumption.

Last, let us study the running time of our algorithm: Iterating over all subsets (of size at most~$k$) of~$\mathcal{Q}$, as well as iterating over all pairs of such subsets, has a running time in~$\mathcal{O}(2^{2^n})$. Verifying JR has a running time of~$\mathcal{O}(n \cdot m)$, and together with the observation that the lines 10-23 can be implemented in~$\mathcal{O}(m)$, we obtain a running time lying in~$\mathcal{O}(2^{2^n} \cdot nm)$.

For EJR+, we again adapt our ILP from earlier to use equivalence classes. However, now we cannot use $x_j$ and $a_j$ as indicator variables for whether or not a single candidate from this class is left over (and thus the voters approving this candidate need to be fulfilled according to EJR+). Instead we add auxiliary binary variables $\hat{x_j}$ (and $\hat{a_j}$) with $\hat{x_j}=1$ if and only if $x_j = \lvert [c_j]\rvert$ holds, ensured by constraints (63) and (64). To see this, first observe that if $x_j = \lvert [c_j]\rvert$, then the constraints postulate that $0 \le (1-\hat{x_j}) \le 0$. On the other hand, if $x_j < \lvert c_j\rvert$ holds, then constraint (64) requires that $-(1 - \hat{x_j})$ is negative (and thus $\hat{x_j} = 0$).

\begin{align}
	\text{maximize } \sum_{[c_j] \in \mathcal{Q}} z_j \\
    \sum_{[c_j] \in \mathcal{Q}} x_j &= k \\
    \sum_{[c_j] \in \mathcal{Q}} a_j &= k \\
    \ell \cdot y_{i, \ell} &\leq \sum_{[c_j] \in A_i} x_j & \quad \forall i \in N, \ell \in [k] \\
    \ell \cdot b_{i, \ell} &\leq \sum_{[c_j] \in A_i} a_j & \quad \forall i \in N, \ell \in [k] \\
     \lvert [c_j]\rvert(1 - \hat{x_j}) &\ge x_j - \lvert [c_j]\rvert & \quad \forall [c_j] \in \mathcal{Q} \\
      -\lvert [c_j]\rvert(1 - \hat{x_j}) &\le x_j - \lvert [c_j]\rvert & \quad \forall [c_j] \in \mathcal{Q} \\
      \lvert [c_j]\rvert(1 - \hat{a_j}) &\ge a_j - \lvert [c_j]\rvert & \quad \forall [c_j] \in \mathcal{Q} \\
      -\lvert [c_j]\rvert(1 - \hat{a_j}) &\le a_j - \lvert [c_j]\rvert & \quad \forall [c_j] \in \mathcal{Q} \\
    \sum_{i \in N: [c_j] \in A_i} (1 - y_{i, \ell})  &< \ell \cdot \nicefrac{n}{k} + n\cdot \hat{x_j} & \quad \forall [c_j] \in \mathcal{Q}, \ell \in [k] \\
    \sum_{i \in N: [c_j] \in A_i} (1 - b_{i, \ell}) &< \ell \cdot \nicefrac{n}{k} + n\cdot \hat{a_j} & \quad \forall [c_j] \in \mathcal{Q}, \ell \in [k] \\
    z_j &\leq x_j & \quad \forall [c_j] \in \mathcal{Q} \\
    z_j &\leq \lvert [c_j]\rvert - a_j & \quad \forall [c_j] \in \mathcal{Q} \\
    x_j,  a_j,  z_j &\in \mathbb{N}  &\quad \forall i \in N, [c_j] \in \mathcal{Q}, \ell \in [k]   \\
    y_{i, \ell}, b_{i, \ell}, \hat{x_j}, \hat{a_j} &\in \{0, 1\}  &\quad \forall i \in N, [c_j] \in \mathcal{Q}, \ell \in [k]   
\end{align}
\end{proof}

\subsection{Missing proofs for the \textsc{$p$-Candidate-X} Problem}
\newcommand{\twocandprop}{\textsc{2-Candidates-JR}}
\newcommand{\smallcom}{\textsc{Small-Prop-Com}}
\newcommand{\countjr}{\textsc{\#JR-committees}}

\restatehere{twocanhard}
\begin{proof}
Given a graph~$G=(V,E)$ with a vertex partition~$V= V_1 \cup \ldots \cup V_h$. Without loss of generality, we assume~$|V_i| = q$ for all~$i \in [h]$, for some~$q \in \mathbb{N}$. To construct an instance of \twocandprop\ we introduce the following voters:

\begin{itemize}
\item For each vertex~$u \in V(G)$, we add a \textit{vertex-voter} $v_u$.
\item We add~$q-4$ \textit{filling-voters}.
\item We add~$q-2h-1$ \textit{dummy-voters}.
\end{itemize}

Next, we add the following candidates:

\begin{itemize}
\item For each color~$i \in [h]$ and each vertex~$u \in V_i$, we introduce a \textit{vertex-candidate}~$c_u$, approved by all vertex-voters corresponding to the vertices~$V_i \setminus \{u\}$.
\item For each edge~$e = \{u, u^\prime\}$, we introduce an \textit{edge-candidate}~$c_e$, approved by~$v_u, v_{u^\prime}$ and all filling-voters.
\item We add two \textit{dummy-candidates} $d_1$ and~$d_2$, approved by all dummy-voters.
\end{itemize}

We say that vertex-voter~$v_u$, resp.\ vertex-candidate~$c_u$, is of color~$i$ if~$u \in V_i$. Further, we say that edge-candidate~$c_e$ with~$e = \{u, u^\prime\}$ has color~$i$ if~$u \in V_i$ or~$u^\prime \in V_i$. Let~$n$ be the total number of voters and set~$k := h+2$. To identify the cohesive groups in our election we observe that 

\begin{align*}
\frac{n}{k} & = \frac{h \cdot q + q -4 + q -2h -1}{h+2} = q - \frac{2h+5}{h+2}
\end{align*}

holds, yielding~$ q-3 < \frac{n}{k} < q-2$. As a consequence, we observe 1-cohesive groups of three types in our constructed instance:
\begingroup
\renewcommand{\theenumi}{\arabic{enumi}}
\renewcommand{\labelenumi}{\theenumi.}
\begin{enumerate}
\item Each vertex-candidate~$c_u$ induces one 1-cohesive group of size~$q-1$.
\item Each vertex-candidate~$c_u$ induces~$(q-1)$ 1-cohesive groups of size~$q-2$ (by excluding exactly one vertex-voter in the above 1-cohesive group each time).
\item Each edge-candidate~$c_e$ with~$e = \{u, u^\prime\}$ induces one 1-cohesive group consisting of~$v_u, v_{u^\prime}$ and the filling-voters (this group has~$q-2$ voters).
\end{enumerate}
\endgroup

$(\Rightarrow)$ To show correctness of our reduction, we transform an~$h$-colorful independent set~$V^\prime =\{u_1, \ldots , u_h\}$, with~$u_i \in V_i$, into a JR-committee~$W$ of size~$h+2$ with~$d_1, d_2 \in W$. In particular, we claim that the~$h$ vertex-candidates, corresponding to the vertices in~$V^\prime$, form together with the two dummy candidates a JR-committee. For each~$i \in [h]$, every vertex-candidate of color~$i$ is approved by all vertex-voters of color~$i$ but one, thus, by~$q-1$ vertex-voters. Clearly, each cohesive group among vertex-voters of color~$i$ (of type 1 and 2) contains at least one voter approving~$c_{u_i}$. It remains to argue that also the 1-cohesive groups induced by edge-candidates are represented. Hence, by contradiction, assume there is a 1-cohesive group induced by some edge~$e = \{u, u^\prime\}$ violating JR. By definition, the cohesive group consists of the filling voters, and the two vertex-voters~$v_u$ and~$v_{u^\prime}$. By construction, the vertex-voter~$v_u$ approves all vertex-candidates of color~$i$ but~$c_u$, and the vertex-voter~$v_{u^\prime}$ approves all vertex-candidates of color~$i$ but~$c_{u^\prime}$. Thus, both,~$c_u$ and~$c_{u^\prime}$, lie in~$W$. However, this means that~$u$ and~$u^\prime$ are in~$V^\prime$, contradicting the assumption that~$V^\prime$ is an independent set.

$(\Leftarrow)$ For the reverse direction assume there is a JR committee~$W$ of size~$h+2$ including~$d_1$ and~$d_2$. We claim that if no 1-cohesive group among the vertex-voters of some color~$i$ violates JR for a given committee~$W^\prime$, then~$W^\prime$ includes a vertex-candidate of color~$i$, or at least three edge-candidates having color~$i$. We omit proving the claim, and instead refer the reader to the hardness proof for \smallcom, containing the proof of the claim (which still goes through for our modified election)~\citep{BFNK19a}.

We now show that~$W$ includes a vertex-candidate of each color. To this end, assume that~$W$ consists of~$j$ edge-candidates and~$h-j$ vertex-candidates. By the above claim we know that we need to have~$\frac{3j}{2} + h -j +2$ candidates in~$W$, implying~$j=0$. Thus, we know that~$W$ consists of~$h$ vertex-candidates and the two dummy-candidates. Since~$W$ satisfies JR, it contains a vertex-candidate of each color. Further, these candidates also need to ensure that no 1-cohesive group induced by an edge violates JR. Thus, no two vertex-candidates, derived from vertices that share an edge in~$G$, are allowed to be in~$W$ at the same time. We conclude that the vertex set~$\{u \mid c_u \in W\}$ forms an~$h$-colorful independent set in~$G$. 

Note that we may apply the same argument that we used for the previous W[1]-hardness conclusion: Since the committee-size is again a function only depending on the number of colors in our reduction, and \mis\ is W[1]-hard when parameterized by the number of colors, we may again conclude W[1]-hardness with respect to~$k$.
\end{proof}

Note that our election does not contain a group of voters of size at least~$\ell \frac{n}{k}$ jointly approving a candidate for any~$\ell \geq 2$. Hence, EJR and EJR+ coincide with JR in our constructed election, resulting in hardness of finding an EJR, resp. EJR+, committee containing two given candidates.
\restatehere{canjrfpt}
\begin{proof}
For JR we give a simple, self-contained proof, while for EJR+ we again use Lenstra's algorithm.

\paragraph{JR.} We again first construct the quotient~$\mathcal{Q} = \nicefrac{C}{\sim}$.  If we obtain at most~$k-p$ equivalence classes, we choose a representative of each equivalence class and check whether they satisfy, together with~$C^\prime$, the axiom JR. In case they do, we found a justifying group and  add arbitrary~$k - p -| \mathcal{Q} | $ candidates to it to form a JR-committee. Otherwise, we answer the question negatively. 

In case~$ |\mathcal{Q}| \geq k - p$ holds, we iterate over all size~$k-p$ subsets of~$\mathcal{Q}$, and check for each subset whether it satisfies, together with~$C^\prime$, the axiom JR. If one of them does, we answer the question positively, otherwise negatively.

To complete the proof, let us look at the running time: Clearly, the partitioning can be done in~$\mathcal{O}(n m)$. As we have~$\binom{2^n}{k-p} \leq 2^{2^n}$ subsets of~$\mathcal{Q}$ of size-$(k-p)$, and checking whether a committee satisfies JR has a running time of~$\mathcal{O}(n m)$, the total running time of the above procedure lies in~$\mathcal{O}(2^{2^n} \cdot n m)$.

\paragraph{EJR+.}
To construct the ILP, we assume that $p([c_j])$ is the number of candidates that should at least be chosen from the equivalence class $[c_j]$. We again follow the previous constructions and replace candidates by their respective equivalence classes, leading to an ILP with size FPT in $n$.
\begin{align}
    \sum_{c_j \in \mathcal{Q}} x_j &= k \\
    \ell \cdot y_{i, \ell} & \le  \sum_{[c_j] \in A_i} x_j & \quad \forall i \in N, \ell \in [k] \\
    \lvert [c_j]\rvert(1 - \hat{x_j}) &\ge x_j - \lvert [c_j]\rvert & \quad \forall [c_j] \in \mathcal{Q} \\
      -\lvert [c_j]\rvert(1 - \hat{x_j}) &\le x_j - \lvert [c_j]\rvert & \quad \forall [c_j] \in \mathcal{Q} \\
    \sum_{i \in N : c_j \in A_i} (1 - y_{i, \ell}) & < \ell \nicefrac{n}{k} + n\cdot \hat{x_j} & \quad \forall [c_j] \in \mathcal{Q}, \ell \in [k] \\
    x_j &\ge p([c_j]) & \quad \forall [c_j] \in \mathcal{Q} \\
    x_j &\in \mathbb{N} & \quad \forall [c_j] \in \mathcal{Q}\\    
    y_{i, \ell} & \quad \forall i \in N, \ell \in [k]
\end{align}
\end{proof}

\subsection{Counting JR-Committees}
\restatehere{countingjr}
In the following, we give an algorithm in which we use the same idea as before: We iterate over all subsets of the quotient~$\mathcal{Q}= \nicefrac{C}{\sim}$ that have size at most~$k$, check for each one whether it satisfies JR, and if so, count the number of possibilities we can build a committee of size~$k$, using only elements from the corresponding equivalence classes, but such that each equivalence class has at least one representative in the committee. Then, summing up all possibilities yields the total number of JR-committees.

\begin{proof}
We claim that~\Cref{alg3} proves our theorem.

\begin{algorithm}
\caption{}\label{alg3}
\textbf{Input:} An instance~$(C, V, k)$ of \countjr
\begin{algorithmic}[1]
\State Construct the quotient~$\mathcal{Q}= C / \sim$
\State{$counter \gets 0$}
\For{$j \in 1, \ldots, k$}
	\For{$ X \in \binom{\mathcal{Q}}{j}$}
		\If{$X$ satisfies JR}
			\For{$[c_j] \in X$}
				\State{$a_j \gets |[c_j]|$}
			\EndFor 
			\State{$T[j^\prime][k^\prime] \gets 0$ for all~$(j^\prime, k^\prime) \in [j] \times [k]$} 
			\For{$j^\prime \leq j$}
				\For{$k^\prime < j^\prime$}
					\State{$T[j^\prime][k^\prime] \gets 0$}
				\EndFor
				\State{$T[j^\prime][j^\prime] \gets \prod_{i =1}^{j^\prime} a_i$}
			\EndFor
			\For{$k^\prime  \leq k$}
				\State{$T[1][k^\prime] \gets \binom{a_1}{k^\prime}$}
			\EndFor
			\For{$1 < k^\prime \leq k$}
				\For{$j^\prime \leq \min \{k^\prime -1, j\}$}
					\State{$ T[j^\prime][k^\prime]= \sum_{i =1} ^{\min \{k^\prime - j^\prime +1, a_{j^\prime}\}} \binom{a_{j^\prime}}{i} \cdot T[j^\prime -1][k^\prime -i] $}
				\EndFor
			\EndFor
			\State{$counter = counter + T[j][k]$}
		\EndIf
	\EndFor
\EndFor
\State{\Return $counter$}	
\end{algorithmic}
\end{algorithm}

In the algorithm, we first build the quotient~$\mathcal{Q}$ and introduce an integer variable, initialized with zero, to count the total number of JR-committees. Next, we iterate for all~$j \in [k]$ over all subsets~$\{[c_1], \ldots, [c_j]\}$ of~$\mathcal{Q}$, that is, we iterate over all possible sets of equivalence classes of size at most~$k$. For each such subset, we check if it satisfies JR. If it does, we would like to compute the number of committees that only consist of candidates of the selected equivalence classes, but also contain at least one representative of each one (to avoid double-counting), and increase the counter accordingly. Let us define this number by~$b([c_1], \ldots, [c_j])$. Assuming we know how to compute these numbers, it is easy to see that in the end the counter gives us the total number of JR-committees since we simply iterate over a decomposition of the set of JR-committees. 

By using a dynamic programming approach, we can calculate each number~$b([c_1], \ldots, [c_j])$ in~$\mathcal{O}(k^3)$: Consider a subset of size~$j$ of~$\mathcal{Q}$ that satisfies JR. For each such subset, we set up a 2-dimensional table of size~$j \times k$, with the intended meaning that for~$k^\prime \in [k]$ the entry in~$T[j^\prime][k^\prime]$ states the number of sets of size~$k^\prime$,  only consisting of candidates of the first~$j^\prime$ equivalence classes~$[c_1], \ldots, [c_{j^\prime}]$, but again such that each equivalence class is represented by at least one representative.

We start by initializing the `easy' entries of the table: Clearly, if~$k^\prime < j^\prime$ holds, then we set~$T[j^\prime][k^\prime]=0$ as we have strictly less spots than equivalence classes that have to appear in the set. Moreover, if~$j^\prime = k^\prime$, then we have exactly one spot for each equivalence class, and thus, set~$T[j^\prime][j^\prime] = \prod_{i =1}^{j^\prime} a_i$, with~$a_i$ being the number of candidates in the equivalence class~$[c_i]$. Further, for~$j^\prime =1$, we set~$T[1][k^\prime] = \binom{a_1}{k^\prime}$. Now, we may compute the remaining entries dynamically. To this end, observe that for a fixed~$k^\prime$, we may assume that the rows~$T[i]$ for all~$i < k^\prime$ are already filled as we compute the entries~$T[j^\prime][k^\prime]$ for~$j^\prime < k^\prime$ successively for increasing~$k^\prime$.
Let us now argue how we compute an arbitrary entry~$T[j^\prime][k^\prime]$ for~$1 < j^\prime < k^\prime$: For~$i \in [\min\{k^\prime - j^\prime + 1, a_{j^\prime}\}]$, we sum up the number of possibilities of choosing~$i$ candidates from~$[c_{j^\prime}]$  and the remaining ones from the first~$j^\prime-1$ equivalence classes~$[x_1], \ldots, [x_{j^\prime-1}]$, that is, exactly~$\binom{a_{j^\prime}}{i} \cdot T[j^\prime -1][k^\prime -i]$. Thus, in the end, we have $T[j][k]= b([c_1], \ldots, [c_j])$.

We claim that the running time of the above algorithm lies in~$\mathcal{O}(2^{2^n} \cdot (nm + k^3))$: In the beginning, we iterate over all subsets of~$\mathcal{Q}$ of size at most~$k$, which can be upper-bounded by~$2^{2^n}$.  As argued before, checking whether a subset~$X$ satisfies JR, can be done in~$\mathcal{O}(mn)$, and filling the 2-dimensional table~$T$ of the dynamic program, associated with~$X$, has a cubic running time in~$k$.
\end{proof}

\begin{observation}
Note that \Cref{alg3} can be easily adapted to also solve \textsc{\#$p$-Candidates-JR}, the counting problem associated to \textsc{$p$-Candidates-JR}, by iterating over all subsets of~$\mathcal{Q}_p:=\nicefrac{C \setminus \{c_1, \ldots, c_p\}}{\sim}$ of size at most~$k-p$. 
\end{observation}

\begin{corollary}
\textsc{\#$p$-Candidates-JR} lies in FPT, when parameterized by~$n$.
\end{corollary}
\restatehere{proplearn}
\begin{proof}
With~$r:=\frac{\ln( \nicefrac{2}{ \delta})}{2 \varepsilon^2}$, we claim that~\Cref{alg1} proves our proposition.

\begin{algorithm}
\caption{}\label{alg1}
\textbf{Input:} An instance~$(C, V, k, C^\prime)$ of \textsc{$p$-Candidates-JR}
\begin{algorithmic}
\State $X, X^\prime \gets 0$
\While{$X < r$}
	\State choose a subset~$W \subseteq C$ with~$|W| =k$ uniformly at random 
	\If{W satisfies JR}
		\State{$X = X +1$}
		\If{$C^\prime \subseteq W$}
			\State{$X^\prime = X^\prime +1$}
		\EndIf
	\EndIf
\EndWhile
\State{\Return $\nicefrac{X'}{X}$}
\end{algorithmic}
\end{algorithm}

We first argue why our algorithm is correct. To this end, consider the~$i^{th}$ time the algorithm generates a committee~$W$ that satisfies JR, that is, the committee~$W$ is not rejected. For simplicity, we call this iteration the \textit{$i^{th}$-iteration} in the following. We view the sampling of the JR-committees as a Bernoulli trial which is successful if and only if~$W \in A_p$. Let~$X_i^\prime$ be the random variable with $X_i^\prime = 1$ if we choose a committee from~$A_p$ in the~$i^{th}$-iteration, and~$X_i^\prime=0$ otherwise. Clearly,~$E[X_i^\prime]= \alpha$ and our algorithm computes the random variable~$X^\prime = \sum_{i=1}^r X_i^\prime$. It is easy to see that~$X^\prime$ has a binomial distribution of~$r$ trials with probability~$\alpha$ success, and hence,~$X^\prime \sim B(r, \alpha)$. It is known that~$\nicefrac{X^\prime}{r}$ is an unbiased, maximum likelihood estimator for~$\alpha$, and in particular~$E[X^\prime]= \alpha \cdot r$. To show that our algorithm achieves the desired accuracy and confidence, we have to show that the number of samples, that is, the number of JR-committees we generate, is sufficient. By Hoeffding’s inequality, we have
\begin{align*}
P\Big( \Bigl| \frac{X^\prime}{r} - \frac{E(X^\prime)}{r} \Bigr| \geq \varepsilon\Big) \leq 2 e^{-2r^2 \varepsilon^2}.
\end{align*}
Substituting~$r$ by~$\frac{\ln ( \nicefrac{2}{\delta})}{2 \varepsilon^2}$ yields indeed
\begin{align*}
P\Big(\Bigl| \frac{X^\prime}{r} - \frac{E(X^\prime)}{r} \Bigr| \geq \varepsilon\Big) \leq \delta.
\end{align*}

We claim that the above algorithm has an expected running time of~$\mathcal{O}(\nicefrac{r}{\beta} \cdot |C| \cdot |V|)$. Clearly, we can choose a random committee of cardinality~$k$ from~$C$ in time~$\mathcal{O}(|C|)$. Further, checking whether a given committee satisfies JR can be done in time $\mathcal{O}(|C| \cdot |V|)$. Since we only increase the variable~$X$ if the generated subset satisfies JR, and reject the committee otherwise, the expected number of committees we have to generate until we reach our desired number of JR-committees is~$\nicefrac{r}{\beta}$. By definition of~$r$, this yields the above claimed running time. 
\end{proof}

\section{Appendix for Experiments} \label{app:exp}
\paragraph{Additional Datasets.}
To generate our synthetic datasets, we consider the resampling model and a variant of the Euclidean model, and sample elections with $n=150$, $m=50$, and $k=10$. For $p, \phi \in [0,1 ]$, in the $(p, \phi)$-resampling model, we generate a uniformly at random sampled central ballot $A^*$ containing  $\lfloor p \cdot m \rfloor$ candidates.
To sample a ballot $A$, we start with $A:=A^*$. Subsequently, for each candidate $c \in C$, with probability $1 - \phi$ we do not touch $c$, i.e., we leave the membership of $c$ in the set $A$ unchanged. Otherwise, we ``resample'' $c$, i.e., $c$ gets approved independently with probability $p$ by each voter, and disapproved otherwise.
In our $(r,d)$-Euclidean model, voters and candidates correspond to randomly drawn points in a $d$-dimensional Euclidean space. Each voter approves all candidates who lie within Euclidean distance $r$ from the voter. In our case, the Euclidean space is the $d$-dimensional uniform cube.
To generate the elections, we use the models \texttt{resampling} and \texttt{euclidean\_vcr} in the prefsampling library \citep{BFJ+24a}. For the resampling model, we generate, for each parameter combination with~$p \in \{0.1,0.3\}$ and~$\phi \in \{0.6, 0.7, 0.8, 0.9\}$, 50 elections, using the seeds~$[0, \ldots, 49]$. For the Euclidean model, we generate for each parameter combinations with~$r \in \{0.04, 0.08, \ldots, 0.32\}$ and~$d \in \{1,2\}$, also 50 elections, again for the seeds~$[0, \ldots, 49]$.\footnote{In our first run, we generated elections for a broader range of parameters. However, for our final datasets we restricted the parameters to the values given above, as those, generally, generate elections that are interesting from the perspective of proportionality, that is, elections where the EJR+-fraction is not close to one. It is important to keep this 'prefiltering' in mind throughout the following analysis.}

\subsection{How Restrictive Are Proportionality Axioms?}

\begin{figure*}[htbp]
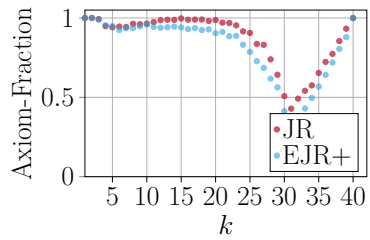
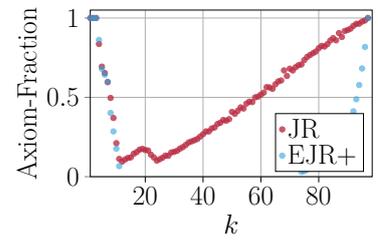

    \centering
    \begin{subfigure}[t]{0.3\textwidth}
        \centering
\begin{tikzpicture}[scale=0.5]

\definecolor{darkgray176}{RGB}{176,176,176}
\input{tikz_figures/color-def}

\begin{axis}[
tick align=outside,
tick pos=left,
x grid style={darkgray176},
xlabel={$k$},
xmajorgrids,
xmin=1, xmax=68,
xtick style={color=black, font=\huge},
y grid style={darkgray176},
ylabel={Axiom-Fraction},
ymajorgrids,
ymin=0, ymax=1.05,
legend style={at={(0.97,0.03)}, anchor=south east, legend columns=1, draw=black, font=\huge},
legend cell align={left},
label style={font =\huge},
ticklabel style={color=black, font=\huge},
height =6cm,
width=9cm,
]
\addplot [draw=blue, fill=blue, mark=*, only marks, opacity=0.8]
table{%
x  y
1 1
2 1
3 1
4 1
5 0.647249190938511
6 0.352360817477097
7 0.224719101123595
8 0.156764383132152
9 0.140252454417952
10 0.130446125750065
11 0.130992926381975
12 0.125078173858662
13 0.146391450739277
14 0.145518044237485
15 0.159083678014636
16 0.178699070764832
17 0.212314225053079
18 0.232828870779977
19 0.260416666666667
20 0.311235605353252
21 0.347705146036161
22 0.381242851696531
23 0.417188151856487
24 0.462748727441
25 0.518941359626362
26 0.572409845449342
27 0.638569604086845
28 0.672494956287828
29 0.705716302046577
30 0.765110941086457
31 0.791139240506329
32 0.843170320404722
33 0.865051903114187
34 0.898472596585804
35 0.919117647058823
36 0.937207122774133
37 0.963391136801541
38 0.956937799043062
39 0.970873786407767
40 0.982318271119843
41 0.984251968503937
42 0.99009900990099
43 0.99601593625498
44 0.99601593625498
45 0.999000999000999
46 1
47 1
48 0.998003992015968
49 1
50 1
51 1
52 1
53 1
54 1
55 1
56 1
57 1
58 1
59 1
60 1
61 1
62 1
63 1
64 1
65 1
66 1
67 1
};
\addlegendentry{JR}
\addplot [draw=green, fill=green, mark=*, only marks, opacity=0.6]
table{%
x  y
1 1
2 1
3 1
4 1
5 0.645161290322581
6 0.366837857666911
7 0.209907640638119
8 0.170940170940171
9 0.137912012136257
10 0.0751540658349617
11 0.0420981729392944
12 0.0298329355608592
13 0.0255147602888271
14 0.0206825232678387
15 0.0171004480317384
16 0.0135080372821829
17 0.0113050556208737
18 0.00867897345102021
19 0.00711637406508636
20 0.00658861355804898
21 0.00574560748307919
22 0.0037370185318749
24 0.00373672528343061
25 0.00392178394107912

41 0.00451322600881884
42 0.00597186059288632
43 0.00793291923494927
44 0.00609128398175051
45 0.00776331213949119
46 0.0103303650751018
47 0.0125909697564906
48 0.0184012954511998
49 0.0219625757708864
50 0.0281175312807536
51 0.0355138859293984
52 0.0464964895150416
53 0.0586957797734343
54 0.0565898930451021
55 0.0726585773450556
56 0.0944019635608421
57 0.120729204394543
58 0.148676776687481
59 0.1573564122738
60 0.17950098725543
61 0.222568439795237
62 0.247954376394743
63 0.28587764436821
64 0.377500943752359
65 0.561482313307131
66 0.723589001447178
67 1
};
\addlegendentry{EJR+}
\end{axis}

\end{tikzpicture}
        \caption{Amsterdam 622 (2022)}
        \label{fig:k-dependency-5}
    \end{subfigure}
    \hfill
    \begin{subfigure}[t]{0.3\textwidth}
        \centering
\begin{tikzpicture}[scale=0.5]

\definecolor{darkgray176}{RGB}{176,176,176}
\input{tikz_figures/color-def}

\begin{axis}[
tick align=outside,
tick pos=left,
x grid style={darkgray176},
xlabel={$k$},
xmajorgrids,
xmin=1, xmax=52,
xtick style={color=black, font=\huge},
y grid style={darkgray176},
ylabel={Axiom-Fraction},
ymajorgrids,
ymin=0, ymax=1.05,
legend style={at={(0.97,0.03)}, anchor=south east, legend columns=1, draw=black, font=\huge},
legend cell align={left},
label style={font =\huge},
ticklabel style={color=black, font=\huge},
height =6cm,
width=9cm,
]
\addplot [draw=blue, fill=blue, mark=*, only marks, opacity=0.8]
table{%
x  y
1 1
2 1
3 1
4 0.999000999000999
5 0.975609756097561
6 0.908265213442325
7 0.846023688663283
8 0.744601638123604
9 0.701754385964912
10 0.658327847267939
11 0.600600600600601
12 0.568828213879408
13 0.513347022587269
14 0.464252553389044
15 0.431034482758621
16 0.385208012326656
17 0.328299409061064
18 0.300751879699248
19 0.263504611330698
20 0.251635631605435
21 0.239005736137667
22 0.230627306273063
23 0.228885328450446
24 0.225428313796213
25 0.230467849734962
26 0.257400257400257
27 0.276014352746343
28 0.296471983397569
29 0.314465408805031
30 0.355239786856128
31 0.379650721336371
32 0.389256520046711
33 0.415282392026578
34 0.445831475702185
35 0.459770114942529
36 0.352858151023289
37 0.377786173026067
38 0.395882818685669
39 0.426985482493595
40 0.470809792843691
41 0.417536534446764
42 0.456412596987677
43 0.493827160493827
44 0.539956803455723
45 0.595238095238095
46 0.633312222925903
47 0.689655172413793
48 0.737463126843658
49 0.808407437348424
50 0.910746812386157
51 1
};
\addlegendentry{JR}
\addplot [draw=green, fill=green, mark=*, only marks, opacity=0.6]
table{%
x  y
1 1
2 1
3 1
4 0.99304865938431
5 0.965250965250965
6 0.903342366757001
7 0.841750841750842
8 0.76219512195122
9 0.693000693000693
10 0.62111801242236
11 0.600240096038415
12 0.586166471277843
13 0.51975051975052
14 0.479616306954436
15 0.423190859077444
16 0.373692077727952
17 0.332446808510638
18 0.293513354857646
19 0.267094017094017
20 0.233100233100233
21 0.216872695727608
22 0.215889464594128
23 0.213265088505012
24 0.210659363808721
25 0.224820143884892
26 0.231803430690774
27 0.244798041615667
28 0.259403372243839
29 0.280977802753582
30 0.29385836027035
31 0.332667997338656
32 0.352733686067019
33 0.346500346500347
34 0.380372765310004
35 0.372856077554064
36 0.286450873675165
37 0.318877551020408
38 0.350262697022767
39 0.378501135503407
40 0.394011032308905
41 0.367242012486228
42 0.396667988893296
43 0.430848772081
44 0.478697941598851
45 0.528541226215645
46 0.600961538461538
47 0.651465798045603
48 0.702247191011236
49 0.779423226812159
50 0.892060660124888
51 1
};
\addlegendentry{EJR+}
\end{axis}

\end{tikzpicture}
        \caption{Wawer (2017)}
        \label{fig:k-dependency-6}
    \end{subfigure}
    \hfill
    \begin{subfigure}[t]{0.3\textwidth}
        \centering
\begin{tikzpicture}[scale=0.5]

\definecolor{darkgray176}{RGB}{176,176,176}
\input{tikz_figures/color-def}

\begin{axis}[
tick align=outside,
tick pos=left,
x grid style={darkgray176},
xlabel={$k$},
xmajorgrids,
xmin=1, xmax=49,
xtick style={color=black, font=\huge},
y grid style={darkgray176},
ylabel={Axiom-Fraction},
ymajorgrids,
ymin=0, ymax=1.05,
legend style={at={(0.97,0.03)}, anchor=south east, legend columns=1, draw=black, font=\huge},
legend cell align={left},
label style={font =\huge},
ticklabel style={color=black, font=\huge},
height =6cm,
width=9cm,
]
\addplot [draw=blue, fill=blue, mark=*, only marks, opacity=0.8]
table{%
x  y
1 1
2 1
3 1
4 0.999000999000999
5 0.973709834469328
6 0.962463907603465
7 0.953288846520496
8 0.965250965250965
9 0.963391136801541
10 0.9765625
11 0.980392156862745
12 0.991080277502478
13 0.992063492063492
14 0.992063492063492
15 0.99601593625498
16 0.984251968503937
17 0.973709834469328
18 0.949667616334283
19 0.929368029739777
20 0.873362445414847
21 0.823723228995058
22 0.783699059561129
23 0.772797527047913
24 0.718390804597701
25 0.709219858156028
26 0.720461095100865
27 0.727802037845706
28 0.786163522012579
29 0.783085356303837
30 0.805152979066023
31 0.838222967309304
32 0.870322019147084
33 0.816993464052288
34 0.738007380073801
35 0.70323488045007
36 0.754716981132076
37 0.760456273764259
38 0.788643533123028
39 0.814332247557003
40 0.814332247557003
41 0.830564784053156
42 0.899280575539568
43 0.877963125548727
44 0.919117647058823
45 0.931966449207829
46 0.956937799043062
47 0.974658869395711
48 1
};
\addlegendentry{JR}
\addplot [draw=green, fill=green, mark=*, only marks, opacity=0.6]
table{%
x  y
1 1
2 1
3 1
4 0.999000999000999
5 0.968992248062015
6 0.954198473282443
7 0.949667616334283
8 0.953288846520496
9 0.944287063267233
10 0.94876660341556
11 0.936329588014981
12 0.92678405931418
13 0.917431192660551
14 0.907441016333938
15 0.907441016333938
16 0.893655049151028
17 0.892060660124888
18 0.851788756388416
19 0.834028356964137
20 0.768639508070715
21 0.731528895391368
22 0.681198910081744
23 0.637755102040816
24 0.604229607250755
25 0.571428571428571
26 0.58173356602676
27 0.610873549175321
28 0.625
29 0.6426735218509
30 0.675675675675676
31 0.695410292072323
32 0.683994528043776
33 0.666222518321119
34 0.615384615384615
35 0.550357732526142
36 0.595947556615018
37 0.615384615384615
38 0.668449197860963
39 0.703729767769177
40 0.719424460431655
41 0.723589001447178
42 0.7627765064836
43 0.79428117553614
44 0.831255195344971
45 0.894454382826476
46 0.909090909090909
47 0.962463907603465
48 1
};
\addlegendentry{EJR+}
\end{axis}

\end{tikzpicture}
        \caption{Białołęka -- Obszar 1 (2018)}
        \label{fig:k-dependency-7}
    \end{subfigure}

    \vspace{1em} 

    \begin{subfigure}[t]{0.3\textwidth}
        \centering
\begin{tikzpicture}[scale=0.5]

\definecolor{darkgray176}{RGB}{176,176,176}
\input{tikz_figures/color-def}

\begin{axis}[
tick align=outside,
tick pos=left,
x grid style={darkgray176},
xlabel={$k$},
xmajorgrids,
xmin=1, xmax=44,
xtick style={color=black, font=\huge},
y grid style={darkgray176},
ylabel={Axiom-Fraction},
ymajorgrids,
ymin=0, ymax=1.05,
legend style={at={(0.97,0.03)}, anchor=south east, legend columns=1, draw=black, font=\huge},
legend cell align={left},
label style={font =\huge},
ticklabel style={color=black, font=\huge},
height =6cm,
width=9cm,
]
\addplot [draw=blue, fill=blue, mark=*, only marks, opacity=0.8]
table{%
x  y
1 1
2 1
3 0.99601593625498
4 0.947867298578199
5 0.911577028258888
6 0.868809730668983
7 0.859845227858985
8 0.846740050804403
9 0.848176420695505
10 0.841750841750842
11 0.825763831544178
12 0.848896434634975
13 0.8424599831508
14 0.820344544708778
15 0.730994152046784
16 0.451467268623025
17 0.326583932070542
18 0.348189415041783
19 0.362713093942691
20 0.398247710075667
21 0.392310710082385
22 0.403388463089956
23 0.436490615451768
24 0.434971726837756
25 0.4510599909788
26 0.464900046490005
27 0.484496124031008
28 0.473260766682442
29 0.477554918815664
30 0.333778371161549
31 0.362056480811006
32 0.393855848759354
33 0.441696113074205
34 0.482625482625483
35 0.549752611324904
36 0.574712643678161
37 0.620732464307883
38 0.697350069735007
39 0.7627765064836
40 0.786163522012579
41 0.818330605564648
42 0.861326442721792
43 1
};
\addlegendentry{JR}
\addplot [draw=green, fill=green, mark=*, only marks, opacity=0.6]
table{%
x  y
1 1
2 1
3 0.99403578528827
4 0.950570342205323
5 0.904159132007233
6 0.864304235090752
7 0.846740050804403
8 0.859845227858985
9 0.815660685154976
10 0.8110300081103
11 0.810372771474878
12 0.803212851405622
13 0.803858520900322
14 0.802568218298555
15 0.698324022346369
16 0.433651344319167
17 0.30571690614491
18 0.321853878339234
19 0.350017500875044
20 0.377786173026067
21 0.392464678178964
22 0.408663669799755
23 0.4293688278231
24 0.436109899694723
25 0.448833034111311
26 0.454959053685168
27 0.452079566003617
28 0.497760079641613
29 0.471475719000471
30 0.348310693138279
31 0.363372093023256
32 0.394166338194718
33 0.456829602558246
34 0.474608448030375
35 0.527148128624143
36 0.554631170271769
37 0.643915003219575
38 0.68259385665529
39 0.725689404934688
40 0.816326530612245
41 0.813008130081301
42 0.88339222614841
43 1
};
\addlegendentry{EJR+}
\end{axis}

\end{tikzpicture}
        \caption{Bielany (2018)}
        \label{fig:k-dependency-8}
    \end{subfigure}
    \hfill
    \begin{subfigure}[t]{0.3\textwidth}
        \centering
\begin{tikzpicture}[scale=0.5]

\definecolor{darkgray176}{RGB}{176,176,176}
\input{tikz_figures/color-def}

\begin{axis}[
tick align=outside,
tick pos=left,
x grid style={darkgray176},
xlabel={$k$},
xmajorgrids,
xmin=1, xmax=56,
xtick style={color=black, font=\huge},
y grid style={darkgray176},
ylabel={Axiom-Fraction},
ymajorgrids,
ymin=0, ymax=1.05,
legend style={at={(0.97,0.03)}, anchor=south east, legend columns=1, draw=black, font=\huge},
legend cell align={left},
label style={font =\huge},
ticklabel style={color=black, font=\huge},
height =6cm,
width=9cm,
]
\addplot [draw=blue, fill=blue, mark=*, only marks, opacity=0.8]
table{%
x  y
1 1
2 1
3 1
4 1
5 1
6 0.997008973080758
7 0.982318271119843
8 0.956937799043062
9 0.935453695042095
10 0.820344544708778
11 0.17120356103407
12 0.177022481855196
13 0.194024058983314
14 0.205888408482602
15 0.230467849734962
16 0.22271714922049
17 0.247586036147561
18 0.262398320650748
19 0.284900284900285
20 0.28137310073157
21 0.289268151576511
22 0.316355583676052
23 0.337040781934614
24 0.348189415041783
25 0.365363536719035
26 0.361663652802893
27 0.398247710075667
28 0.396825396825397
29 0.418935902806871
30 0.422832980972516
31 0.442282176028306
32 0.437445319335083
33 0.449034575662326
34 0.466853408029879
35 0.495540138751239
36 0.512557662737058
37 0.522739153162572
38 0.549450549450549
39 0.550660792951542
40 0.571755288736421
41 0.590318772136954
42 0.609756097560976
43 0.601684717208183
44 0.632511068943707
45 0.682128240109141
46 0.672494956287828
47 0.730994152046784
48 0.76219512195122
49 0.802568218298555
50 0.833333333333333
51 0.870322019147084
52 0.880281690140845
53 0.921658986175115
54 0.968054211035818
55 1
};
\addlegendentry{JR}
\addplot [draw=green, fill=green, mark=*, only marks, opacity=0.6]
table{%
x  y
1 1
2 1
3 1
4 1
5 1
6 0.99601593625498
7 0.98135426889107
8 0.965250965250965
9 0.940733772342427
10 0.819672131147541
11 0.17247326664367
12 0.183318056828598
13 0.197316495659037
14 0.206825232678387
15 0.212856534695615
16 0.227427791676143
17 0.238435860753457
18 0.243249817562637
19 0.263088660878716
20 0.274047684297068
21 0.291290416545296
22 0.286779466590192
23 0.314169022934339
24 0.319182891797
25 0.341646737273659
26 0.341646737273659
27 0.37037037037037
28 0.37037037037037
29 0.377500943752359
30 0.385059684251059
31 0.401606425702811
32 0.4359197907585
33 0.43917435221783
34 0.456829602558246
35 0.481695568400771
36 0.493339911198816
37 0.515463917525773
38 0.527983104540655
39 0.53475935828877
40 0.554938956714761
41 0.5720823798627
42 0.595238095238095
43 0.619578686493185
44 0.620347394540943
45 0.647249190938511
46 0.675675675675676
47 0.734753857457752
48 0.753579502637528
49 0.791139240506329
50 0.836820083682008
51 0.873362445414847
52 0.876424189307625
53 0.929368029739777
54 0.970873786407767
55 1
};
\addlegendentry{EJR+}
\end{axis}

\end{tikzpicture}
        \caption{Wawer (2018)}
        \label{fig:k-dependency-9}
    \end{subfigure}
    \hfill
    \begin{subfigure}[t]{0.3\textwidth}
        \centering
\begin{tikzpicture}[scale=0.5]

\definecolor{darkgray176}{RGB}{176,176,176}
\input{tikz_figures/color-def}

\begin{axis}[
tick align=outside,
tick pos=left,
x grid style={darkgray176},
xlabel={$k$},
xmajorgrids,
xmin=1, xmax=62,
xtick style={color=black, font=\huge},
y grid style={darkgray176},
ylabel={Axiom-Fraction},
ymajorgrids,
ymin=0, ymax=1.05,
legend style={at={(0.97,0.03)}, anchor=south east, legend columns=1, draw=black, font=\huge},
legend cell align={left},
label style={font =\huge},
ticklabel style={color=black, font=\huge},
height =6cm,
width=9cm,
]
\addplot [draw=blue, fill=blue, mark=*, only marks, opacity=0.8]
table{%
x  y
1 1
2 1
3 1
4 0.894454382826476
5 0.695410292072323
6 0.550660792951542
7 0.493583415597236
8 0.426985482493595
9 0.424628450106157
10 0.410340582683627
11 0.437254044599913
12 0.471475719000471
13 0.505305709954523
14 0.505305709954523
15 0.547945205479452
16 0.585137507314219
17 0.631313131313131
18 0.645161290322581
19 0.697350069735007
20 0.729394602479942
21 0.775795190069822
22 0.771010023130301
23 0.792393026941363
24 0.805152979066023
25 0.8110300081103
26 0.788643533123028
27 0.76103500761035
28 0.66577896138482
29 0.580046403712297
30 0.488042947779405
31 0.386548125241593
32 0.361141206211629
33 0.321130378933847
34 0.335683115139308
35 0.318674314850223
36 0.337154416722859
37 0.361271676300578
38 0.383288616328095
39 0.394011032308905
40 0.43010752688172
41 0.470809792843691
42 0.473036896877956
43 0.488997555012225
44 0.50761421319797
45 0.538213132400431
46 0.559284116331096
47 0.568828213879408
48 0.600240096038415
49 0.647668393782383
50 0.648088139987038
51 0.661375661375661
52 0.695410292072323
53 0.701262272089762
54 0.717875089734386
55 0.721500721500722
56 0.7627765064836
57 0.825763831544178
58 0.846740050804403
59 0.897666068222621
60 0.949667616334283
61 1
};
\addlegendentry{JR}
\addplot [draw=green, fill=green, mark=*, only marks, opacity=0.6]
table{%
x  y
1 1
2 1
3 1
4 0.908265213442325
5 0.708215297450425
6 0.547345374931582
7 0.467726847521048
8 0.415454923140839
9 0.329055610398157
10 0.297176820208024
11 0.260213374967473
12 0.248508946322068
13 0.245098039215686
14 0.249376558603491
15 0.251067034898318
16 0.257334019557385
17 0.272702481592582
18 0.284252416145537
19 0.298864315600717
20 0.30902348578492
21 0.323939099449304
22 0.33921302578019
23 0.353232073472271
24 0.365230094959825
25 0.379218809252939
26 0.357270453733476
27 0.371885459278542
28 0.335120643431635
29 0.308737264587836
30 0.270343336036767
31 0.221336874723329
32 0.177367860943597
33 0.157728706624606
34 0.140904607580668
35 0.148853825543316
36 0.155424308361828
37 0.155134967421657
38 0.169750466813784
39 0.184094256259205
40 0.19782393669634
41 0.216450216450216
42 0.231535077564251
43 0.240963855421687
44 0.258331180573495
45 0.287686996547756
46 0.292997363023733
47 0.318369945877109
48 0.336134453781513
49 0.370782350760104
50 0.407000407000407
51 0.41101520756268
52 0.460405156537753
53 0.465116279069767
54 0.469483568075117
55 0.528541226215645
56 0.583090379008746
57 0.656598818122127
58 0.745156482861401
59 0.797448165869219
60 0.900900900900901
61 1
};
\addlegendentry{EJR+}
\end{axis}

\end{tikzpicture}
        \caption{Tarchomin, Nowodwory, Kępa Tarchomińska -- Obszar 1 (2019)}
        \label{fig:k-dependency-10}
    \end{subfigure}

    \vspace{1em}

    \begin{subfigure}[t]{0.3\textwidth}
        \centering
\begin{tikzpicture}[scale=0.5]

\definecolor{darkgray176}{RGB}{176,176,176}
\input{tikz_figures/color-def}

\begin{axis}[
tick align=outside,
tick pos=left,
x grid style={darkgray176},
xlabel={$k$},
xmajorgrids,
xmin=1, xmax=106.5,
xtick style={color=black, font=\huge},
y grid style={darkgray176},
ylabel={Axiom-Fraction},
ymajorgrids,
ymin=0, ymax=1.05,
legend style={at={(0.97,0.03)}, anchor=south east, legend columns=1, draw=black, font=\huge},
legend cell align={left},
label style={font =\huge},
ticklabel style={color=black, font=\huge},
height =6cm,
width=9cm,
]
\addplot [draw=blue, fill=blue, mark=*, only marks, opacity=0.8]
table{%
x  y
1 1
2 1
3 1
4 0.978473581213307
5 0.847457627118644
6 0.741289844329133
7 0.713266761768902
8 0.697350069735007
9 0.723589001447178
10 0.753579502637528
11 0.758150113722517
12 0.802568218298555
13 0.846023688663283
14 0.853242320819113
15 0.885739592559787
16 0.911577028258888
17 0.936329588014981
18 0.944287063267233
19 0.965250965250965
20 0.973709834469328
21 0.977517106549365
22 0.968054211035818
23 0.956937799043062
24 0.945179584120983
25 0.919963201471941
26 0.908265213442325
27 0.871080139372822
28 0.802568218298555
29 0.767459708365311
30 0.735835172921266
31 0.70323488045007
32 0.687757909215956
33 0.698812019566737
34 0.690131124913734
35 0.703729767769177
36 0.70871722182849
37 0.727802037845706
38 0.731528895391368
39 0.730460189919649
40 0.718907260963336
41 0.693962526023595
42 0.702247191011236
43 0.694444444444444
44 0.671591672263264
45 0.683994528043776
46 0.69060773480663
47 0.717875089734386
48 0.687757909215956
49 0.728332119446468
50 0.714285714285714
51 0.737463126843658
52 0.731528895391368
53 0.76219512195122
54 0.758725341426404
55 0.78003120124805
56 0.783699059561129
57 0.772200772200772
58 0.808407437348424
59 0.824402308326463
60 0.815660685154976
61 0.813669650122051
62 0.837520938023451
63 0.856164383561644
64 0.852514919011083
65 0.840336134453782
66 0.846023688663283
67 0.840336134453782
68 0.8424599831508
69 0.832639467110741
70 0.803858520900322
71 0.78125
72 0.753012048192771
73 0.728862973760933
74 0.758150113722517
75 0.713266761768902
76 0.702247191011236
77 0.700770847932726
78 0.673854447439353
79 0.692041522491349
80 0.703729767769177
81 0.744047619047619
82 0.740740740740741
83 0.745156482861401
84 0.779423226812159
85 0.782472613458529
86 0.769822940723634
87 0.789889415481833
88 0.795544948289578
89 0.8058017727639
90 0.794912559618442
91 0.796178343949045
92 0.813669650122051
93 0.794912559618442
94 0.765110941086457
95 0.796178343949045
96 0.823045267489712
97 0.859845227858985
98 0.863557858376511
99 0.884173297966401
100 0.922509225092251
101 0.930232558139535
102 0.942507068803016
103 0.978473581213307
104 0.975609756097561
105 1
};
\addlegendentry{JR}
\addplot [draw=green, fill=green, mark=*, only marks, opacity=0.6]
table{%
x  y
1 1
2 1
3 1
4 0.9765625
5 0.843170320404722
6 0.74019245003701
7 0.712250712250712
8 0.686813186813187
9 0.588928150765607
10 0.513610683102208
11 0.482160077145612
12 0.4293688278231
13 0.336134453781513
14 0.296823983377857
15 0.259268861809697
16 0.195541650371529
17 0.151584053357587
18 0.135080372821829
19 0.107480653482373
20 0.0785607667530835
21 0.0646788694133626
93 0.0252838107759602
94 0.0283631619252914
95 0.0384674565317741
96 0.0551419906258616
97 0.061020258725897
98 0.0866551126516465
99 0.122594090964815
100 0.180440274269217
101 0.250815149235014
102 0.366703337000367
103 0.492125984251969
104 0.703729767769177
105 1
};
\addlegendentry{EJR+}
\end{axis}

\end{tikzpicture}
        \caption{Ursynów (2021)}
        \label{fig:k-dependency-11}
    \end{subfigure}
    \hfill
    \begin{subfigure}[t]{0.3\textwidth}
        \centering
\begin{tikzpicture}[scale=0.5]

\definecolor{darkgray176}{RGB}{176,176,176}
\input{tikz_figures/color-def}

\begin{axis}[
tick align=outside,
tick pos=left,
x grid style={darkgray176},
xlabel={$k$},
xmajorgrids,
xmin=1, xmax=42,
xtick style={color=black, font=\huge},
y grid style={darkgray176},
ylabel={Axiom-Fraction},
ymajorgrids,
ymin=0, ymax=1.05,
legend style={at={(0.97,0.03)}, anchor=south east, legend columns=1, draw=black, font=\huge},
legend cell align={left},
label style={font =\huge},
ticklabel style={color=black, font=\huge},
height =6cm,
width=9cm,
]
\addplot [draw=blue, fill=blue, mark=*, only marks, opacity=0.8]
table{%
x  y
1 1
2 1
3 0.99304865938431
4 0.94876660341556
5 0.941619585687382
6 0.946073793755913
7 0.943396226415094
8 0.963391136801541
9 0.960614793467819
10 0.964320154291225
11 0.974658869395711
12 0.986193293885602
13 0.989119683481701
14 0.988142292490119
15 0.997008973080758
16 0.989119683481701
17 0.99009900990099
18 0.99009900990099
19 0.98135426889107
20 0.987166831194472
21 0.970873786407767
22 0.968054211035818
23 0.953288846520496
24 0.915750915750916
25 0.905797101449275
26 0.836820083682008
27 0.831255195344971
28 0.73909830007391
29 0.642260757867694
30 0.507356671740233
31 0.428816466552316
32 0.491883915395967
33 0.541418516513265
34 0.575373993095512
35 0.65402223675605
36 0.723589001447178
37 0.773993808049536
38 0.854700854700855
39 0.932835820895522
40 1
};
\addlegendentry{JR}
\addplot [draw=green, fill=green, mark=*, only marks, opacity=0.6]
table{%
x  y
1 1
2 1
3 0.99009900990099
4 0.954198473282443
5 0.949667616334283
6 0.924214417744917
7 0.935453695042095
8 0.936329588014981
9 0.946969696969697
10 0.958772770853308
11 0.944287063267233
12 0.938967136150235
13 0.941619585687382
14 0.946073793755913
15 0.940733772342427
16 0.930232558139535
17 0.936329588014981
18 0.924214417744917
19 0.928505106778087
20 0.904159132007233
21 0.912408759124088
22 0.884955752212389
23 0.887311446317657
24 0.831946755407654
25 0.786782061369001
26 0.727272727272727
27 0.685871056241427
28 0.618046971569839
29 0.563063063063063
30 0.412541254125413
31 0.349040139616056
32 0.3813882532418
33 0.429922613929493
34 0.496524329692155
35 0.568181818181818
36 0.641848523748395
37 0.719424460431655
38 0.805152979066023
39 0.879507475813544
40 1
};
\addlegendentry{EJR+}
\end{axis}

\end{tikzpicture}
        \caption{Wesoła (2021)}
        \label{fig:k-dependency-12}
    \end{subfigure}
    \hfill
    \begin{subfigure}[t]{0.3\textwidth}
        \centering
\begin{tikzpicture}[scale=0.5]

\definecolor{darkgray176}{RGB}{176,176,176}
\input{tikz_figures/color-def}

\begin{axis}[
tick align=outside,
tick pos=left,
x grid style={darkgray176},
xlabel={$k$},
xmajorgrids,
xmin=1, xmax=98.5,
xtick style={color=black, font=\huge},
y grid style={darkgray176},
ylabel={Axiom-Fraction},
ymajorgrids,
ymin=0, ymax=1.05,
legend style={at={(0.97,0.03)}, anchor=south east, legend columns=1, draw=black, font=\huge},
legend cell align={left},
label style={font =\huge},
ticklabel style={color=black, font=\huge},
height =6cm,
width=9cm,
]
\addplot [draw=blue, fill=blue, mark=*, only marks, opacity=0.8]
table{%
x  y
1 1
2 1
3 1
4 0.836120401337793
5 0.693481276005548
6 0.653167864141084
7 0.595592614651578
8 0.496277915632754
9 0.369959304476508
10 0.21137180300148
11 0.111994624258036
12 0.0952562392836731
13 0.108754758020663
14 0.117966261649168
15 0.121802679658953
16 0.144383482529599
17 0.154655119084442
18 0.170183798502383
19 0.175407823188914
20 0.168095478231636
21 0.164826108455579
22 0.138350857775318
23 0.120656370656371
24 0.101194090265129
25 0.110180696342001
26 0.119602918311207
27 0.133209004928733
28 0.131769666622743
29 0.152648450618226
30 0.156470035988108
31 0.162972620599739
32 0.176772140710624
33 0.186219739292365
34 0.199600798403194
35 0.2154243860405
36 0.222469410456062
37 0.229568411386593
38 0.237247924080664
39 0.252908447142135
40 0.267594327000268
41 0.285388127853881
42 0.285388127853881
43 0.303951367781155
44 0.310655483069276
45 0.332225913621262
46 0.340367597004765
47 0.357270453733476
48 0.352733686067019
49 0.362581580855693
50 0.408329930583912
51 0.396039603960396
52 0.4149377593361
53 0.436681222707424
54 0.429737859905458
55 0.460617227084293
56 0.473260766682442
57 0.470145745181006
58 0.499750124937531
59 0.504795557799091
60 0.517330574236937
61 0.530222693531283
62 0.564334085778781
63 0.562429696287964
64 0.553709856035437
65 0.582072176949942
66 0.600240096038415
67 0.606428138265615
68 0.669792364367046
69 0.634920634920635
70 0.68073519400953
71 0.669344042838019
72 0.689655172413793
73 0.702740688685875
74 0.707213578500707
75 0.734753857457752
76 0.754716981132076
77 0.766871165644172
78 0.778210116731518
79 0.788643533123028
80 0.809716599190283
81 0.824402308326463
82 0.820344544708778
83 0.835421888053467
84 0.858369098712446
85 0.873362445414847
86 0.859106529209622
87 0.904977375565611
88 0.880281690140845
89 0.918273645546373
90 0.924214417744917
91 0.932835820895522
92 0.950570342205323
93 0.963391136801541
94 0.964320154291225
95 0.977517106549365
96 0.984251968503937
97 1
};
\addlegendentry{JR}
\addplot [draw=green, fill=green, mark=*, only marks, opacity=0.6]
table{%
x  y
1 1
2 1
3 1
4 0.862068965517241
5 0.679809653297077
6 0.643086816720257
7 0.598444045481747
8 0.401929260450161
9 0.285388127853881
10 0.176397953783736
11 0.0664982045484772
74 0.0305278261135025
75 0.0347608453837597
76 0.0454235748353395
77 0.0545821734621473
78 0.0615801465607488
79 0.0727802037845706
80 0.0872448089338684
81 0.103487529752665
82 0.107607876896589
83 0.117247039512252
84 0.134517083669626
85 0.157059839798963
86 0.185288123031314
87 0.211595429538722
88 0.227066303360581
89 0.266453503863576
90 0.327439423706614
91 0.375093773443361
92 0.411692054343351
93 0.48780487804878
94 0.577700751010976
95 0.682128240109141
96 0.817661488143908
97 1
};
\addlegendentry{EJR+}
\end{axis}

\end{tikzpicture}
        \caption{Mokotów (2022)}
        \label{fig:k-dependency-13}
    \end{subfigure}

    \vspace{1em}

    \begin{subfigure}[t]{0.3\textwidth}
        \centering
\begin{tikzpicture}[scale=0.5]

\definecolor{darkgray176}{RGB}{176,176,176}
\input{tikz_figures/color-def}

\begin{axis}[
tick align=outside,
tick pos=left,
x grid style={darkgray176},
xlabel={$k$},
xmajorgrids,
xmin=1, xmax=52,
xtick style={color=black, font=\huge},
y grid style={darkgray176},
ylabel={Axiom-Fraction},
ymajorgrids,
ymin=0, ymax=1.05,
legend style={at={(0.97,0.03)}, anchor=south east, legend columns=1, draw=black, font=\huge},
legend cell align={left},
label style={font =\huge},
ticklabel style={color=black, font=\huge},
height =6cm,
width=9cm,
]
\addplot [draw=blue, fill=blue, mark=*, only marks, opacity=0.8]
table{%
x  y
1 1
2 1
3 0.982318271119843
4 0.899280575539568
5 0.869565217391304
6 0.878734622144112
7 0.853242320819113
8 0.87260034904014
9 0.877963125548727
10 0.874125874125874
11 0.823723228995058
12 0.817661488143908
13 0.744601638123604
14 0.659195781147001
15 0.568828213879408
16 0.547945205479452
17 0.558971492453885
18 0.594530321046373
19 0.616903146206046
20 0.633713561470216
21 0.66006600660066
22 0.702247191011236
23 0.702247191011236
24 0.705218617771509
25 0.757575757575758
26 0.768049155145929
27 0.799360511590727
28 0.793021411578113
29 0.819672131147541
30 0.843170320404722
31 0.834724540901502
32 0.857632933104631
33 0.880281690140845
34 0.904159132007233
35 0.893655049151028
36 0.917431192660551
37 0.935453695042095
38 0.910746812386157
39 0.907441016333938
40 0.9000900090009
41 0.835421888053467
42 0.809061488673139
43 0.832639467110741
44 0.863557858376511
45 0.865051903114187
46 0.902527075812274
47 0.912408759124088
48 0.959692898272553
49 0.969932104752667
50 0.979431929480901
51 1
};
\addlegendentry{JR}
\addplot [draw=green, fill=green, mark=*, only marks, opacity=0.6]
table{%
x  y
1 1
2 1
3 0.985221674876847
4 0.924214417744917
5 0.861326442721792
6 0.878734622144112
7 0.874125874125874
8 0.851788756388416
9 0.831946755407654
10 0.838222967309304
11 0.773993808049536
12 0.75187969924812
13 0.701262272089762
14 0.596658711217184
15 0.50709939148073
16 0.467508181393174
17 0.435540069686411
18 0.455166135639508
19 0.453514739229025
20 0.459770114942529
21 0.44762757385855
22 0.444247001332741
23 0.433275563258232
24 0.386847195357834
25 0.396667988893296
26 0.373831775700935
27 0.35486160397445
28 0.351246926589392
29 0.333444481493831
30 0.341413451689997
31 0.31328320802005
32 0.322893122376493
33 0.291545189504373
34 0.253164556962025
35 0.249500998003992
36 0.238322211630124
37 0.274499039253363
38 0.280269058295964
39 0.303766707168894
40 0.311332503113325
41 0.293255131964809
42 0.300210147102972
43 0.342817963661296
44 0.377073906485671
45 0.455580865603645
46 0.455996352029184
47 0.493096646942801
48 0.547945205479452
49 0.626959247648903
50 0.783699059561129
51 1
};
\addlegendentry{EJR+}
\end{axis}

\end{tikzpicture}
        \caption{Śródmieście (2024)}
        \label{fig:k-dependency-14}
    \end{subfigure}
    \hfill
    \begin{subfigure}[t]{0.3\textwidth}
        \centering
\begin{tikzpicture}[scale=0.5]

\definecolor{darkgray176}{RGB}{176,176,176}
\input{tikz_figures/color-def}

\begin{axis}[
tick align=outside,
tick pos=left,
x grid style={darkgray176},
xlabel={$k$},
xmajorgrids,
xmin=1, xmax=47,
xtick style={color=black, font=\huge},
y grid style={darkgray176},
ylabel={Axiom-Fraction},
ymajorgrids,
ymin=0, ymax=1.05,
legend style={at={(0.97,0.03)}, anchor=south east, legend columns=1, draw=black, font=\huge},
legend cell align={left},
label style={font =\huge},
ticklabel style={color=black, font=\huge},
height =6cm,
width=9cm,
]
\addplot [draw=blue, fill=blue, mark=*, only marks, opacity=0.8]
table{%
x  y
1 1
2 1
3 0.903342366757001
4 0.863557858376511
5 0.905797101449275
6 0.912408759124088
7 0.938086303939963
8 0.953288846520496
9 0.975609756097561
10 0.975609756097561
11 0.868809730668983
12 0.616142945163278
13 0.410340582683627
14 0.355366027007818
15 0.33500837520938
16 0.362713093942691
17 0.354233085370174
18 0.377500943752359
19 0.416146483562214
20 0.428449014567266
21 0.452898550724638
22 0.488519785051295
23 0.492368291482029
24 0.512820512820513
25 0.535331905781585
26 0.563380281690141
27 0.571428571428571
28 0.600961538461538
29 0.637755102040816
30 0.647249190938511
31 0.672043010752688
32 0.695410292072323
33 0.720980533525595
34 0.726216412490922
35 0.769822940723634
36 0.78003120124805
37 0.796178343949045
38 0.833333333333333
39 0.846023688663283
40 0.875656742556918
41 0.902527075812274
42 0.929368029739777
43 0.941619585687382
44 0.953288846520496
45 0.984251968503937
46 1
};
\addlegendentry{JR}
\addplot [draw=green, fill=green, mark=*, only marks, opacity=0.6]
table{%
x  y
1 1
2 1
3 0.920810313075506
4 0.87719298245614
5 0.836120401337793
6 0.815660685154976
7 0.79428117553614
8 0.737463126843658
9 0.734214390602056
10 0.658327847267939
11 0.568181818181818
12 0.32520325203252
13 0.186950831931202
14 0.138888888888889
15 0.1127649977447
16 0.10482180293501
17 0.0964506172839506
18 0.0924470740501063
19 0.100745516824501
20 0.0863632438034373
21 0.0903179190751445
22 0.0870019140421089
23 0.0982897582071948
24 0.0946521533364884
25 0.107828337287039
26 0.105652403592182
27 0.121212121212121
28 0.124890720619458
29 0.146563095412575
30 0.141843971631206
31 0.152928582352042
32 0.159667890787163
33 0.15921031682853
34 0.158002844051193
35 0.149521531100478
36 0.151883353584447
37 0.152532031726663
38 0.189933523266857
39 0.224769611148573
40 0.251130085384229
41 0.335683115139308
42 0.419815281276238
43 0.535045478865704
44 0.614628149969269
45 0.78003120124805
46 1
};
\addlegendentry{EJR+}
\end{axis}

\end{tikzpicture}
        \caption{Żoliborz (2024)}
        \label{fig:k-dependency-15}
    \end{subfigure}
    \hfill
    \begin{subfigure}[t]{0.3\textwidth}
        \centering
\begin{tikzpicture}[scale=0.5]

\definecolor{darkgray176}{RGB}{176,176,176}
\input{tikz_figures/color-def}

\begin{axis}[
tick align=outside,
tick pos=left,
x grid style={darkgray176},
xlabel={$k$},
xmajorgrids,
xmin=1, xmax=54,
xtick style={color=black, font=\huge},
y grid style={darkgray176},
ylabel={Axiom-Fraction},
ymajorgrids,
ymin=0, ymax=1.05,
legend style={at={(0.97,0.03)}, anchor=south east, legend columns=1, draw=black, font=\huge},
legend cell align={left},
label style={font =\huge},
ticklabel style={color=black, font=\huge},
height =6cm,
width=9cm,
]
\addplot [draw=blue, fill=blue, mark=*, only marks, opacity=0.8]
table{%
x  y
1 1
2 1
3 1
4 0.972762645914397
5 0.946073793755913
6 0.954198473282443
7 0.950570342205323
8 0.949667616334283
9 0.966183574879227
10 0.938086303939963
11 0.941619585687382
12 0.945179584120983
13 0.93984962406015
14 0.938086303939963
15 0.942507068803016
16 0.940733772342427
17 0.940733772342427
18 0.937207122774133
19 0.931098696461825
20 0.946969696969697
21 0.946969696969697
22 0.956937799043062
23 0.959692898272553
24 0.962463907603465
25 0.968054211035818
26 0.970873786407767
27 0.9765625
28 0.9765625
29 0.984251968503937
30 0.982318271119843
31 0.982318271119843
32 0.982318271119843
33 0.989119683481701
34 0.986193293885602
35 0.989119683481701
36 0.99304865938431
37 0.989119683481701
38 0.99304865938431
39 0.989119683481701
40 0.992063492063492
41 0.99009900990099
42 0.973709834469328
43 0.873362445414847
44 0.814995925020375
45 0.846740050804403
46 0.87260034904014
47 0.884173297966401
48 0.898472596585804
49 0.922509225092251
50 0.900900900900901
51 0.921658986175115
52 0.956937799043062
53 1
};
\addlegendentry{JR}
\addplot [draw=green, fill=green, mark=*, only marks, opacity=0.6]
table{%
x  y
1 1
2 1
3 1
4 0.983284169124877
5 0.950570342205323
6 0.946969696969697
7 0.941619585687382
8 0.930232558139535
9 0.925925925925926
10 0.919963201471941
11 0.894454382826476
12 0.898472596585804
13 0.865800865800866
14 0.847457627118644
15 0.840336134453782
16 0.843881856540084
17 0.801282051282051
18 0.778210116731518
19 0.775795190069822
20 0.755287009063444
21 0.702740688685875
22 0.714796283059328
23 0.673854447439353
24 0.649350649350649
25 0.632111251580278
26 0.581057524694945
27 0.540540540540541
28 0.503524672708963
29 0.467508181393174
30 0.452079566003617
31 0.415627597672485
32 0.382848392036753
33 0.348310693138279
34 0.349162011173184
35 0.325309043591412
36 0.354233085370174
37 0.351741118536757
38 0.370782350760104
39 0.384319754035357
40 0.419991600167997
41 0.429000429000429
42 0.488519785051295
43 0.461467466543609
44 0.455788514129444
45 0.475737392959087
46 0.521648408972353
47 0.507872016251904
48 0.585137507314219
49 0.659630606860158
50 0.698324022346369
51 0.74794315632012
52 0.843881856540084
53 1
};
\addlegendentry{EJR+}
\end{axis}

\end{tikzpicture}
        \caption{Śródmieście (2025)}
        \label{fig:k-dependency-16}
    \end{subfigure}
    \caption{JR-fraction and EJR+-fraction for committee sizes~$k \in [m]$ across selected pabulib instances.}
    \label{fig:jr-ejr-dependency-all}
\end{figure*}

\paragraph{Dependency on~$k$.}
As mentioned in Section 4.1, we encounter a variety of shapes when studying how the JR-fractions, resp. EJR+-fractions, depend on the selected value for~$k$. In \Cref{fig:jr-ejr-dependency-all} we display additional shapes that pabulib instances exhibit. Although this set, together with the shapes shown before, should not be viewed as a classification, we think they give a good overview of the kind of shapes that can be found across the pabulib dataset.\footnote{The missing values for EJR+ in three of the figures are due to the timeout we set, i.e., we did not find 1000 EJR+-committees within 15 minutes for the corresponding values of~$k$, probably due to a very low EJR+ fraction.}

\newpage

\begin{figure}[htbp]
    \centering
    \begin{tikzpicture}
\input{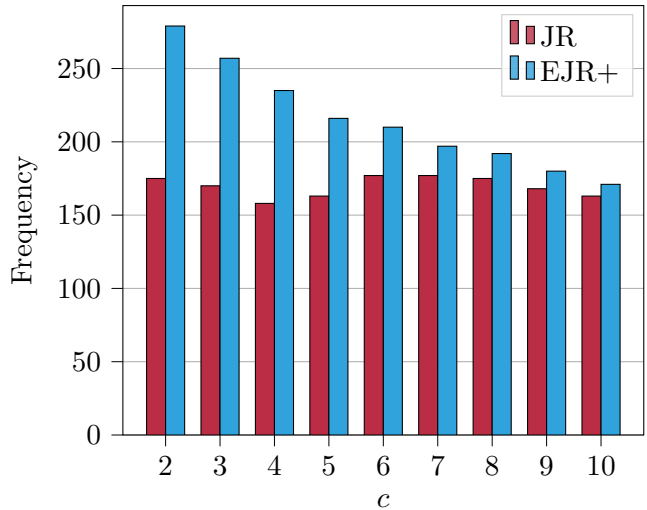}

\begin{axis}[
legend cell align={left},
legend style={fill opacity=0.8, draw opacity=1, text opacity=1, draw=lightgray204},
tick align=outside,
tick pos=left,
x grid style={darkgray176},
xlabel={$c$ },
xmin=-0.785, xmax=8.785,
xtick style={color=black},
xtick={0,1,2,3,4,5,6,7,8},
xticklabels={2,3,4,5,6,7,8,9,10},
y grid style={darkgray176},
ylabel={Frequency},
ymajorgrids,
ymin=0, ymax=292.95,
ytick style={color=black}
]

\addlegendimage{ybar,ybar legend,draw=black,fill=blue}
\addlegendentry{JR}
\addlegendimage{ybar,ybar legend,draw=black,fill=green}
\addlegendentry{EJR+}

\draw[draw=black,fill=blue] (axis cs:-0.35,0) rectangle (axis cs:0,175);
\draw[draw=black,fill=blue] (axis cs:0.65,0) rectangle (axis cs:1,170);
\draw[draw=black,fill=blue] (axis cs:1.65,0) rectangle (axis cs:2,158);
\draw[draw=black,fill=blue] (axis cs:2.65,0) rectangle (axis cs:3,163);
\draw[draw=black,fill=blue] (axis cs:3.65,0) rectangle (axis cs:4,177);
\draw[draw=black,fill=blue] (axis cs:4.65,0) rectangle (axis cs:5,177);
\draw[draw=black,fill=blue] (axis cs:5.65,0) rectangle (axis cs:6,175);
\draw[draw=black,fill=blue] (axis cs:6.65,0) rectangle (axis cs:7,168);
\draw[draw=black,fill=blue] (axis cs:7.65,0) rectangle (axis cs:8,163);

\draw[draw=black,fill=green] (axis cs:2.77555756156289e-17,0) rectangle (axis cs:0.35,279);
\draw[draw=black,fill=green] (axis cs:1,0) rectangle (axis cs:1.35,257);
\draw[draw=black,fill=green] (axis cs:2,0) rectangle (axis cs:2.35,235);
\draw[draw=black,fill=green] (axis cs:3,0) rectangle (axis cs:3.35,216);
\draw[draw=black,fill=green] (axis cs:4,0) rectangle (axis cs:4.35,210);
\draw[draw=black,fill=green] (axis cs:5,0) rectangle (axis cs:5.35,197);
\draw[draw=black,fill=green] (axis cs:6,0) rectangle (axis cs:6.35,192);
\draw[draw=black,fill=green] (axis cs:7,0) rectangle (axis cs:7.35,180);
\draw[draw=black,fill=green] (axis cs:8,0) rectangle (axis cs:8.35,171);

\end{axis}
\end{tikzpicture}
    \caption{Histogram showing, for different values of~$c$, the number of pabulib instances with JR, resp. EJR+, fraction of at most 0.95 when setting~$k = m/c$.}
    \label{fig:best-k-histo}
\end{figure}

Our aim in designing this experiment was for the shapes to uncover a suitable value for~$k$, or more precisely, how to choose the relation between~$m$ and~$k$ such that the majority of pabulib instances are interesting from the perspective of proportionality. However, given the unexpected variety of shapes, it does not become obvious how to choose~$c$ in order to set~$k = m/c$. Hence, we count for each~$c \in [2, \ldots, 10]$ the number of pabulib instances with an EJR+ fraction of at most 0.95 when seeting~$k = m/c$. Examining~\Cref{fig:best-k-histo}, we choose~$k := \lfloor m/2 \rfloor$ to maximize this number.

Although we have chosen~$k$ in advance for our synthetic elections, we are curious what kind of shapes we can generate with our previously described selection of parameter combinations.

\begin{figure*}[htbp]
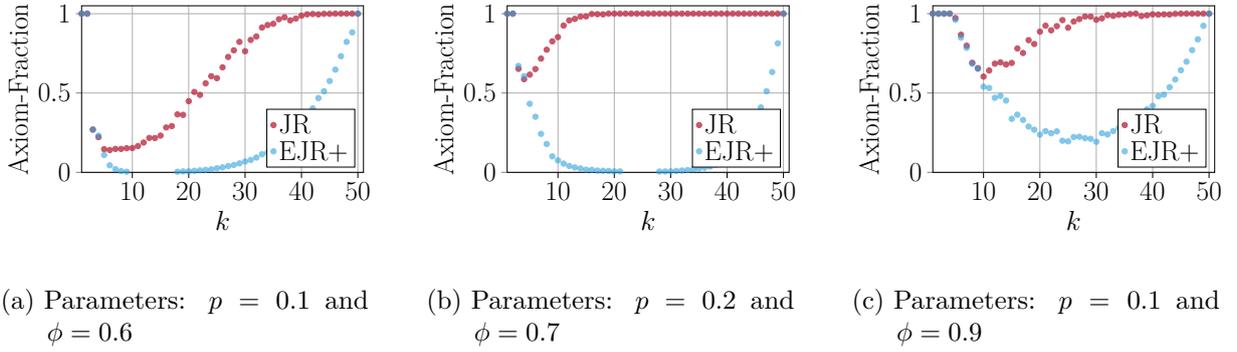

	\centering
	\begin{subfigure}[t]{0.3\textwidth}
	\centering
\begin{tikzpicture}[scale= 0.5]

\definecolor{darkgray176}{RGB}{176,176,176}
\input{tikz_figures/color-def}

\begin{axis}[
tick align=outside,
tick pos=left,
x grid style={darkgray176},
xlabel={$k$},
xmajorgrids,
xmin=1, xmax=51,
xtick style={color=black, font=\huge},
y grid style={darkgray176},
ylabel={Axiom-Fraction},
ymajorgrids,
ymin=0, ymax=1.05,
legend style={at={(0.97,0.03)}, anchor=south east, legend columns=1, draw=black, font=\huge},
legend cell align={left},
label style={font =\huge},
ticklabel style={color=black, font=\huge},
height =6cm,
width=9cm,
]
\addplot [draw=blue, fill=blue, mark=*, only marks, opacity=0.8]
table{%
x  y
1 1
2 1
3 0.269759913676828
4 0.221926320461607
5 0.146198830409357
6 0.140587656403768
7 0.147754137115839
8 0.147754137115839
9 0.152183838076396
10 0.153350713080816
11 0.166002656042497
12 0.18796992481203
13 0.217060994139353
14 0.216356555603635
15 0.231642344220524
16 0.282885431400283
17 0.29171528588098
18 0.365363536719035
19 0.36231884057971
20 0.448631673396142
21 0.506842372022301
22 0.48828125
23 0.561167227833894
24 0.606060606060606
25 0.593471810089021
26 0.661813368630046
27 0.727272727272727
28 0.769230769230769
29 0.822368421052632
30 0.763358778625954
31 0.834724540901502
32 0.856164383561644
33 0.912408759124088
34 0.928505106778087
35 0.936329588014981
36 0.965250965250965
37 0.977517106549365
38 0.957854406130268
39 0.968992248062015
40 0.987166831194472
41 0.99601593625498
42 0.998003992015968
43 0.99601593625498
44 0.999000999000999
45 0.999000999000999
46 1
47 1
48 1
49 1
50 1
};
\addlegendentry{JR}
\addplot [draw=green, fill=green, mark=*, only marks, opacity=0.6]
table{%
x  y
1 1
2 1
3 0.268889486421081
4 0.231160425335183
5 0.109385254867644
6 0.044412861964825
7 0.0205014658548086
8 0.00812156356341723
9 0.00391515085076228
18 0.00392958189248664
19 0.005688443926164
20 0.00685936921240723
21 0.00984872360542074
22 0.0125743458196587
23 0.0159961609213789
24 0.0191296030607365
25 0.0257440016476161
26 0.031324395439168
27 0.0393576826196474
28 0.0467267884678286
29 0.0566411781365052
30 0.0684415851071111
31 0.0780762023735165
32 0.0934055669717915
33 0.114090131203651
34 0.126071608673727
35 0.158177791838026
36 0.184128153194623
37 0.209117524048515
38 0.247954376394743
39 0.277546489036914
40 0.312207305650952
41 0.353606789250354
42 0.40290088638195
43 0.467945718296678
44 0.511247443762781
45 0.575705238917674
46 0.647249190938511
47 0.732064421669107
48 0.822368421052632
49 0.881834215167548
50 1
};
\addlegendentry{EJR+}
\end{axis}

\end{tikzpicture}
 \caption{Parameters: $p=0.1$ and $\phi=0.6$}
        \label{fig:res-1}
\end{subfigure}
\hfill
\begin{subfigure}[t]{0.3\textwidth}
	\centering
\begin{tikzpicture}[scale=0.5]

\definecolor{darkgray176}{RGB}{176,176,176}
\input{tikz_figures/color-def}
\begin{axis}[
tick align=outside,
tick pos=left,
x grid style={darkgray176},
xlabel={$k$},
xmajorgrids,
xmin=1, xmax=51,
xtick style={color=black, font=\huge},
y grid style={darkgray176},
ylabel={Axiom-Fraction},
ymajorgrids,
ymin=0, ymax=1.05,
legend style={at={(0.97,0.03)}, anchor=south east, legend columns=1, draw=black, font=\huge},
legend cell align={left},
label style={font =\huge},
ticklabel style={color=black, font=\huge},
height =6cm,
width=9cm,
]
\addplot [draw=blue, fill=blue, mark=*, only marks, opacity=0.8]
table{%
x  y
1 1
2 1
3 0.652315720808871
4 0.587199060481503
5 0.61576354679803
6 0.651041666666667
7 0.716845878136201
8 0.772797527047913
9 0.825763831544178
10 0.852514919011083
11 0.925069380203515
12 0.957854406130268
13 0.967117988394584
14 0.98135426889107
15 0.983284169124877
16 0.998003992015968
17 0.997008973080758
18 0.99601593625498
19 1
20 1
21 1
22 1
23 1
24 1
25 1
26 1
27 1
28 1
29 1
30 1
31 1
32 1
33 1
34 1
35 1
36 1
37 1
38 1
39 1
40 1
41 1
42 1
43 1
44 1
45 1
46 1
47 1
48 1
49 1
50 1
};
\addlegendentry{JR}
\addplot [draw=green, fill=green, mark=*, only marks, opacity=0.6]
table{%
x  y
1 1
2 1
3 0.670690811535882
4 0.606428138265615
5 0.432152117545376
6 0.350385423966363
7 0.242659548653239
8 0.178571428571429
9 0.100765820233777
10 0.0754147812971342
11 0.0554078014184397
12 0.0401703221659838
13 0.0310867943297687
14 0.0243480801538799
15 0.0171836068390755
16 0.0160451832359926
17 0.0110381367625145
18 0.0100808484042017
19 0.00877446980266217
20 0.00622537087646997
21 0.0059807539338409
28 0.0051017024381036
29 0.00549580395368136
30 0.00657963996210127
31 0.00904518976808133
32 0.0114137010066884
33 0.0143457615447516
34 0.0192041788293133
35 0.0238413122258249
36 0.0290520321896517
37 0.0363596698541977
38 0.0480376615266369
39 0.0639018467633715
40 0.0808603541683513
41 0.108636610537751
42 0.140193466984439
43 0.18138944313441
44 0.250501002004008
45 0.310462589257994
46 0.409165302782324
47 0.51150895140665
48 0.632511068943707
49 0.813008130081301
50 1
};
\addlegendentry{EJR+}
\end{axis}

\end{tikzpicture}
 \caption{Parameters: $p=0.2$ and $\phi =0.7$}
        \label{fig:res-2}
	\end{subfigure}
	\hfill
	\begin{subfigure}[t]{0.3\textwidth}
	\centering
\begin{tikzpicture}[scale=0.5]

\definecolor{darkgray176}{RGB}{176,176,176}
\input{tikz_figures/color-def}

\begin{axis}[
tick align=outside,
tick pos=left,
x grid style={darkgray176},
xlabel={$k$},
xmajorgrids,
xmin=1, xmax=51,
xtick style={color=black, font=\huge},
y grid style={darkgray176},
ylabel={Axiom-Fraction},
ymajorgrids,
ymin=0, ymax=1.05,
legend style={at={(0.97,0.03)}, anchor=south east, legend columns=1, draw=black, font=\huge},
legend cell align={left},
label style={font =\huge},
ticklabel style={color=black, font=\huge},
height =6cm,
width=9cm,
]
\addplot [draw=blue, fill=blue, mark=*, only marks, opacity=0.8]
table{%
x  y
1 1
2 1
3 1
4 1
5 0.972762645914397
6 0.867302688638335
7 0.798084596967279
8 0.69060773480663
9 0.656598818122127
10 0.603500301750151
11 0.641848523748395
12 0.684931506849315
13 0.693000693000693
14 0.679809653297077
15 0.689655172413793
16 0.78064012490242
17 0.752445447705041
18 0.833333333333333
19 0.809061488673139
20 0.886524822695035
21 0.924214417744917
22 0.895255147717099
23 0.921658986175115
24 0.959692898272553
25 0.911577028258888
26 0.950570342205323
27 0.966183574879227
28 0.982318271119843
29 0.98135426889107
30 0.961538461538462
31 0.970873786407767
32 0.991080277502478
33 0.986193293885602
34 0.99304865938431
35 0.99304865938431
36 0.998003992015968
37 1
38 0.985221674876847
39 0.99009900990099
40 0.99601593625498
41 0.99403578528827
42 0.99601593625498
43 0.995024875621891
44 0.999000999000999
45 1
46 1
47 1
48 1
49 1
50 1
};
\addlegendentry{JR}
\addplot [draw=green, fill=green, mark=*, only marks, opacity=0.6]
table{%
x  y
1 1
2 1
3 1
4 1
5 0.961538461538462
6 0.848896434634975
7 0.784929356357928
8 0.683994528043776
9 0.651890482398957
10 0.539374325782093
11 0.531632110579479
12 0.469704086425552
13 0.481927710843373
14 0.452488687782805
15 0.337495781302734
16 0.363372093023256
17 0.330578512396694
18 0.288683602771363
19 0.268672756582483
20 0.237981913374584
21 0.260010400416017
22 0.246669955599408
23 0.258331180573495
24 0.199203187250996
25 0.195350654424692
26 0.222617987533393
27 0.224416517055655
28 0.21551724137931
29 0.211505922165821
30 0.19293845263361
31 0.246852628980499
32 0.23906287353574
33 0.260213374967473
34 0.280112044817927
35 0.28184892897407
36 0.28328611898017
37 0.338066260987153
38 0.335683115139308
39 0.397772474144789
40 0.419111483654652
41 0.479386385426654
42 0.489715964740451
43 0.536480686695279
44 0.584795321637427
45 0.643915003219575
46 0.697350069735007
47 0.770416024653313
48 0.838222967309304
49 0.923361034164358
50 1
};
\addlegendentry{EJR+}
\end{axis}

\end{tikzpicture}
 \caption{Parameters: $p=0.1$ and $\phi = 0.9$}
        \label{fig:res-3}
	\end{subfigure}
    \caption{JR-fraction and EJR+-fraction for committee sizes~$k \in [m]$ across selected instances from the resampling model.}
    \label{fig:shapes-res} 
\end{figure*}

Both the resampling and the Euclidean dataset create each a set of shapes that is more homogeneous, compared to our pabulib dataset\footnote{To approximate the JR, resp. EJR+, fraction of an instance, we sample, as for pabulib, until we found 1000 committees satisfying the axiom.}. However, instancewise, we make similar, unexpected observations as for the pabulib dataset, see Section 4.1. In the following, we want to point out the most eye-catching differences: In the resampling dataset, most (nontrivial) shapes  are quite flat, especially the ones for EJR+. Moreover, each considered election, once it has reached its first local minimum behaves afterwards, roughly speaking, quite monotonously, i.e., only allows 'small zigzags'\footnote{At first glance, one might suspect those are due to sampling errors. However, by closer inspection, one observes that their appearance highly correlates with the number of candidates that are relevant for cohesiveness.}; for some examples, see~\Cref{fig:shapes-res}.

\begin{figure*}[h]
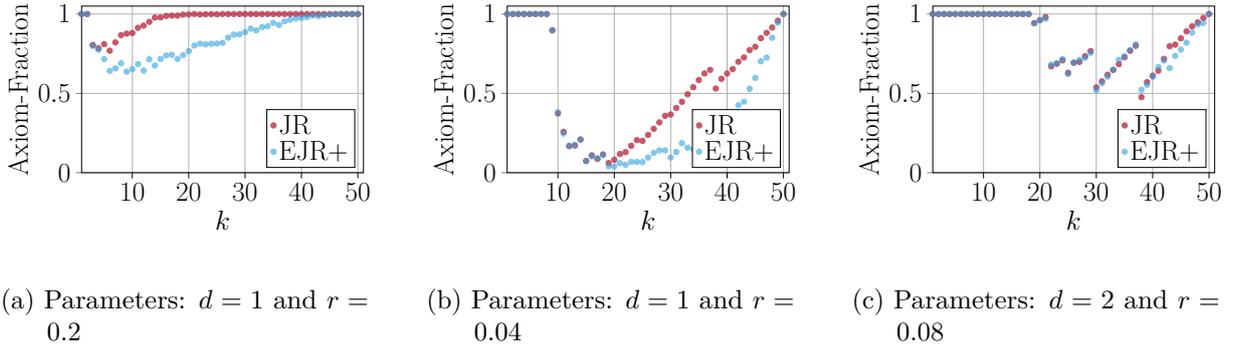

	\centering
	\begin{subfigure}[t]{0.3\textwidth}
	\centering
\begin{tikzpicture}[scale=0.5]

\definecolor{darkgray176}{RGB}{176,176,176}
\input{tikz_figures/color-def}

\begin{axis}[
tick align=outside,
tick pos=left,
x grid style={darkgray176},
xlabel={$k$},
xmajorgrids,
xmin=1, xmax=51,
xtick style={color=black, font=\huge},
y grid style={darkgray176},
ylabel={Axiom-Fraction},
ymajorgrids,
ymin=0, ymax=1.05,
legend style={at={(0.97,0.03)}, anchor=south east, legend columns=1, draw=black, font=\huge},
legend cell align={left},
label style={font =\huge},
ticklabel style={color=black, font=\huge},
height =6cm,
width=9cm,
]
\addplot [draw=blue, fill=blue, mark=*, only marks, opacity=0.8]
table{%
x  y
1 1
2 1
3 0.805152979066023
4 0.783699059561129
5 0.809716599190283
6 0.768639508070715
7 0.821018062397373
8 0.866551126516464
9 0.876424189307625
10 0.880281690140845
11 0.91324200913242
12 0.92678405931418
13 0.950570342205323
14 0.977517106549365
15 0.978473581213307
16 0.988142292490119
17 0.99009900990099
18 0.99009900990099
19 0.99601593625498
20 0.997008973080758
21 1
22 0.997008973080758
23 0.998003992015968
24 0.999000999000999
25 0.998003992015968
26 1
27 1
28 1
29 1
30 1
31 1
32 1
33 1
34 1
35 1
36 1
37 1
38 1
39 1
40 1
41 1
42 1
43 1
44 1
45 1
46 1
47 1
48 1
49 1
50 1
};
\addlegendentry{JR}

\addplot [draw=green, fill=green, mark=*, only marks, opacity=0.6]
table{%
x  y
1 1
2 1
3 0.798722044728435
4 0.775193798449612
5 0.715307582260372
6 0.641848523748395
7 0.657462195923734
8 0.689655172413793
9 0.63653723742839
10 0.651890482398957
11 0.683994528043776
12 0.643500643500644
13 0.714285714285714
14 0.675675675675676
15 0.716845878136201
16 0.738552437223043
17 0.742942050520059
18 0.716845878136201
19 0.740740740740741
20 0.766283524904215
21 0.801282051282051
22 0.813008130081301
23 0.808407437348424
24 0.812347684809098
25 0.813669650122051
26 0.818330605564648
27 0.851063829787234
28 0.870322019147084
29 0.871839581517001
30 0.885739592559787
31 0.908265213442325
32 0.896860986547085
33 0.916590284142988
34 0.922509225092251
35 0.944287063267233
36 0.931098696461825
37 0.954198473282443
38 0.964320154291225
39 0.971817298347911
40 0.975609756097561
41 0.98135426889107
42 0.99009900990099
43 0.987166831194472
44 0.991080277502478
45 0.998003992015968
46 0.998003992015968
47 1
48 1
49 1
50 1
};
\addlegendentry{EJR+}
\end{axis}

\end{tikzpicture}
 \caption{Parameters: $d=1$ and $r=0.2$}
        \label{fig:eukl-1}
\end{subfigure}
\hfill
\begin{subfigure}[t]{0.3\textwidth}
	\centering
\begin{tikzpicture}[scale=0.5]

\definecolor{darkgray176}{RGB}{176,176,176}
\input{tikz_figures/color-def}

\begin{axis}[
tick align=outside,
tick pos=left,
x grid style={darkgray176},
xlabel={$k$},
xmajorgrids,
xmin=1, xmax=51,
xtick style={color=black, font=\huge},
y grid style={darkgray176},
ylabel={Axiom-Fraction},
ymajorgrids,
ymin=0, ymax=1.05,
legend style={at={(0.97,0.03)}, anchor=south east, legend columns=1, draw=black, font=\huge},
legend cell align={left},
label style={font =\huge},
ticklabel style={color=black, font=\huge},
height =6cm,
width=9cm,
]
\addplot [draw=blue, fill=blue, mark=*, only marks, opacity=0.8]
table{%
x  y
1 1
2 1
3 1
4 1
5 1
6 1
7 1
8 1
9 0.896860986547085
10 0.373552484124019
11 0.257466529351184
12 0.167084377610693
13 0.170998632010944
14 0.209467951403435
15 0.0737245650250664
16 0.109553023663453
17 0.0866776458351391
18 0.115393491807062
19 0.0611583389395144
20 0.0818464560484531
21 0.117882824472474
22 0.129516901955705
23 0.169952413324269
24 0.205212394828648
25 0.200240288346015
26 0.238549618320611
27 0.276548672566372
28 0.317158261972724
29 0.358551452133381
30 0.366703337000367
31 0.408329930583912
32 0.446827524575514
33 0.494071146245059
34 0.53850296176629
35 0.583771161704612
36 0.625390869293308
37 0.648088139987038
38 0.530503978779841
39 0.591715976331361
40 0.625
41 0.653167864141084
42 0.699300699300699
43 0.726216412490922
44 0.772797527047913
45 0.793650793650794
46 0.847457627118644
47 0.881057268722467
48 0.914913083257091
49 0.957854406130268
50 1
};
\addlegendentry{JR}
\addplot [draw=green, fill=green, mark=*, only marks, opacity=0.6]
table{%
x  y
1 1
2 1
3 1
4 1
5 1
6 1
7 1
8 1
9 0.897666068222621
10 0.380517503805175
11 0.246913580246914
12 0.170561146170902
13 0.176087339320303
14 0.208550573514077
15 0.0731582412758797
16 0.102796052631579
17 0.0916842394792335
18 0.111982082866741
19 0.0402220255812083
20 0.0394026557389968
21 0.0603281853281853
22 0.049649967727521
23 0.0681709727997818
24 0.0679855870555442
25 0.0666755567408988
26 0.0958772770853308
27 0.124750499001996
28 0.140508641281439
29 0.141322781232335
30 0.0958221540820238
31 0.13051422605064
32 0.18628912071535
33 0.158353127474268
34 0.147514382652309
35 0.170881749829118
36 0.215146299483649
37 0.276778300581234
38 0.225733634311512
39 0.287604256542997
40 0.271665308340125
41 0.342114266164899
42 0.426075841499787
43 0.447427293064877
44 0.528820729772607
45 0.597371565113501
46 0.702247191011236
47 0.725689404934688
48 0.850340136054422
49 0.945179584120983
50 1
};
\addlegendentry{EJR+}
\end{axis}

\end{tikzpicture}
 \caption{Parameters: $d=1$ and $r=0.04$}
        \label{fig:eukl-2}
	\end{subfigure}
	\hfill
	\begin{subfigure}[t]{0.3\textwidth}
	\centering
\begin{tikzpicture}[scale=0.5]

\definecolor{darkgray176}{RGB}{176,176,176}
\input{tikz_figures/color-def}

\begin{axis}[
tick align=outside,
tick pos=left,
x grid style={darkgray176},
xlabel={$k$},
xmajorgrids,
xmin=1, xmax=51,
xtick style={color=black, font=\huge},
y grid style={darkgray176},
ylabel={Axiom-Fraction},
ymajorgrids,
ymin=0, ymax=1.05,
legend style={at={(0.97,0.03)}, anchor=south east, legend columns=1, draw=black, font=\huge},
legend cell align={left},
label style={font =\huge},
ticklabel style={color=black, font=\huge},
height =6cm,
width=9cm,
]
\addplot [draw=blue, fill=blue, mark=*, only marks, opacity=0.8]
table{%
x  y
1 1
2 1
3 1
4 1
5 1
6 1
7 1
8 1
9 1
10 1
11 1
12 1
13 1
14 1
15 1
16 1
17 1
18 1
19 0.942507068803016
20 0.961538461538462
21 0.980392156862745
22 0.669792364367046
23 0.686341798215511
24 0.707714083510262
25 0.629326620516048
26 0.693481276005548
27 0.698324022346369
28 0.733675715333822
29 0.766283524904215
30 0.537345513164965
31 0.575705238917674
32 0.617283950617284
33 0.647249190938511
34 0.683526999316473
35 0.727272727272727
36 0.768639508070715
37 0.8
38 0.475737392959087
39 0.570450656018254
40 0.611620795107034
41 0.643086816720257
42 0.717360114777618
43 0.796812749003984
44 0.807102502017756
45 0.846740050804403
46 0.890471950133571
47 0.922509225092251
48 0.946969696969697
49 0.975609756097561
50 1
};
\addlegendentry{JR}
\addplot [draw=green, fill=green, mark=*, only marks, opacity=0.6]
table{%
x  y
1 1
2 1
3 1
4 1
5 1
6 1
7 1
8 1
9 1
10 1
11 1
12 1
13 1
14 1
15 1
16 1
17 1
18 1
19 0.941619585687382
20 0.960614793467819
21 0.969932104752667
22 0.681663258350375
23 0.689179875947622
24 0.716332378223496
25 0.62266500622665
26 0.692041522491349
27 0.710732054015636
28 0.723589001447178
29 0.755857898715042
30 0.521104742053153
31 0.563380281690141
32 0.605326876513317
33 0.650195058517555
34 0.711743772241993
35 0.734214390602056
36 0.772200772200772
37 0.809061488673139
38 0.522739153162572
39 0.553709856035437
40 0.604960677555959
41 0.66577896138482
42 0.706713780918728
43 0.659195781147001
44 0.735835172921266
45 0.775795190069822
46 0.819672131147541
47 0.88261253309797
48 0.934579439252336
49 0.943396226415094
50 1
};
\addlegendentry{EJR+}
\end{axis}

\end{tikzpicture}
 \caption{Parameters: $d=2$ and $r=0.08$}
        \label{fig:eukl-3}
	\end{subfigure}
    \caption{JR-fraction and EJR+-fraction for committee sizes~$k \in [m]$ across selected instances from the Euclidean model.}
    \label{fig:shapes-eukl} 
\end{figure*}

Although the observations we made above for the resampling dataset, hold for the majority of the Euclidean instances as well, see~\Cref{fig:shapes-eukl}, there is one striking phenomenon that we only observe, at least to this extent, on the Euclidean dataset: In \Cref{fig:eukl-3} we observe that for both plots, for JR and EJR+, the shape contains some piecewise linear patterns--which are almost parallel to each other--with rapid jumps in between two. At least for those instances which exhibit this behaviour clearly, we observe that, restricted to a single linear piece, the number of candidates that are relevant for cohesiveness does not change, and differs to the next one. However, although we think that this behaviour is far from accidental, one should be careful to reduce one on the other without any further investigation.

\paragraph{JR/EJR+-fractions.}

\begin{figure*}[htbp]
	\centering
\begin{subfigure}[t]{0.47\textwidth}
	\centering
    \resizebox{\linewidth}{!}{
\begin{tikzpicture}

\definecolor{darkgray176}{RGB}{176,176,176}
\definecolor{green}{RGB}{0,128,0}
\definecolor{lightgray204}{RGB}{204,204,204}
\input{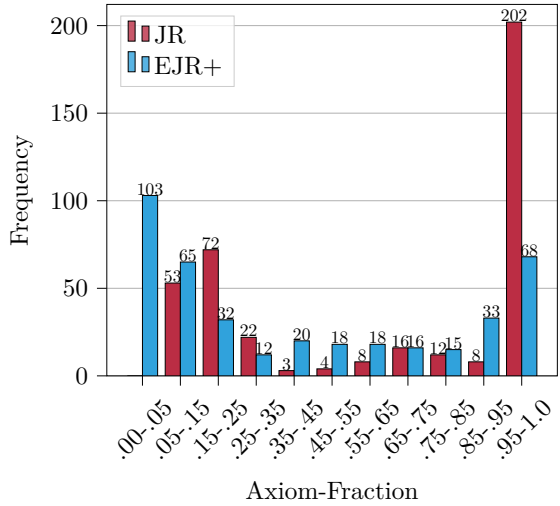}

\begin{axis}[
legend cell align={left},
legend style={
  fill opacity=0.8,
  draw opacity=1,
  text opacity=1,
  at={(0.03,0.97)},
  anchor=north west,
  draw=lightgray204
},
tick align=outside,
tick pos=left,
x grid style={darkgray176},
xlabel={Axiom-Fraction},
xmin=-0.94, xmax=10.94,
xtick style={color=black},
xtick={0,1,2,3,4,5,6,7,8,9,10},
xticklabel style={rotate=45.0},
xticklabels={
  .00-.05,
  .05-.15,
  .15-.25,
  .25-.35,
  .35-.45,
  .45-.55,
  .55-.65,
  .65-.75,
  .75-.85,
  .85-.95,
  .95-1.0
},
y grid style={darkgray176},
ylabel={Frequency},
ymajorgrids,
ymin=0, ymax=212.1,
ytick style={color=black}
]
\draw[draw=black,fill=blue] (axis cs:-0.4,0) rectangle (axis cs:0,0);
\addlegendimage{ybar,ybar legend,draw=black,fill=blue}
\addlegendentry{JR}

\draw[draw=black,fill=blue] (axis cs:0.6,0) rectangle (axis cs:1,53);
\draw[draw=black,fill=blue] (axis cs:1.6,0) rectangle (axis cs:2,72);
\draw[draw=black,fill=blue] (axis cs:2.6,0) rectangle (axis cs:3,22);
\draw[draw=black,fill=blue] (axis cs:3.6,0) rectangle (axis cs:4,3);
\draw[draw=black,fill=blue] (axis cs:4.6,0) rectangle (axis cs:5,4);
\draw[draw=black,fill=blue] (axis cs:5.6,0) rectangle (axis cs:6,8);
\draw[draw=black,fill=blue] (axis cs:6.6,0) rectangle (axis cs:7,16);
\draw[draw=black,fill=blue] (axis cs:7.6,0) rectangle (axis cs:8,12);
\draw[draw=black,fill=blue] (axis cs:8.6,0) rectangle (axis cs:9,8);
\draw[draw=black,fill=blue] (axis cs:9.6,0) rectangle (axis cs:10,202);
\draw[draw=black,fill=green] (axis cs:-2.77555756156289e-17,0) rectangle (axis cs:0.4,103);
\addlegendimage{ybar,ybar legend,draw=black,fill=green}
\addlegendentry{EJR+}

\draw[draw=black,fill=green] (axis cs:1,0) rectangle (axis cs:1.4,65);
\draw[draw=black,fill=green] (axis cs:2,0) rectangle (axis cs:2.4,32);
\draw[draw=black,fill=green] (axis cs:3,0) rectangle (axis cs:3.4,12);
\draw[draw=black,fill=green] (axis cs:4,0) rectangle (axis cs:4.4,20);
\draw[draw=black,fill=green] (axis cs:5,0) rectangle (axis cs:5.4,18);
\draw[draw=black,fill=green] (axis cs:6,0) rectangle (axis cs:6.4,18);
\draw[draw=black,fill=green] (axis cs:7,0) rectangle (axis cs:7.4,16);
\draw[draw=black,fill=green] (axis cs:8,0) rectangle (axis cs:8.4,15);
\draw[draw=black,fill=green] (axis cs:9,0) rectangle (axis cs:9.4,33);
\draw[draw=black,fill=green] (axis cs:10,0) rectangle (axis cs:10.4,68);
\draw (axis cs:0.2,103.5) node[
  scale=0.7,
  anchor=base,
  text=black,
  rotate=0.0
]{103};
\draw (axis cs:0.8,53.5) node[
  scale=0.7,
  anchor=base,
  text=black,
  rotate=0.0
]{53};
\draw (axis cs:1.2,65.5) node[
  scale=0.7,
  anchor=base,
  text=black,
  rotate=0.0
]{65};
\draw (axis cs:1.8,72.5) node[
  scale=0.7,
  anchor=base,
  text=black,
  rotate=0.0
]{72};
\draw (axis cs:2.2,32.5) node[
  scale=0.7,
  anchor=base,
  text=black,
  rotate=0.0
]{32};
\draw (axis cs:2.8,22.5) node[
  scale=0.7,
  anchor=base,
  text=black,
  rotate=0.0
]{22};
\draw (axis cs:3.2,12.5) node[
  scale=0.7,
  anchor=base,
  text=black,
  rotate=0.0
]{12};
\draw (axis cs:3.8,3.5) node[
  scale=0.7,
  anchor=base,
  text=black,
  rotate=0.0
]{3};
\draw (axis cs:4.2,20.5) node[
  scale=0.7,
  anchor=base,
  text=black,
  rotate=0.0
]{20};
\draw (axis cs:4.8,4.5) node[
  scale=0.7,
  anchor=base,
  text=black,
  rotate=0.0
]{4};
\draw (axis cs:5.2,18.5) node[
  scale=0.7,
  anchor=base,
  text=black,
  rotate=0.0
]{18};
\draw (axis cs:5.8,8.5) node[
  scale=0.7,
  anchor=base,
  text=black,
  rotate=0.0
]{8};
\draw (axis cs:6.2,18.5) node[
  scale=0.7,
  anchor=base,
  text=black,
  rotate=0.0
]{18};
\draw (axis cs:6.8,16.5) node[
  scale=0.7,
  anchor=base,
  text=black,
  rotate=0.0
]{16};
\draw (axis cs:7.2,16.5) node[
  scale=0.7,
  anchor=base,
  text=black,
  rotate=0.0
]{16};
\draw (axis cs:7.8,12.5) node[
  scale=0.7,
  anchor=base,
  text=black,
  rotate=0.0
]{12};
\draw (axis cs:8.2,15.5) node[
  scale=0.7,
  anchor=base,
  text=black,
  rotate=0.0
]{15};
\draw (axis cs:8.8,8.5) node[
  scale=0.7,
  anchor=base,
  text=black,
  rotate=0.0
]{8};
\draw (axis cs:9.2,33.5) node[
  scale=0.7,
  anchor=base,
  text=black,
  rotate=0.0
]{33};
\draw (axis cs:9.8,202.5) node[
  scale=0.7,
  anchor=base,
  text=black,
  rotate=0.0
]{202};
\draw (axis cs:10.2,68.5) node[
  scale=0.7,
  anchor=base,
  text=black,
  rotate=0.0
]{68};
\end{axis}

\end{tikzpicture}}
    \caption{Resampling dataset (400 instances)}
        \label{fig:histgram-ratio-resampling}
	\end{subfigure}
	\hfill
	\begin{subfigure}[t]{0.47\textwidth}
	\centering
\resizebox{\linewidth}{!}{
\begin{tikzpicture}

\definecolor{darkgray176}{RGB}{176,176,176}
\definecolor{green}{RGB}{0,128,0}
\definecolor{lightgray204}{RGB}{204,204,204}
\input{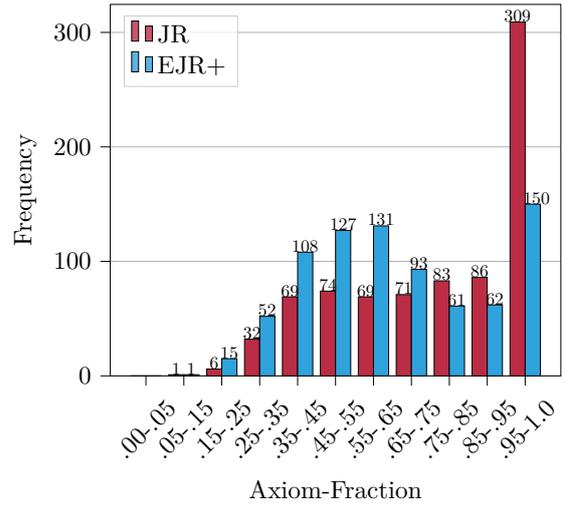}

\begin{axis}[
legend cell align={left},
legend style={
  fill opacity=0.8,
  draw opacity=1,
  text opacity=1,
  at={(0.03,0.97)},
  anchor=north west,
  draw=lightgray204,
},
tick align=outside,
tick pos=left,
x grid style={darkgray176},
xlabel={Axiom-Fraction},
xmin=-0.94, xmax=10.94,
xtick style={color=black},
xtick={0,1,2,3,4,5,6,7,8,9,10},
xticklabel style={rotate=45.0},
xticklabels={
  .00-.05,
  .05-.15,
  .15-.25,
  .25-.35,
  .35-.45,
  .45-.55,
  .55-.65,
  .65-.75,
  .75-.85,
  .85-.95,
  .95-1.0
},
y grid style={darkgray176},
ylabel={Frequency},
ymajorgrids,
ymin=0, ymax=324.45,
ytick style={color=black}
]
\draw[draw=black,fill=blue] (axis cs:-0.4,0) rectangle (axis cs:0,0);
\addlegendimage{ybar,ybar legend,draw=black,fill=blue}
\addlegendentry{JR}

\draw[draw=black,fill=blue] (axis cs:0.6,0) rectangle (axis cs:1,1);
\draw[draw=black,fill=blue] (axis cs:1.6,0) rectangle (axis cs:2,6);
\draw[draw=black,fill=blue] (axis cs:2.6,0) rectangle (axis cs:3,32);
\draw[draw=black,fill=blue] (axis cs:3.6,0) rectangle (axis cs:4,69);
\draw[draw=black,fill=blue] (axis cs:4.6,0) rectangle (axis cs:5,74);
\draw[draw=black,fill=blue] (axis cs:5.6,0) rectangle (axis cs:6,69);
\draw[draw=black,fill=blue] (axis cs:6.6,0) rectangle (axis cs:7,71);
\draw[draw=black,fill=blue] (axis cs:7.6,0) rectangle (axis cs:8,83);
\draw[draw=black,fill=blue] (axis cs:8.6,0) rectangle (axis cs:9,86);
\draw[draw=black,fill=blue] (axis cs:9.6,0) rectangle (axis cs:10,309);
\draw[draw=black,fill=green] (axis cs:-2.77555756156289e-17,0) rectangle (axis cs:0.4,0);
\addlegendimage{ybar,ybar legend,draw=black,fill=green}
\addlegendentry{EJR+}

\draw[draw=black,fill=green] (axis cs:1,0) rectangle (axis cs:1.4,1);
\draw[draw=black,fill=green] (axis cs:2,0) rectangle (axis cs:2.4,15);
\draw[draw=black,fill=green] (axis cs:3,0) rectangle (axis cs:3.4,52);
\draw[draw=black,fill=green] (axis cs:4,0) rectangle (axis cs:4.4,108);
\draw[draw=black,fill=green] (axis cs:5,0) rectangle (axis cs:5.4,127);
\draw[draw=black,fill=green] (axis cs:6,0) rectangle (axis cs:6.4,131);
\draw[draw=black,fill=green] (axis cs:7,0) rectangle (axis cs:7.4,93);
\draw[draw=black,fill=green] (axis cs:8,0) rectangle (axis cs:8.4,61);
\draw[draw=black,fill=green] (axis cs:9,0) rectangle (axis cs:9.4,62);
\draw[draw=black,fill=green] (axis cs:10,0) rectangle (axis cs:10.4,150);
\draw (axis cs:0.8,1.5) node[
  scale=0.7,
  anchor=base,
  text=black,
  rotate=0.0
]{1};
\draw (axis cs:1.2,1.5) node[
  scale=0.7,
  anchor=base,
  text=black,
  rotate=0.0
]{1};
\draw (axis cs:1.8,6.5) node[
  scale=0.7,
  anchor=base,
  text=black,
  rotate=0.0
]{6};
\draw (axis cs:2.2,15.5) node[
  scale=0.7,
  anchor=base,
  text=black,
  rotate=0.0
]{15};
\draw (axis cs:2.8,32.5) node[
  scale=0.7,
  anchor=base,
  text=black,
  rotate=0.0
]{32};
\draw (axis cs:3.2,52.5) node[
  scale=0.7,
  anchor=base,
  text=black,
  rotate=0.0
]{52};
\draw (axis cs:3.8,69.5) node[
  scale=0.7,
  anchor=base,
  text=black,
  rotate=0.0
]{69};
\draw (axis cs:4.2,108.5) node[
  scale=0.7,
  anchor=base,
  text=black,
  rotate=0.0
]{108};
\draw (axis cs:4.8,74.5) node[
  scale=0.7,
  anchor=base,
  text=black,
  rotate=0.0
]{74};
\draw (axis cs:5.2,127.5) node[
  scale=0.7,
  anchor=base,
  text=black,
  rotate=0.0
]{127};
\draw (axis cs:5.8,69.5) node[
  scale=0.7,
  anchor=base,
  text=black,
  rotate=0.0
]{69};
\draw (axis cs:6.2,131.5) node[
  scale=0.7,
  anchor=base,
  text=black,
  rotate=0.0
]{131};
\draw (axis cs:6.8,71.5) node[
  scale=0.7,
  anchor=base,
  text=black,
  rotate=0.0
]{71};
\draw (axis cs:7.2,93.5) node[
  scale=0.7,
  anchor=base,
  text=black,
  rotate=0.0
]{93};
\draw (axis cs:7.8,83.5) node[
  scale=0.7,
  anchor=base,
  text=black,
  rotate=0.0
]{83};
\draw (axis cs:8.2,61.5) node[
  scale=0.7,
  anchor=base,
  text=black,
  rotate=0.0
]{61};
\draw (axis cs:8.8,86.5) node[
  scale=0.7,
  anchor=base,
  text=black,
  rotate=0.0
]{86};
\draw (axis cs:9.2,62.5) node[
  scale=0.7,
  anchor=base,
  text=black,
  rotate=0.0
]{62};
\draw (axis cs:9.8,309.5) node[
  scale=0.7,
  anchor=base,
  text=black,
  rotate=0.0
]{309};
\draw (axis cs:10.3,150.5) node[
  scale=0.7,
  anchor=base,
  text=black,
  rotate=0.0
]{150};
\end{axis}

\end{tikzpicture}}
 \caption{Euclidean dataset (800 instances)}
        \label{histogram-ratio-Euclidean}
	\end{subfigure}
    \caption{Frequency histograms displaying the distribution of JR and EJR+ fractions across instances in our two synthetic datasets.}
    \label{fig:histogram-ratio-synth}
\end{figure*}

For each election--defined by a specific parameter combination and seed--we sample committees until we have found 20000 JR, resp. EJR+, committees. Using the sampled data, we approximate the JR and EJR+ fractions, as well as the candidates' prevalences and power indices under these axioms.\\
\\
In \Cref{fig:histogram-ratio-synth}, we give two frequency histograms for the resampling and the Euclidean dataset, showing the distribution of their JR and EJR+ fractions. As for pabulib, we observe quite a diverse distribution: On the one hand, many instances exhibit high fractions, but on the other hand, there are also numerous instances where only a small fraction of committees satisfy these axioms; this applies in particular for EJR+ in the resampling dataset.

\paragraph{Distances Between Committees.}

\begin{figure*}[htbp]
	\centering
\begin{subfigure}[t]{0.47\textwidth}
	\centering
\resizebox{\linewidth}{!}{\input{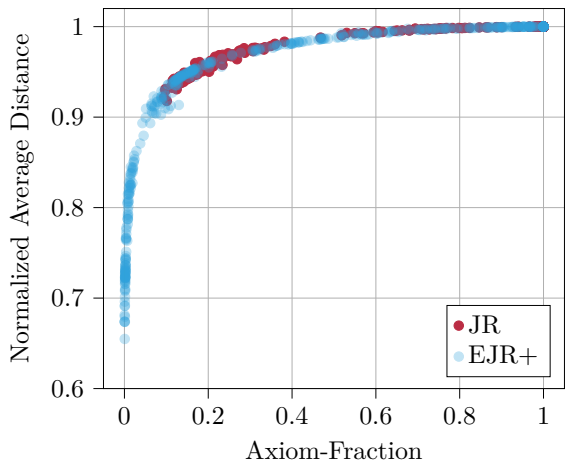}}
 \caption{Resampling dataset (400 instances)}
        \label{fig:corr-resampling}
	\end{subfigure}
	\hfill
	\begin{subfigure}[t]{0.47\textwidth}
	\centering
\resizebox{\linewidth}{!}{\input{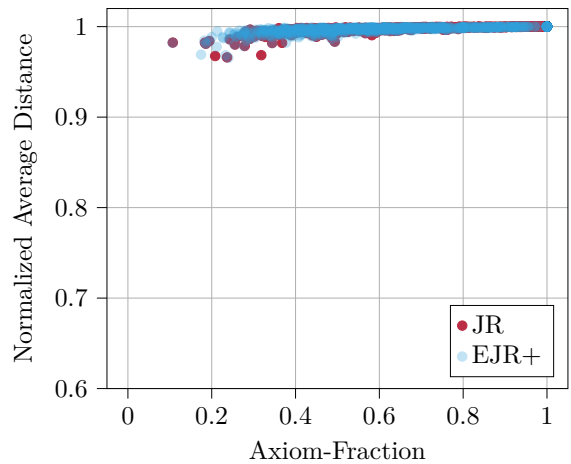}}
 \caption{Euclidean dataset (800 instances)}
        \label{corr-Euclidean}
	\end{subfigure}
    \caption{Each point represents one instance and one axiom (JR in red, EJR+ in blue). The plot shows the correlation between the axiom-fraction and the normalized average distance among committees fulfilling the axiom.}
    \label{fig:histogram-corr-synth}
\end{figure*}

In \Cref{fig:histogram-corr-synth} we show a plot illustrating the correlation between the JR-, resp. EJR+-, fraction and the normalized average distance for our resampling and Euclidean dataset, analogously to~\Cref{fig:ratio-vs-distance}. Recall that for each instance we compute the normalized average distance by computing the distance for each pair of sampled JR, resp EJR+, committees, taking the average and dividing the result by the expected distance of two randomly drawn committees. We observe that the resampling instances induce almost a logarithmic curve: Instances for JR and EJR+ lie indeed on the same one, but the ones for JR form a strict subgraph of the curve induced by EJR+. Note that also instances with rather small JR, resp. EJR+, fraction  exhibit quite high average distances. This phenomenon is particularly strong on the Euclidean dataset: All average distances are higher than 0.96 for both, JR and EJR+. Nevertheless, by rescaling the axes, we could also observe a logarithmic behaviour for this dataset. 

To finish this discussion, we want to take a small detour: Note that, theoretically, also normalized average distances greater than one are possible (or at least we are not aware of any theoretical result excluding them). However, we did not observe them\footnote{To be precise, we did observe values in the order of magnitude of $1 + 10^{-5}$, but we assume these are due to sampling errors.}, which does not seem surprising to us: Intuitively, the existence of values larger than one should indicate that the set of JR, resp. EJR+-committees of an election can be clustered such that committees within one cluster are quite similar to each other, and committees from different clusters have a somehow large distance from each other.

\begin{figure}[htbp]
    \centering
    \begin{tikzpicture}
\definecolor{darkgray176}{RGB}{176,176,176}
\definecolor{green}{RGB}{0,128,0}
\definecolor{orange}{RGB}{255,165,0}
\input{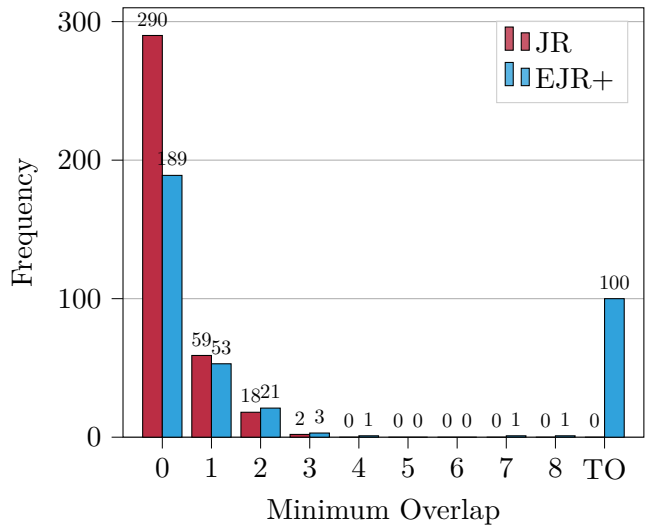}

\begin{axis}[
legend cell align={left},
legend style={
  fill opacity=0.8,
  draw opacity=1,
  text opacity=1,
  at={(0.97,0.97)},  
  anchor=north east,
  draw=lightgray204
},
tick align=outside,
tick pos=left,
x grid style={darkgray176},
xlabel={Minimum Overlap},
xmin=-0.8, xmax=9.8,
xtick style={color=black},
xtick={0,1,2,3,4,5,6,7,8,9},
xticklabel style={},
xticklabels={0, 1, 2,3,4,5,6,7,8, TO},
y grid style={darkgray176},
ylabel={Frequency},
ymajorgrids,
ymin=0, ymax=310,
ytick style={color=black}
]
\draw[draw=black,fill=blue] (axis cs:-0.4,0) rectangle (axis cs:0,290);
\addlegendimage{ybar,ybar legend,draw=black,fill=blue}
\addlegendentry{JR}

\draw[draw=black,fill=blue] (axis cs:0.6,0) rectangle (axis cs:1,59);
\draw[draw=black,fill=blue] (axis cs:1.6,0) rectangle (axis cs:2,18);
\draw[draw=black,fill=blue] (axis cs:2.6,0) rectangle (axis cs:3,2);
\draw[draw=black,fill=blue] (axis cs:3.6,0) rectangle (axis cs:4,0);
\draw[draw=black,fill=blue] (axis cs:4.6,0) rectangle (axis cs:5,0);
\draw[draw=black,fill=blue] (axis cs:5.6,0) rectangle (axis cs:6,0);
\draw[draw=black,fill=blue] (axis cs:6.6,0) rectangle (axis cs:7,0);
\draw[draw=black,fill=blue] (axis cs:7.6,0) rectangle (axis cs:8,0);
\draw[draw=black,fill=blue] (axis cs:8.6,0) rectangle (axis cs:9,0);
\draw[draw=black,fill=green] (axis cs:-2.77555756156289e-17,0) rectangle (axis cs:0.4,189);
\addlegendimage{ybar,ybar legend,draw=black,fill=green}
\addlegendentry{EJR+}

\draw[draw=black,fill=green] (axis cs:1,0) rectangle (axis cs:1.4,53);
\draw[draw=black,fill=green] (axis cs:2,0) rectangle (axis cs:2.4,21);
\draw[draw=black,fill=green] (axis cs:3,0) rectangle (axis cs:3.4,3);
\draw[draw=black,fill=green] (axis cs:4,0) rectangle (axis cs:4.4,1);
\draw[draw=black,fill=green] (axis cs:5,0) rectangle (axis cs:5.4,0);
\draw[draw=black,fill=green] (axis cs:6,0) rectangle (axis cs:6.4,0);
\draw[draw=black,fill=green] (axis cs:7,0) rectangle (axis cs:7.4,1);
\draw[draw=black,fill=green] (axis cs:8,0) rectangle (axis cs:8.4,1);
\draw[draw=black,fill=green] (axis cs:9,0) rectangle (axis cs:9.4,100);
\draw (axis cs:-0.2,292) node[
  scale=0.7,
  anchor=south,
  text=black,
  rotate=0.0
]{290};
\draw (axis cs:0.8,61) node[
  scale=0.7,
  anchor=south,
  text=black,
  rotate=0.0
]{59};
\draw (axis cs:1.8,20) node[
  scale=0.7,
  anchor=south,
  text=black,
  rotate=0.0
]{18};
\draw (axis cs:2.8,4) node[
  scale=0.7,
  anchor=south,
  text=black,
  rotate=0.0
]{2};
\draw (axis cs:3.8,2) node[
  scale=0.7,
  anchor=south,
  text=black,
  rotate=0.0
]{0};
\draw (axis cs:4.8,2) node[
  scale=0.7,
  anchor=south,
  text=black,
  rotate=0.0
]{0};
\draw (axis cs:5.8,2) node[
  scale=0.7,
  anchor=south,
  text=black,
  rotate=0.0
]{0};
\draw (axis cs:6.8,2) node[
  scale=0.7,
  anchor=south,
  text=black,
  rotate=0.0
]{0};
\draw (axis cs:7.8,2) node[
  scale=0.7,
  anchor=south,
  text=black,
  rotate=0.0
]{0};
\draw (axis cs:8.8,2) node[
  scale=0.7,
  anchor=south,
  text=black,
  rotate=0.0
]{0};
\draw (axis cs:0.2,191) node[
  scale=0.7,
  anchor=south,
  text=black,
  rotate=0.0
]{189};
\draw (axis cs:1.2,55) node[
  scale=0.7,
  anchor=south,
  text=black,
  rotate=0.0
]{53};
\draw (axis cs:2.2,23) node[
  scale=0.7,
  anchor=south,
  text=black,
  rotate=0.0
]{21};
\draw (axis cs:3.2,5) node[
  scale=0.7,
  anchor=south,
  text=black,
  rotate=0.0
]{3};
\draw (axis cs:4.2,3) node[
  scale=0.7,
  anchor=south,
  text=black,
  rotate=0.0
]{1};
\draw (axis cs:5.2,2) node[
  scale=0.7,
  anchor=south,
  text=black,
  rotate=0.0
]{0};
\draw (axis cs:6.2,2) node[
  scale=0.7,
  anchor=south,
  text=black,
  rotate=0.0
]{0};
\draw (axis cs:7.2,3) node[
  scale=0.7,
  anchor=south,
  text=black,
  rotate=0.0
]{1};
\draw (axis cs:8.2,3) node[
  scale=0.7,
  anchor=south,
  text=black,
  rotate=0.0
]{1};
\draw (axis cs:9.2,102) node[
  scale=0.7,
  anchor=south,
  text=black,
  rotate=0.0
]{100};
\end{axis}
\end{tikzpicture}
    \caption{Histogram depicting the distribution of the minimum overlaps between pairs of JR, resp. EJR+, committees in the pabulib dataset. The bar labeled ‘TO’ represents instances that were not solved within the time limit.}
    \label{fig:min-overlap-histo}
\end{figure}

As discussed in Section 4.1, we also computed, via an ILP formulation, the minimum overlap two JR, resp. EJR+, committees can have for each pabulib instance for~$k=\lfloor m/2 \rfloor$, see~\Cref{fig:min-overlap-histo}. Our results confirm our impression from the preceding discussion regarding the normalized average distance, that is, that JR, resp. EJR+, committees can be quite diverse. In particular, for a large majority of instances, we find two disjoint JR-committees. For EJR+ we encounter one instance with minimum overlap of 7 (Amsterdam 622, $m=26$) and one with minimum overlap of 8 (Amsterdam 285, $m=40$). Consistently, we observe that both elections have a very low EJR+ fraction (0.003 resp. 0.002), and there are 7, resp. 8, fixed candidates that are contained in every EJR+ committee sampled by us. For 100 instances, we did not find an optimal solution within the 30-minute time limit.

\subsection{What Makes a Candidate Important for Proportionality?}\label{sec:CI}
Analogously to pabulib in Section 4.2, we only consider instances with an EJR+-fraction of~$\leq 0.95\%$ in this part, resulting in a dataset of 650 Euclidean instances, and a dataset of 332 resampling instances. 

\paragraph{Correlation Between Measures.}

In \Cref{tab:pcc-correlations} we give an overview over the correlation between our different candidates' importance measures, for each of our three datasets. To this end, we compute the Pearson correlation coefficient (PCC) between each pair of measures across all candidates within one instance. 

For all three of them we observe that the PCC between the EJR+-prevalence and power index is not only remarkably high in average, but also for the 25\% quartile. While the approval score for elections, following the resampling model, seems to be quite a good measure to predict whether a candidate is likely to be included in an EJR+-committee, this does not hold for the pabulib and Euclidean dataset. Intuitively, this makes sense, since in the resampling model candidates have, roughly speaking, either 'high' or 'low' approval scores, (assuming relatively high values of~$\phi$), whereas we usually do not find such a partition of approval scores in Euclidean elections. Similarly, for real-world data, such as pabulib, it is likely that the approval scores of candidates (or projects in our case) are not either 'high' or 'low'. Last, we want to comment on the fact that, compared to EJR+, the average PCCs, but also all quartiles, are often significantly lower for JR. This can be, at least partially, explained by our filtering: We excluded all instances with an EJR+ fraction greater than 0.95. However, as seen in~\Cref{fig:histogram-ratio-synth} and~\Cref{freq-dist}, many instances remain with a JR-fraction above 0.95--indicating that JR imposes only mild constraints for them. In addition, we explain the differences in the first two columns (pcc(app, prev) and pcc(app, pw)) between JR and EJR+ by the specific nature of the EJR+ definition: When forming an EJR+ committee, it is difficult to avoid including candidates with high approval scores (assuming EJR+ does not imposes only trivial constraints in the given election).

\begin{table*}[htbp]
\centering
\caption{Summary statistics of the Pearson correlation coefficients (PCC) computed between all pairs of candidates' importance measures.
The table reports the mean and quartiles [$Q_{0.25},Q_{0.5},Q_{0.75}$] of the PCC values for both JR and EJR+, across our three datasets: pabulib (267 instances), Euclidean (650 instances), and resampling (332 instances).}
\label{tab:pcc-correlations}
\begin{adjustbox}{width=\textwidth}
\begin{tabular}{llccc}
\toprule
\textbf{Dataset} & \textbf{Axiom} & \textbf{pcc(app, prev)} & \textbf{pcc(app, pw)} & \textbf{pcc(prev, pw)} \\
\midrule
\multirow{2}{*}{pabulib} 
    & JR      & 0.22, [0.01, 0.19, 0.41] & 0.42, [0.15, 0.38, 0.75] & 0.72, [0.42, 0.98, 1.00] \\
    & EJR+    & 0.65, [0.54, 0.75, 0.84] & 0.66, [0.57, 0.75, 0.85] & 0.99, [0.99, 1.00, 1.00] \\
\midrule
\multirow{2}{*}{Euclidean} 
    & JR      & 0.10, [-0.23, 0.08, 0.45] & 0.07, [-0.32, 0.07, 0.47] & 0.84, [0.80, 0.97, 0.99] \\
    & EJR+   & 0.56, [0.40, 0.62, 0.78] & 0.58, [0.42, 0.64, 0.81] & 0.98, [0.98, 0.99, 1.00] \\
\midrule
\multirow{2}{*}{resampling} 
    & JR     & 0.62, [0.23, 0.80, 0.96] & 0.83, [0.78, 0.95, 0.97] & 0.68, [0.25, 0.99, 1.00] \\
    & EJR+   & 0.92, [0.90, 0.95, 0.98] & 0.92, [0.90, 0.95, 0.98] & 0.99, [1.00, 1.00, 1.00] \\
\bottomrule
\end{tabular}
\end{adjustbox}
\end{table*}

\paragraph{Measures and Proportional Voting Rules.}

\begin{figure*}[b!]
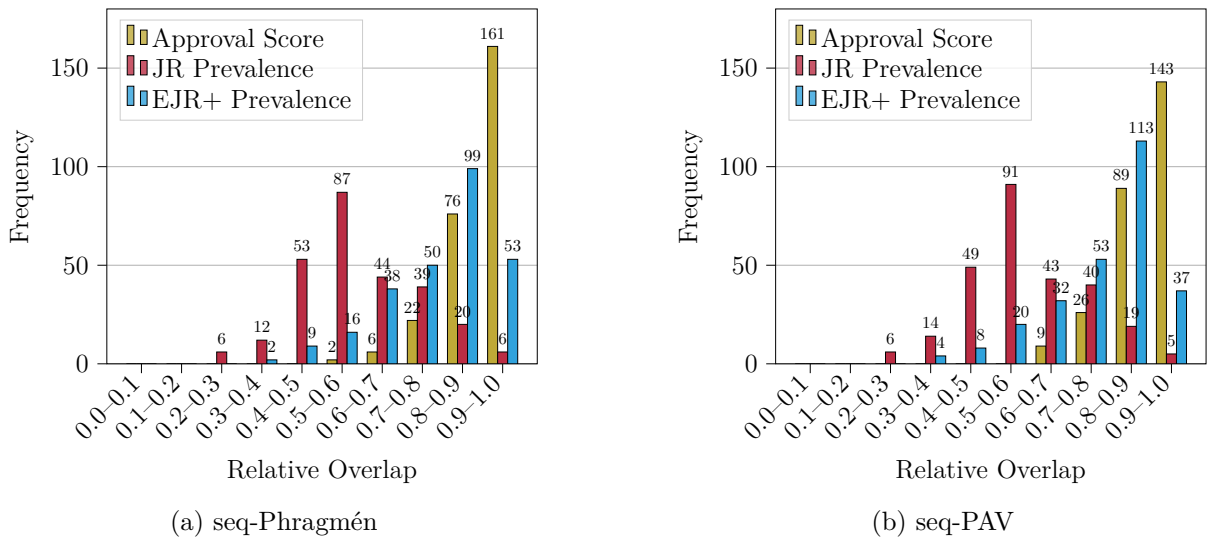

	\centering
\begin{subfigure}[t]{0.45\textwidth}
	\centering
\resizebox{\linewidth}{!}{
\begin{tikzpicture}[scale =0.7]

\definecolor{darkgray176}{RGB}{176,176,176}
\input{tikz_figures/color-def}

\begin{axis}[
legend cell align={left},
legend style={
  fill opacity=0.8,
  draw opacity=1,
  text opacity=1,
  at={(0.03,0.97)},
  anchor=north west,
  draw=lightgray204,
  font = \large
},
tick align=outside,
tick pos=left,
x grid style={darkgray176},
xlabel={Relative Overlap},
xmin=-0.8625, xmax=9.8625,
xtick style={color=black},
xtick={0,1,2,3,4,5,6,7,8,9},
xticklabel style={rotate=45.0,anchor=east},
xticklabels={0.0–0.1,0.1–0.2,0.2–0.3,0.3–0.4,0.4–0.5,0.5–0.6,0.6–0.7,0.7–0.8,0.8–0.9,0.9–1.0},
y grid style={darkgray176},
ylabel={Frequency},
ymajorgrids,
ymin=0, ymax=180,
ytick style={color=black},
tick label style={color=black, font=\large},
label style={font =\large},
]
\draw[draw=black,fill=orange] (axis cs:-0.375,0) rectangle (axis cs:-0.125,0);
\addlegendimage{ybar,ybar legend,draw=black,fill=orange}
\addlegendentry{Approval Score}

\draw[draw=black,fill=orange] (axis cs:0.625,0) rectangle (axis cs:0.875,0);
\draw[draw=black,fill=orange] (axis cs:1.625,0) rectangle (axis cs:1.875,0);
\draw[draw=black,fill=orange] (axis cs:2.625,0) rectangle (axis cs:2.875,0);
\draw[draw=black,fill=orange] (axis cs:3.625,0) rectangle (axis cs:3.875,0);
\draw[draw=black,fill=orange] (axis cs:4.625,0) rectangle (axis cs:4.875,2);
\draw[draw=black,fill=orange] (axis cs:5.625,0) rectangle (axis cs:5.875,6);
\draw[draw=black,fill=orange] (axis cs:6.625,0) rectangle (axis cs:6.875,22);
\draw[draw=black,fill=orange] (axis cs:7.625,0) rectangle (axis cs:7.875,76);
\draw[draw=black,fill=orange] (axis cs:8.625,0) rectangle (axis cs:8.875,161);
\draw[draw=black,fill=blue] (axis cs:-0.125,0) rectangle (axis cs:0.125,0);
\addlegendimage{ybar,ybar legend,draw=black,fill=blue}
\addlegendentry{JR Prevalence}

\draw[draw=black,fill=blue] (axis cs:0.875,0) rectangle (axis cs:1.125,0);
\draw[draw=black,fill=blue] (axis cs:1.875,0) rectangle (axis cs:2.125,6);
\draw[draw=black,fill=blue] (axis cs:2.875,0) rectangle (axis cs:3.125,12);
\draw[draw=black,fill=blue] (axis cs:3.875,0) rectangle (axis cs:4.125,53);
\draw[draw=black,fill=blue] (axis cs:4.875,0) rectangle (axis cs:5.125,87);
\draw[draw=black,fill=blue] (axis cs:5.875,0) rectangle (axis cs:6.125,44);
\draw[draw=black,fill=blue] (axis cs:6.875,0) rectangle (axis cs:7.125,39);
\draw[draw=black,fill=blue] (axis cs:7.875,0) rectangle (axis cs:8.125,20);
\draw[draw=black,fill=blue] (axis cs:8.875,0) rectangle (axis cs:9.125,6);
\draw[draw=black,fill=green] (axis cs:0.125,0) rectangle (axis cs:0.375,0);
\addlegendimage{ybar,ybar legend,draw=black,fill=green}
\addlegendentry{EJR+ Prevalence}

\draw[draw=black,fill=green] (axis cs:1.125,0) rectangle (axis cs:1.375,0);
\draw[draw=black,fill=green] (axis cs:2.125,0) rectangle (axis cs:2.375,0);
\draw[draw=black,fill=green] (axis cs:3.125,0) rectangle (axis cs:3.375,2);
\draw[draw=black,fill=green] (axis cs:4.125,0) rectangle (axis cs:4.375,9);
\draw[draw=black,fill=green] (axis cs:5.125,0) rectangle (axis cs:5.375,16);
\draw[draw=black,fill=green] (axis cs:6.125,0) rectangle (axis cs:6.375,38);
\draw[draw=black,fill=green] (axis cs:7.125,0) rectangle (axis cs:7.375,50);
\draw[draw=black,fill=green] (axis cs:8.125,0) rectangle (axis cs:8.375,99);
\draw[draw=black,fill=green] (axis cs:9.125,0) rectangle (axis cs:9.375,53);
\draw (axis cs:4.75,3) node[
  scale=0.7,
  anchor=south,
  text=black,
  rotate=0.0
]{2};
\draw (axis cs:5.75,7) node[
  scale=0.7,
  anchor=south,
  text=black,
  rotate=0.0
]{6};
\draw (axis cs:6.75,23) node[
  scale=0.7,
  anchor=south,
  text=black,
  rotate=0.0
]{22};
\draw (axis cs:7.75,77) node[
  scale=0.7,
  anchor=south,
  text=black,
  rotate=0.0
]{76};
\draw (axis cs:8.75,162) node[
  scale=0.7,
  anchor=south,
  text=black,
  rotate=0.0
]{161};
\draw (axis cs:2,7) node[
  scale=0.7,
  anchor=south,
  text=black,
  rotate=0.0
]{6};
\draw (axis cs:3,13) node[
  scale=0.7,
  anchor=south,
  text=black,
  rotate=0.0
]{12};
\draw (axis cs:4,54) node[
  scale=0.7,
  anchor=south,
  text=black,
  rotate=0.0
]{53};
\draw (axis cs:5,88) node[
  scale=0.7,
  anchor=south,
  text=black,
  rotate=0.0
]{87};
\draw (axis cs:6,45) node[
  scale=0.7,
  anchor=south,
  text=black,
  rotate=0.0
]{44};
\draw (axis cs:7,40) node[
  scale=0.7,
  anchor=south,
  text=black,
  rotate=0.0
]{39};
\draw (axis cs:8,21) node[
  scale=0.7,
  anchor=south,
  text=black,
  rotate=0.0
]{20};
\draw (axis cs:9,7) node[
  scale=0.7,
  anchor=south,
  text=black,
  rotate=0.0
]{6};
\draw (axis cs:3.25,3) node[
  scale=0.7,
  anchor=south,
  text=black,
  rotate=0.0
]{2};
\draw (axis cs:4.25,10) node[
  scale=0.7,
  anchor=south,
  text=black,
  rotate=0.0
]{9};
\draw (axis cs:5.25,17) node[
  scale=0.7,
  anchor=south,
  text=black,
  rotate=0.0
]{16};
\draw (axis cs:6.25,39) node[
  scale=0.7,
  anchor=south,
  text=black,
  rotate=0.0
]{38};
\draw (axis cs:7.25,51) node[
  scale=0.7,
  anchor=south,
  text=black,
  rotate=0.0
]{50};
\draw (axis cs:8.25,100) node[
  scale=0.7,
  anchor=south,
  text=black,
  rotate=0.0
]{99};
\draw (axis cs:9.25,54) node[
  scale=0.7,
  anchor=south,
  text=black,
  rotate=0.0
]{53};
\end{axis}

\end{tikzpicture}}
 \caption{seq-Phragmén}
        \label{fig:histgram-seqphrag-pabulib}
	\end{subfigure}
	\hfill
    \begin{subfigure}[t]{0.45\textwidth}
	\centering
\resizebox{\linewidth}{!}{
\begin{tikzpicture}[scale =0.7]

\definecolor{darkgray176}{RGB}{176,176,176}
\input{tikz_figures/color-def}

\begin{axis}[
legend cell align={left},
legend style={
  fill opacity=0.8,
  draw opacity=1,
  text opacity=1,
  at={(0.03,0.97)},
  anchor=north west,
  draw=lightgray204,
  font = \large
},
tick align=outside,
tick pos=left,
x grid style={darkgray176},
xlabel={Relative Overlap},
xmin=-0.8625, xmax=9.8625,
xtick style={color=black},
xtick={0,1,2,3,4,5,6,7,8,9},
xticklabel style={rotate=45.0,anchor=east},
xticklabels={0.0–0.1,0.1–0.2,0.2–0.3,0.3–0.4,0.4–0.5,0.5–0.6,0.6–0.7,0.7–0.8,0.8–0.9,0.9–1.0},
y grid style={darkgray176},
ylabel={Frequency},
ymajorgrids,
ymin=0, ymax=180,
ytick style={color=black},
tick label style={color=black, font=\large},
label style={font =\large},
]
\draw[draw=black,fill=orange] (axis cs:-0.375,0) rectangle (axis cs:-0.125,0);
\addlegendimage{ybar,ybar legend,draw=black,fill=orange}
\addlegendentry{Approval Score}

\draw[draw=black,fill=orange] (axis cs:0.625,0) rectangle (axis cs:0.875,0);
\draw[draw=black,fill=orange] (axis cs:1.625,0) rectangle (axis cs:1.875,0);
\draw[draw=black,fill=orange] (axis cs:2.625,0) rectangle (axis cs:2.875,0);
\draw[draw=black,fill=orange] (axis cs:3.625,0) rectangle (axis cs:3.875,0);
\draw[draw=black,fill=orange] (axis cs:4.625,0) rectangle (axis cs:4.875,0);
\draw[draw=black,fill=orange] (axis cs:5.625,0) rectangle (axis cs:5.875,9);
\draw[draw=black,fill=orange] (axis cs:6.625,0) rectangle (axis cs:6.875,26);
\draw[draw=black,fill=orange] (axis cs:7.625,0) rectangle (axis cs:7.875,89);
\draw[draw=black,fill=orange] (axis cs:8.625,0) rectangle (axis cs:8.875,143);
\draw[draw=black,fill=blue] (axis cs:-0.125,0) rectangle (axis cs:0.125,0);
\addlegendimage{ybar,ybar legend,draw=black,fill=blue}
\addlegendentry{JR Prevalence}

\draw[draw=black,fill=blue] (axis cs:0.875,0) rectangle (axis cs:1.125,0);
\draw[draw=black,fill=blue] (axis cs:1.875,0) rectangle (axis cs:2.125,6);
\draw[draw=black,fill=blue] (axis cs:2.875,0) rectangle (axis cs:3.125,14);
\draw[draw=black,fill=blue] (axis cs:3.875,0) rectangle (axis cs:4.125,49);
\draw[draw=black,fill=blue] (axis cs:4.875,0) rectangle (axis cs:5.125,91);
\draw[draw=black,fill=blue] (axis cs:5.875,0) rectangle (axis cs:6.125,43);
\draw[draw=black,fill=blue] (axis cs:6.875,0) rectangle (axis cs:7.125,40);
\draw[draw=black,fill=blue] (axis cs:7.875,0) rectangle (axis cs:8.125,19);
\draw[draw=black,fill=blue] (axis cs:8.875,0) rectangle (axis cs:9.125,5);
\draw[draw=black,fill=green] (axis cs:0.125,0) rectangle (axis cs:0.375,0);
\addlegendimage{ybar,ybar legend,draw=black,fill=green}
\addlegendentry{EJR+ Prevalence}

\draw[draw=black,fill=green] (axis cs:1.125,0) rectangle (axis cs:1.375,0);
\draw[draw=black,fill=green] (axis cs:2.125,0) rectangle (axis cs:2.375,0);
\draw[draw=black,fill=green] (axis cs:3.125,0) rectangle (axis cs:3.375,4);
\draw[draw=black,fill=green] (axis cs:4.125,0) rectangle (axis cs:4.375,8);
\draw[draw=black,fill=green] (axis cs:5.125,0) rectangle (axis cs:5.375,20);
\draw[draw=black,fill=green] (axis cs:6.125,0) rectangle (axis cs:6.375,32);
\draw[draw=black,fill=green] (axis cs:7.125,0) rectangle (axis cs:7.375,53);
\draw[draw=black,fill=green] (axis cs:8.125,0) rectangle (axis cs:8.375,113);
\draw[draw=black,fill=green] (axis cs:9.125,0) rectangle (axis cs:9.375,37);
\draw (axis cs:5.75,10) node[
  scale=0.7,
  anchor=south,
  text=black,
  rotate=0.0
]{9};
\draw (axis cs:6.75,27) node[
  scale=0.7,
  anchor=south,
  text=black,
  rotate=0.0
]{26};
\draw (axis cs:7.75,90) node[
  scale=0.7,
  anchor=south,
  text=black,
  rotate=0.0
]{89};
\draw (axis cs:8.75,144) node[
  scale=0.7,
  anchor=south,
  text=black,
  rotate=0.0
]{143};
\draw (axis cs:2,7) node[
  scale=0.7,
  anchor=south,
  text=black,
  rotate=0.0
]{6};
\draw (axis cs:3,15) node[
  scale=0.7,
  anchor=south,
  text=black,
  rotate=0.0
]{14};
\draw (axis cs:4,50) node[
  scale=0.7,
  anchor=south,
  text=black,
  rotate=0.0
]{49};
\draw (axis cs:5,92) node[
  scale=0.7,
  anchor=south,
  text=black,
  rotate=0.0
]{91};
\draw (axis cs:6,44) node[
  scale=0.7,
  anchor=south,
  text=black,
  rotate=0.0
]{43};
\draw (axis cs:7,41) node[
  scale=0.7,
  anchor=south,
  text=black,
  rotate=0.0
]{40};
\draw (axis cs:8,20) node[
  scale=0.7,
  anchor=south,
  text=black,
  rotate=0.0
]{19};
\draw (axis cs:9,6) node[
  scale=0.7,
  anchor=south,
  text=black,
  rotate=0.0
]{5};
\draw (axis cs:3.25,5) node[
  scale=0.7,
  anchor=south,
  text=black,
  rotate=0.0
]{4};
\draw (axis cs:4.25,9) node[
  scale=0.7,
  anchor=south,
  text=black,
  rotate=0.0
]{8};
\draw (axis cs:5.25,21) node[
  scale=0.7,
  anchor=south,
  text=black,
  rotate=0.0
]{20};
\draw (axis cs:6.25,33) node[
  scale=0.7,
  anchor=south,
  text=black,
  rotate=0.0
]{32};
\draw (axis cs:7.25,54) node[
  scale=0.7,
  anchor=south,
  text=black,
  rotate=0.0
]{53};
\draw (axis cs:8.25,114) node[
  scale=0.7,
  anchor=south,
  text=black,
  rotate=0.0
]{113};
\draw (axis cs:9.25,38) node[
  scale=0.7,
  anchor=south,
  text=black,
  rotate=0.0
]{37};
\end{axis}

\end{tikzpicture}}
 \caption{seq-PAV}
        \label{histogram-seqpav-pabulib}
	\end{subfigure}
    \caption{Histograms showing the distribution of relative overlaps between committees selected by different voting rules and the top $k$ candidates according to our three candidates' importance measures for the pabulib dataset (267 instances).}
    \label{fig:histogram-vot-pabulib}
\end{figure*}

\begin{figure*}[h]
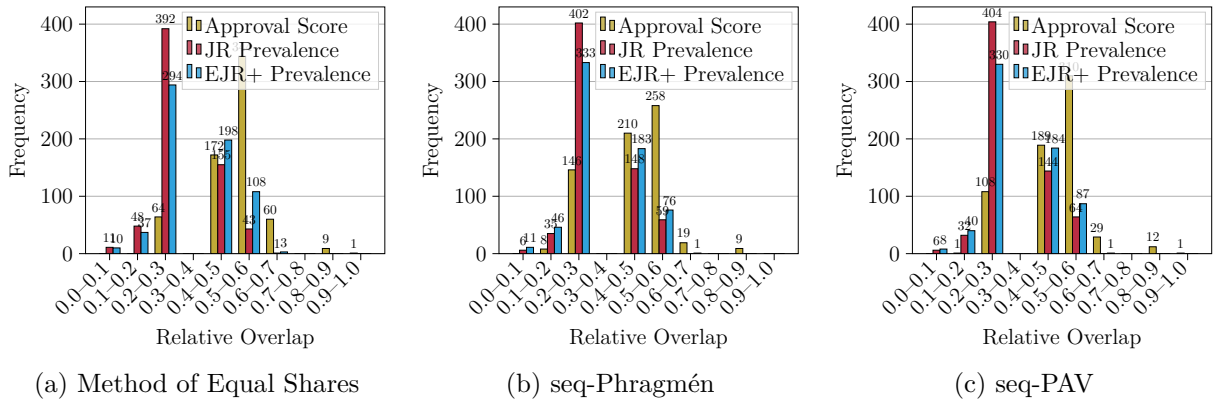

	\centering
    \begin{subfigure}[t]{0.32\textwidth}
	\centering
\resizebox{\linewidth}{!}{
\begin{tikzpicture}[scale=.7]

\definecolor{darkgray176}{RGB}{176,176,176}
\definecolor{lightgray204}{RGB}{204,204,204}
\input{tikz_figures/color-def}

\begin{axis}[
legend cell align={left},
legend style={fill opacity=0.8, draw opacity=1, text opacity=1, draw=lightgray204, font = \Large},
tick align=outside,
tick pos=left,
x grid style={darkgray176},
xlabel={Relative Overlap},
xmin=-0.8625, xmax=9.8625,
xtick style={color=black},
xtick={0,1,2,3,4,5,6,7,8,9},
xticklabel style={rotate=45.0,anchor=east},
xticklabels={0.0–0.1,0.1–0.2,0.2–0.3,0.3–0.4,0.4–0.5,0.5–0.6,0.6–0.7,0.7–0.8,0.8–0.9,0.9–1.0},
y grid style={darkgray176},
ylabel={Frequency},
ymajorgrids,
ymin=0, ymax=430,
ytick style={color=black},
tick label style={color=black, font=\Large},
label style={font =\Large},
]
\draw[draw=black,fill=orange] (axis cs:-0.375,0) rectangle (axis cs:-0.125,0);
\addlegendimage{ybar,ybar legend,draw=black,fill=orange}
\addlegendentry{Approval Score}

\draw[draw=black,fill=orange] (axis cs:0.625,0) rectangle (axis cs:0.875,0);
\draw[draw=black,fill=orange] (axis cs:1.625,0) rectangle (axis cs:1.875,64);
\draw[draw=black,fill=orange] (axis cs:2.625,0) rectangle (axis cs:2.875,0);
\draw[draw=black,fill=orange] (axis cs:3.625,0) rectangle (axis cs:3.875,172);
\draw[draw=black,fill=orange] (axis cs:4.625,0) rectangle (axis cs:4.875,344);
\draw[draw=black,fill=orange] (axis cs:5.625,0) rectangle (axis cs:5.875,60);
\draw[draw=black,fill=orange] (axis cs:6.625,0) rectangle (axis cs:6.875,0);
\draw[draw=black,fill=orange] (axis cs:7.625,0) rectangle (axis cs:7.875,9);
\draw[draw=black,fill=orange] (axis cs:8.625,0) rectangle (axis cs:8.875,1);
\draw[draw=black,fill=blue] (axis cs:-0.125,0) rectangle (axis cs:0.125,11);
\addlegendimage{ybar,ybar legend,draw=black,fill=blue}
\addlegendentry{JR Prevalence}

\draw[draw=black,fill=blue] (axis cs:0.875,0) rectangle (axis cs:1.125,48);
\draw[draw=black,fill=blue] (axis cs:1.875,0) rectangle (axis cs:2.125,392);
\draw[draw=black,fill=blue] (axis cs:2.875,0) rectangle (axis cs:3.125,0);
\draw[draw=black,fill=blue] (axis cs:3.875,0) rectangle (axis cs:4.125,155);
\draw[draw=black,fill=blue] (axis cs:4.875,0) rectangle (axis cs:5.125,43);
\draw[draw=black,fill=blue] (axis cs:5.875,0) rectangle (axis cs:6.125,1);
\draw[draw=black,fill=blue] (axis cs:6.875,0) rectangle (axis cs:7.125,0);
\draw[draw=black,fill=blue] (axis cs:7.875,0) rectangle (axis cs:8.125,0);
\draw[draw=black,fill=blue] (axis cs:8.875,0) rectangle (axis cs:9.125,0);
\draw[draw=black,fill=green] (axis cs:0.125,0) rectangle (axis cs:0.375,10);
\addlegendimage{ybar,ybar legend,draw=black,fill=green}
\addlegendentry{EJR+ Prevalence}

\draw[draw=black,fill=green] (axis cs:1.125,0) rectangle (axis cs:1.375,37);
\draw[draw=black,fill=green] (axis cs:2.125,0) rectangle (axis cs:2.375,294);
\draw[draw=black,fill=green] (axis cs:3.125,0) rectangle (axis cs:3.375,0);
\draw[draw=black,fill=green] (axis cs:4.125,0) rectangle (axis cs:4.375,198);
\draw[draw=black,fill=green] (axis cs:5.125,0) rectangle (axis cs:5.375,108);
\draw[draw=black,fill=green] (axis cs:6.125,0) rectangle (axis cs:6.375,3);
\draw[draw=black,fill=green] (axis cs:7.125,0) rectangle (axis cs:7.375,0);
\draw[draw=black,fill=green] (axis cs:8.125,0) rectangle (axis cs:8.375,0);
\draw[draw=black,fill=green] (axis cs:9.125,0) rectangle (axis cs:9.375,0);
\draw (axis cs:1.75,65) node[
  scale=0.8,
  anchor=south,
  text=black,
  rotate=0.0
]{64};
\draw (axis cs:3.75,173) node[
  scale=0.8,
  anchor=south,
  text=black,
  rotate=0.0
]{172};
\draw (axis cs:4.75,345) node[
  scale=0.8,
  anchor=south,
  text=black,
  rotate=0.0
]{344};
\draw (axis cs:5.75,61) node[
  scale=0.8,
  anchor=south,
  text=black,
  rotate=0.0
]{60};
\draw (axis cs:7.75,10) node[
  scale=0.8,
  anchor=south,
  text=black,
  rotate=0.0
]{9};
\draw (axis cs:8.75,2) node[
  scale=0.8,
  anchor=south,
  text=black,
  rotate=0.0
]{1};
\draw (axis cs:0,12) node[
  scale=0.8,
  anchor=south,
  text=black,
  rotate=0.0
]{11};
\draw (axis cs:1,49) node[
  scale=0.8,
  anchor=south,
  text=black,
  rotate=0.0
]{48};
\draw (axis cs:2,393) node[
  scale=0.8,
  anchor=south,
  text=black,
  rotate=0.0
]{392};
\draw (axis cs:4,156) node[
  scale=0.8,
  anchor=south,
  text=black,
  rotate=0.0
]{155};
\draw (axis cs:5,44) node[
  scale=0.8,
  anchor=south,
  text=black,
  rotate=0.0
]{43};
\draw (axis cs:6,2) node[
  scale=0.8,
  anchor=south,
  text=black,
  rotate=0.0
]{1};
\draw (axis cs:0.25,11) node[
  scale=0.8,
  anchor=south,
  text=black,
  rotate=0.0
]{10};
\draw (axis cs:1.25,38) node[
  scale=0.8,
  anchor=south,
  text=black,
  rotate=0.0
]{37};
\draw (axis cs:2.25,295) node[
  scale=0.8,
  anchor=south,
  text=black,
  rotate=0.0
]{294};
\draw (axis cs:4.25,199) node[
  scale=0.8,
  anchor=south,
  text=black,
  rotate=0.0
]{198};
\draw (axis cs:5.25,109) node[
  scale=0.8,
  anchor=south,
  text=black,
  rotate=0.0
]{108};
\draw (axis cs:6.25,4) node[
  scale=0.8,
  anchor=south,
  text=black,
  rotate=0.0
]{3};
\end{axis}
\end{tikzpicture}}
 \caption{Method of Equal Shares}
        \label{fig:histgram-mes-Euclidean}
	\end{subfigure}
	\hfill
\begin{subfigure}[t]{0.32\textwidth}
	\centering
\resizebox{\linewidth}{!}{
\begin{tikzpicture}[scale =0.7]

\definecolor{darkgray176}{RGB}{176,176,176}
\input{tikz_figures/color-def}

\begin{axis}[
legend cell align={left},
legend style={fill opacity=0.8, draw opacity=1, text opacity=1, draw=lightgray204, font=\Large},
tick align=outside,
tick pos=left,
x grid style={darkgray176},
xlabel={Relative Overlap},
xmin=-0.8625, xmax=9.8625,
xtick style={color=black},
xtick={0,1,2,3,4,5,6,7,8,9},
xticklabel style={rotate=45.0,anchor=east},
xticklabels={0.0–0.1,0.1–0.2,0.2–0.3,0.3–0.4,0.4–0.5,0.5–0.6,0.6–0.7,0.7–0.8,0.8–0.9,0.9–1.0},
y grid style={darkgray176},
ylabel={Frequency},
ymajorgrids,
ymin=0, ymax=430,
ytick style={color=black},
tick label style={color=black, font=\Large},
label style={font =\Large},
]
\draw[draw=black,fill=orange] (axis cs:-0.375,0) rectangle (axis cs:-0.125,0);
\addlegendimage{ybar,ybar legend,draw=black,fill=orange}
\addlegendentry{Approval Score}

\draw[draw=black,fill=orange] (axis cs:0.625,0) rectangle (axis cs:0.875,8);
\draw[draw=black,fill=orange] (axis cs:1.625,0) rectangle (axis cs:1.875,146);
\draw[draw=black,fill=orange] (axis cs:2.625,0) rectangle (axis cs:2.875,0);
\draw[draw=black,fill=orange] (axis cs:3.625,0) rectangle (axis cs:3.875,210);
\draw[draw=black,fill=orange] (axis cs:4.625,0) rectangle (axis cs:4.875,258);
\draw[draw=black,fill=orange] (axis cs:5.625,0) rectangle (axis cs:5.875,19);
\draw[draw=black,fill=orange] (axis cs:6.625,0) rectangle (axis cs:6.875,0);
\draw[draw=black,fill=orange] (axis cs:7.625,0) rectangle (axis cs:7.875,9);
\draw[draw=black,fill=orange] (axis cs:8.625,0) rectangle (axis cs:8.875,0);
\draw[draw=black,fill=blue] (axis cs:-0.125,0) rectangle (axis cs:0.125,6);
\addlegendimage{ybar,ybar legend,draw=black,fill=blue}
\addlegendentry{JR Prevalence}

\draw[draw=black,fill=blue] (axis cs:0.875,0) rectangle (axis cs:1.125,35);
\draw[draw=black,fill=blue] (axis cs:1.875,0) rectangle (axis cs:2.125,402);
\draw[draw=black,fill=blue] (axis cs:2.875,0) rectangle (axis cs:3.125,0);
\draw[draw=black,fill=blue] (axis cs:3.875,0) rectangle (axis cs:4.125,148);
\draw[draw=black,fill=blue] (axis cs:4.875,0) rectangle (axis cs:5.125,59);
\draw[draw=black,fill=blue] (axis cs:5.875,0) rectangle (axis cs:6.125,0);
\draw[draw=black,fill=blue] (axis cs:6.875,0) rectangle (axis cs:7.125,0);
\draw[draw=black,fill=blue] (axis cs:7.875,0) rectangle (axis cs:8.125,0);
\draw[draw=black,fill=blue] (axis cs:8.875,0) rectangle (axis cs:9.125,0);
\draw[draw=black,fill=green] (axis cs:0.125,0) rectangle (axis cs:0.375,11);
\addlegendimage{ybar,ybar legend,draw=black,fill=green}
\addlegendentry{EJR+ Prevalence}

\draw[draw=black,fill=green] (axis cs:1.125,0) rectangle (axis cs:1.375,46);
\draw[draw=black,fill=green] (axis cs:2.125,0) rectangle (axis cs:2.375,333);
\draw[draw=black,fill=green] (axis cs:3.125,0) rectangle (axis cs:3.375,0);
\draw[draw=black,fill=green] (axis cs:4.125,0) rectangle (axis cs:4.375,183);
\draw[draw=black,fill=green] (axis cs:5.125,0) rectangle (axis cs:5.375,76);
\draw[draw=black,fill=green] (axis cs:6.125,0) rectangle (axis cs:6.375,1);
\draw[draw=black,fill=green] (axis cs:7.125,0) rectangle (axis cs:7.375,0);
\draw[draw=black,fill=green] (axis cs:8.125,0) rectangle (axis cs:8.375,0);
\draw[draw=black,fill=green] (axis cs:9.125,0) rectangle (axis cs:9.375,0);
\draw (axis cs:0.75,9) node[
  scale=0.8,
  anchor=south,
  text=black,
  rotate=0.0
]{8};
\draw (axis cs:1.75,147) node[
  scale=0.8,
  anchor=south,
  text=black,
  rotate=0.0
]{146};
\draw (axis cs:3.75,211) node[
  scale=0.8,
  anchor=south,
  text=black,
  rotate=0.0
]{210};
\draw (axis cs:4.75,259) node[
  scale=0.8,
  anchor=south,
  text=black,
  rotate=0.0
]{258};
\draw (axis cs:5.75,20) node[
  scale=0.8,
  anchor=south,
  text=black,
  rotate=0.0
]{19};
\draw (axis cs:7.75,10) node[
  scale=0.8,
  anchor=south,
  text=black,
  rotate=0.0
]{9};
\draw (axis cs:0,7) node[
  scale=0.8,
  anchor=south,
  text=black,
  rotate=0.0
]{6};
\draw (axis cs:1,36) node[
  scale=0.8,
  anchor=south,
  text=black,
  rotate=0.0
]{35};
\draw (axis cs:2,403) node[
  scale=0.8,
  anchor=south,
  text=black,
  rotate=0.0
]{402};
\draw (axis cs:4,149) node[
  scale=0.8,
  anchor=south,
  text=black,
  rotate=0.0
]{148};
\draw (axis cs:5,60) node[
  scale=0.8,
  anchor=south,
  text=black,
  rotate=0.0
]{59};
\draw (axis cs:0.25,12) node[
  scale=0.8,
  anchor=south,
  text=black,
  rotate=0.0
]{11};
\draw (axis cs:1.25,47) node[
  scale=0.8,
  anchor=south,
  text=black,
  rotate=0.0
]{46};
\draw (axis cs:2.25,334) node[
  scale=0.8,
  anchor=south,
  text=black,
  rotate=0.0
]{333};
\draw (axis cs:4.25,184) node[
  scale=0.8,
  anchor=south,
  text=black,
  rotate=0.0
]{183};
\draw (axis cs:5.25,77) node[
  scale=0.8,
  anchor=south,
  text=black,
  rotate=0.0
]{76};
\draw (axis cs:6.25,2) node[
  scale=0.8,
  anchor=south,
  text=black,
  rotate=0.0
]{1};
\end{axis}

\end{tikzpicture}}
 \caption{seq-Phragmén}
        \label{fig:histgram-seqphrag-Euclidean}
	\end{subfigure}
	\hfill
	\begin{subfigure}[t]{0.32\textwidth}
	\centering
\resizebox{\linewidth}{!}{
\begin{tikzpicture}[scale =0.7]

\definecolor{darkgray176}{RGB}{176,176,176}
\input{tikz_figures/color-def}

\begin{axis}[
legend cell align={left},
legend style={fill opacity=0.8, draw opacity=1, text opacity=1, draw=lightgray204, font=\Large},
tick align=outside,
tick pos=left,
x grid style={darkgray176},
xlabel={Relative Overlap},
xmin=-0.8625, xmax=9.8625,
xtick style={color=black},
xtick={0,1,2,3,4,5,6,7,8,9},
xticklabel style={rotate=45.0,anchor=east},
xticklabels={0.0–0.1,0.1–0.2,0.2–0.3,0.3–0.4,0.4–0.5,0.5–0.6,0.6–0.7,0.7–0.8,0.8–0.9,0.9–1.0},
y grid style={darkgray176},
ylabel={Frequency},
ymajorgrids,
ymin=0, ymax=430,
ytick style={color=black},
tick label style={color=black, font=\Large},
label style={font =\Large},
]
\draw[draw=black,fill=orange] (axis cs:-0.375,0) rectangle (axis cs:-0.125,0);
\addlegendimage{ybar,ybar legend,draw=black,fill=orange}
\addlegendentry{Approval Score}

\draw[draw=black,fill=orange] (axis cs:0.625,0) rectangle (axis cs:0.875,1);
\draw[draw=black,fill=orange] (axis cs:1.625,0) rectangle (axis cs:1.875,108);
\draw[draw=black,fill=orange] (axis cs:2.625,0) rectangle (axis cs:2.875,0);
\draw[draw=black,fill=orange] (axis cs:3.625,0) rectangle (axis cs:3.875,189);
\draw[draw=black,fill=orange] (axis cs:4.625,0) rectangle (axis cs:4.875,310);
\draw[draw=black,fill=orange] (axis cs:5.625,0) rectangle (axis cs:5.875,29);
\draw[draw=black,fill=orange] (axis cs:6.625,0) rectangle (axis cs:6.875,0);
\draw[draw=black,fill=orange] (axis cs:7.625,0) rectangle (axis cs:7.875,12);
\draw[draw=black,fill=orange] (axis cs:8.625,0) rectangle (axis cs:8.875,1);
\draw[draw=black,fill=blue] (axis cs:-0.125,0) rectangle (axis cs:0.125,6);
\addlegendimage{ybar,ybar legend,draw=black,fill=blue}
\addlegendentry{JR Prevalence}

\draw[draw=black,fill=blue] (axis cs:0.875,0) rectangle (axis cs:1.125,32);
\draw[draw=black,fill=blue] (axis cs:1.875,0) rectangle (axis cs:2.125,404);
\draw[draw=black,fill=blue] (axis cs:2.875,0) rectangle (axis cs:3.125,0);
\draw[draw=black,fill=blue] (axis cs:3.875,0) rectangle (axis cs:4.125,144);
\draw[draw=black,fill=blue] (axis cs:4.875,0) rectangle (axis cs:5.125,64);
\draw[draw=black,fill=blue] (axis cs:5.875,0) rectangle (axis cs:6.125,0);
\draw[draw=black,fill=blue] (axis cs:6.875,0) rectangle (axis cs:7.125,0);
\draw[draw=black,fill=blue] (axis cs:7.875,0) rectangle (axis cs:8.125,0);
\draw[draw=black,fill=blue] (axis cs:8.875,0) rectangle (axis cs:9.125,0);
\draw[draw=black,fill=green] (axis cs:0.125,0) rectangle (axis cs:0.375,8);
\addlegendimage{ybar,ybar legend,draw=black,fill=green}
\addlegendentry{EJR+ Prevalence}

\draw[draw=black,fill=green] (axis cs:1.125,0) rectangle (axis cs:1.375,40);
\draw[draw=black,fill=green] (axis cs:2.125,0) rectangle (axis cs:2.375,330);
\draw[draw=black,fill=green] (axis cs:3.125,0) rectangle (axis cs:3.375,0);
\draw[draw=black,fill=green] (axis cs:4.125,0) rectangle (axis cs:4.375,184);
\draw[draw=black,fill=green] (axis cs:5.125,0) rectangle (axis cs:5.375,87);
\draw[draw=black,fill=green] (axis cs:6.125,0) rectangle (axis cs:6.375,1);
\draw[draw=black,fill=green] (axis cs:7.125,0) rectangle (axis cs:7.375,0);
\draw[draw=black,fill=green] (axis cs:8.125,0) rectangle (axis cs:8.375,0);
\draw[draw=black,fill=green] (axis cs:9.125,0) rectangle (axis cs:9.375,0);
\draw (axis cs:0.75,2) node[
  scale=0.8,
  anchor=south,
  text=black,
  rotate=0.0
]{1};
\draw (axis cs:1.75,109) node[
  scale=0.8,
  anchor=south,
  text=black,
  rotate=0.0
]{108};
\draw (axis cs:3.75,190) node[
  scale=0.8,
  anchor=south,
  text=black,
  rotate=0.0
]{189};
\draw (axis cs:4.75,311) node[
  scale=0.8,
  anchor=south,
  text=black,
  rotate=0.0
]{310};
\draw (axis cs:5.75,30) node[
  scale=0.8,
  anchor=south,
  text=black,
  rotate=0.0
]{29};
\draw (axis cs:7.75,13) node[
  scale=0.8,
  anchor=south,
  text=black,
  rotate=0.0
]{12};
\draw (axis cs:8.75,2) node[
  scale=0.8,
  anchor=south,
  text=black,
  rotate=0.0
]{1};
\draw (axis cs:0,7) node[
  scale=0.8,
  anchor=south,
  text=black,
  rotate=0.0
]{6};
\draw (axis cs:1,33) node[
  scale=0.8,
  anchor=south,
  text=black,
  rotate=0.0
]{32};
\draw (axis cs:2,405) node[
  scale=0.8,
  anchor=south,
  text=black,
  rotate=0.0
]{404};
\draw (axis cs:4,145) node[
  scale=0.8,
  anchor=south,
  text=black,
  rotate=0.0
]{144};
\draw (axis cs:5,65) node[
  scale=0.8,
  anchor=south,
  text=black,
  rotate=0.0
]{64};
\draw (axis cs:0.25,9) node[
  scale=0.8,
  anchor=south,
  text=black,
  rotate=0.0
]{8};
\draw (axis cs:1.25,41) node[
  scale=0.8,
  anchor=south,
  text=black,
  rotate=0.0
]{40};
\draw (axis cs:2.25,331) node[
  scale=0.8,
  anchor=south,
  text=black,
  rotate=0.0
]{330};
\draw (axis cs:4.25,185) node[
  scale=0.8,
  anchor=south,
  text=black,
  rotate=0.0
]{184};
\draw (axis cs:5.25,88) node[
  scale=0.8,
  anchor=south,
  text=black,
  rotate=0.0
]{87};
\draw (axis cs:6.25,2) node[
  scale=0.8,
  anchor=south,
  text=black,
  rotate=0.0
]{1};
\end{axis}

\end{tikzpicture}}
 \caption{seq-PAV}
        \label{histogram-seqpav-Euclidean}
	\end{subfigure}
    \caption{Histograms showing the distribution of relative overlaps between committees selected by different voting rules and the top $k$ candidates according to our three candidates' importance measures for the Euclidean dataset (650 instances).}
    \label{fig:histogram-vot-Euclidean}
\end{figure*}

Recall from Section 4.3 that we compare they~$k$ selected candidates by three voting rules (the Method of Equal Shares, seq-Phragmén, seq-PAV) with the~$k$ candidates that score highest with respect to each of our three candidates' importance measures, by computing the relative overlaps for each instance. We show our results separately for each of our datasets in~\Cref{fig:histogram-mes,fig:histogram-vot-pabulib,fig:histogram-vot-Euclidean,fig:histogram-vot-resampling}. We observe that for each dataset and each voting rule, the committees selected by the voting rule have, in general, the highest overlap with the top~$k$ candidates ranked by approval scores, followed by those ranked by EJR+-prevalence, and finally by JR-prevalence. A further observation consistent across datasets is that, for a given candidates' importance measure, the distribution of overlaps varies only marginally among the three voting rules. However, while the distributions for the pabulib and resampling datasets are quite similar, the Euclidean dataset exhibits different behavior: the overlaps are significantly lower across all voting rules and all three candidates' importance measures.

\begin{figure*}[h]
	\centering
    \begin{subfigure}[t]{0.32\textwidth}
	\centering
\resizebox{\linewidth}{!}{
\begin{tikzpicture}[scale = 0.7]

\definecolor{darkgray176}{RGB}{176,176,176}
\definecolor{green}{RGB}{0,128,0}
\definecolor{lightgray204}{RGB}{204,204,204}
\definecolor{orange}{RGB}{255,165,0}
\input{tikz_figures/color-def}

\begin{axis}[
legend cell align={left},
legend style={
  fill opacity=0.8,
  draw opacity=1,
  text opacity=1,
  at={(0.03,0.97)},
  anchor=north west,
  draw=lightgray204,
  font=\Large
},
tick align=outside,
tick pos=left,
x grid style={darkgray176},
xlabel={Relative Overlap},
xmin=-0.8625, xmax=9.8625,
xtick style={color=black},
xtick={0,1,2,3,4,5,6,7,8,9},
xticklabel style={rotate=45.0,anchor=east},
xticklabels={0.0–0.1,0.1–0.2,0.2–0.3,0.3–0.4,0.4–0.5,0.5–0.6,0.6–0.7,0.7–0.8,0.8–0.9,0.9–1.0},
y grid style={darkgray176},
ylabel={Frequency},
ymajorgrids,
ymin=0, ymax=220,
ytick style={color=black},
tick label style={font=\Large},
label style={font =\Large},
]
\draw[draw=black,fill=orange] (axis cs:-0.375,0) rectangle (axis cs:-0.125,0);
\addlegendimage{ybar,ybar legend,draw=black,fill=orange}
\addlegendentry{Approval Score}

\draw[draw=black,fill=orange] (axis cs:0.625,0) rectangle (axis cs:0.875,0);
\draw[draw=black,fill=orange] (axis cs:1.625,0) rectangle (axis cs:1.875,0);
\draw[draw=black,fill=orange] (axis cs:2.625,0) rectangle (axis cs:2.875,0);
\draw[draw=black,fill=orange] (axis cs:3.625,0) rectangle (axis cs:3.875,0);
\draw[draw=black,fill=orange] (axis cs:4.625,0) rectangle (axis cs:4.875,2);
\draw[draw=black,fill=orange] (axis cs:5.625,0) rectangle (axis cs:5.875,36);
\draw[draw=black,fill=orange] (axis cs:6.625,0) rectangle (axis cs:6.875,0);
\draw[draw=black,fill=orange] (axis cs:7.625,0) rectangle (axis cs:7.875,119);
\draw[draw=black,fill=orange] (axis cs:8.625,0) rectangle (axis cs:8.875,175);
\draw[draw=black,fill=blue] (axis cs:-0.125,0) rectangle (axis cs:0.125,5);
\addlegendimage{ybar,ybar legend,draw=black,fill=blue}
\addlegendentry{JR Prevalence}

\draw[draw=black,fill=blue] (axis cs:0.875,0) rectangle (axis cs:1.125,22);
\draw[draw=black,fill=blue] (axis cs:1.875,0) rectangle (axis cs:2.125,58);
\draw[draw=black,fill=blue] (axis cs:2.875,0) rectangle (axis cs:3.125,0);
\draw[draw=black,fill=blue] (axis cs:3.875,0) rectangle (axis cs:4.125,28);
\draw[draw=black,fill=blue] (axis cs:4.875,0) rectangle (axis cs:5.125,84);
\draw[draw=black,fill=blue] (axis cs:5.875,0) rectangle (axis cs:6.125,75);
\draw[draw=black,fill=blue] (axis cs:6.875,0) rectangle (axis cs:7.125,0);
\draw[draw=black,fill=blue] (axis cs:7.875,0) rectangle (axis cs:8.125,44);
\draw[draw=black,fill=blue] (axis cs:8.875,0) rectangle (axis cs:9.125,16);
\draw[draw=black,fill=green] (axis cs:0.125,0) rectangle (axis cs:0.375,0);
\addlegendimage{ybar,ybar legend,draw=black,fill=green}
\addlegendentry{EJR+ Prevalence}

\draw[draw=black,fill=green] (axis cs:1.125,0) rectangle (axis cs:1.375,0);
\draw[draw=black,fill=green] (axis cs:2.125,0) rectangle (axis cs:2.375,2);
\draw[draw=black,fill=green] (axis cs:3.125,0) rectangle (axis cs:3.375,0);
\draw[draw=black,fill=green] (axis cs:4.125,0) rectangle (axis cs:4.375,2);
\draw[draw=black,fill=green] (axis cs:5.125,0) rectangle (axis cs:5.375,107);
\draw[draw=black,fill=green] (axis cs:6.125,0) rectangle (axis cs:6.375,116);
\draw[draw=black,fill=green] (axis cs:7.125,0) rectangle (axis cs:7.375,0);
\draw[draw=black,fill=green] (axis cs:8.125,0) rectangle (axis cs:8.375,81);
\draw[draw=black,fill=green] (axis cs:9.125,0) rectangle (axis cs:9.375,24);
\draw (axis cs:4.75,3) node[
  scale=0.8,
  anchor=south,
  text=black,
  rotate=0.0
]{2};
\draw (axis cs:5.75,37) node[
  scale=0.8,
  anchor=south,
  text=black,
  rotate=0.0
]{36};
\draw (axis cs:7.75,120) node[
  scale=0.8,
  anchor=south,
  text=black,
  rotate=0.0
]{119};
\draw (axis cs:8.75,176) node[
  scale=0.8,
  anchor=south,
  text=black,
  rotate=0.0
]{175};
\draw (axis cs:0,6) node[
  scale=0.8,
  anchor=south,
  text=black,
  rotate=0.0
]{5};
\draw (axis cs:1,23) node[
  scale=0.8,
  anchor=south,
  text=black,
  rotate=0.0
]{22};
\draw (axis cs:2,59) node[
  scale=0.8,
  anchor=south,
  text=black,
  rotate=0.0
]{58};
\draw (axis cs:4,29) node[
  scale=0.8,
  anchor=south,
  text=black,
  rotate=0.0
]{28};
\draw (axis cs:5,85) node[
  scale=0.8,
  anchor=south,
  text=black,
  rotate=0.0
]{84};
\draw (axis cs:6,76) node[
  scale=0.8,
  anchor=south,
  text=black,
  rotate=0.0
]{75};
\draw (axis cs:8,45) node[
  scale=0.8,
  anchor=south,
  text=black,
  rotate=0.0
]{44};
\draw (axis cs:9,17) node[
  scale=0.8,
  anchor=south,
  text=black,
  rotate=0.0
]{16};
\draw (axis cs:2.25,3) node[
  scale=0.8,
  anchor=south,
  text=black,
  rotate=0.0
]{2};
\draw (axis cs:4.25,3) node[
  scale=0.8,
  anchor=south,
  text=black,
  rotate=0.0
]{2};
\draw (axis cs:5.25,108) node[
  scale=0.8,
  anchor=south,
  text=black,
  rotate=0.0
]{107};
\draw (axis cs:6.25,117) node[
  scale=0.8,
  anchor=south,
  text=black,
  rotate=0.0
]{116};
\draw (axis cs:8.25,82) node[
  scale=0.8,
  anchor=south,
  text=black,
  rotate=0.0
]{81};
\draw (axis cs:9.25,25) node[
  scale=0.8,
  anchor=south,
  text=black,
  rotate=0.0
]{24};
\end{axis}
\end{tikzpicture}}
 \caption{Method of Equal Shares}
        \label{fig:histgram-mes-resampling}
	\end{subfigure}
	\hfill
\begin{subfigure}[t]{0.32\textwidth}
	\centering
\resizebox{\linewidth}{!}{
\begin{tikzpicture}[scale=0.7]

\definecolor{darkgray176}{RGB}{176,176,176}
\input{tikz_figures/color-def}

\begin{axis}[
legend cell align={left},
legend style={
  fill opacity=0.8,
  draw opacity=1,
  text opacity=1,
  at={(0.03,0.97)},
  anchor=north west,
  draw=lightgray204,
  font=\Large
},
tick align=outside,
tick pos=left,
x grid style={darkgray176},
xlabel={Relative Overlap},
xmin=-0.8625, xmax=9.8625,
xtick style={color=black},
xtick={0,1,2,3,4,5,6,7,8,9},
xticklabel style={rotate=45.0,anchor=east},
xticklabels={0.0–0.1,0.1–0.2,0.2–0.3,0.3–0.4,0.4–0.5,0.5–0.6,0.6–0.7,0.7–0.8,0.8–0.9,0.9–1.0},
y grid style={darkgray176},
ylabel={Frequency},
ymajorgrids,
ymin=0, ymax=220,
ytick style={color=black},
tick label style={color=black, font=\Large},
label style={font =\Large}
]
\draw[draw=black,fill=orange] (axis cs:-0.375,0) rectangle (axis cs:-0.125,0);
\addlegendimage{ybar,ybar legend,draw=black,fill=orange}
\addlegendentry{Approval Score}

\draw[draw=black,fill=orange] (axis cs:0.625,0) rectangle (axis cs:0.875,0);
\draw[draw=black,fill=orange] (axis cs:1.625,0) rectangle (axis cs:1.875,0);
\draw[draw=black,fill=orange] (axis cs:2.625,0) rectangle (axis cs:2.875,0);
\draw[draw=black,fill=orange] (axis cs:3.625,0) rectangle (axis cs:3.875,0);
\draw[draw=black,fill=orange] (axis cs:4.625,0) rectangle (axis cs:4.875,1);
\draw[draw=black,fill=orange] (axis cs:5.625,0) rectangle (axis cs:5.875,31);
\draw[draw=black,fill=orange] (axis cs:6.625,0) rectangle (axis cs:6.875,0);
\draw[draw=black,fill=orange] (axis cs:7.625,0) rectangle (axis cs:7.875,97);
\draw[draw=black,fill=orange] (axis cs:8.625,0) rectangle (axis cs:8.875,203);
\draw[draw=black,fill=blue] (axis cs:-0.125,0) rectangle (axis cs:0.125,6);
\addlegendimage{ybar,ybar legend,draw=black,fill=blue}
\addlegendentry{JR Prevalence}

\draw[draw=black,fill=blue] (axis cs:0.875,0) rectangle (axis cs:1.125,22);
\draw[draw=black,fill=blue] (axis cs:1.875,0) rectangle (axis cs:2.125,55);
\draw[draw=black,fill=blue] (axis cs:2.875,0) rectangle (axis cs:3.125,0);
\draw[draw=black,fill=blue] (axis cs:3.875,0) rectangle (axis cs:4.125,27);
\draw[draw=black,fill=blue] (axis cs:4.875,0) rectangle (axis cs:5.125,85);
\draw[draw=black,fill=blue] (axis cs:5.875,0) rectangle (axis cs:6.125,75);
\draw[draw=black,fill=blue] (axis cs:6.875,0) rectangle (axis cs:7.125,0);
\draw[draw=black,fill=blue] (axis cs:7.875,0) rectangle (axis cs:8.125,51);
\draw[draw=black,fill=blue] (axis cs:8.875,0) rectangle (axis cs:9.125,11);
\draw[draw=black,fill=green] (axis cs:0.125,0) rectangle (axis cs:0.375,0);
\addlegendimage{ybar,ybar legend,draw=black,fill=green}
\addlegendentry{EJR+ Prevalence}

\draw[draw=black,fill=green] (axis cs:1.125,0) rectangle (axis cs:1.375,0);
\draw[draw=black,fill=green] (axis cs:2.125,0) rectangle (axis cs:2.375,1);
\draw[draw=black,fill=green] (axis cs:3.125,0) rectangle (axis cs:3.375,0);
\draw[draw=black,fill=green] (axis cs:4.125,0) rectangle (axis cs:4.375,4);
\draw[draw=black,fill=green] (axis cs:5.125,0) rectangle (axis cs:5.375,88);
\draw[draw=black,fill=green] (axis cs:6.125,0) rectangle (axis cs:6.375,113);
\draw[draw=black,fill=green] (axis cs:7.125,0) rectangle (axis cs:7.375,0);
\draw[draw=black,fill=green] (axis cs:8.125,0) rectangle (axis cs:8.375,89);
\draw[draw=black,fill=green] (axis cs:9.125,0) rectangle (axis cs:9.375,37);
\draw (axis cs:4.75,2) node[
  scale=0.8,
  anchor=south,
  text=black,
  rotate=0.0
]{1};
\draw (axis cs:5.75,32) node[
  scale=0.8,
  anchor=south,
  text=black,
  rotate=0.0
]{31};
\draw (axis cs:7.75,98) node[
  scale=0.8,
  anchor=south,
  text=black,
  rotate=0.0
]{97};
\draw (axis cs:8.75,204) node[
  scale=0.8,
  anchor=south,
  text=black,
  rotate=0.0
]{203};
\draw (axis cs:0,7) node[
  scale=0.8,
  anchor=south,
  text=black,
  rotate=0.0
]{6};
\draw (axis cs:1,23) node[
  scale=0.8,
  anchor=south,
  text=black,
  rotate=0.0
]{22};
\draw (axis cs:2,56) node[
  scale=0.8,
  anchor=south,
  text=black,
  rotate=0.0
]{55};
\draw (axis cs:4,28) node[
  scale=0.8,
  anchor=south,
  text=black,
  rotate=0.0
]{27};
\draw (axis cs:5,86) node[
  scale=0.8,
  anchor=south,
  text=black,
  rotate=0.0
]{85};
\draw (axis cs:6,76) node[
  scale=0.8,
  anchor=south,
  text=black,
  rotate=0.0
]{75};
\draw (axis cs:8,52) node[
  scale=0.8,
  anchor=south,
  text=black,
  rotate=0.0
]{51};
\draw (axis cs:9,12) node[
  scale=0.8,
  anchor=south,
  text=black,
  rotate=0.0
]{11};
\draw (axis cs:2.25,2) node[
  scale=0.8,
  anchor=south,
  text=black,
  rotate=0.0
]{1};
\draw (axis cs:4.25,5) node[
  scale=0.8,
  anchor=south,
  text=black,
  rotate=0.0
]{4};
\draw (axis cs:5.25,89) node[
  scale=0.8,
  anchor=south,
  text=black,
  rotate=0.0
]{88};
\draw (axis cs:6.25,114) node[
  scale=0.8,
  anchor=south,
  text=black,
  rotate=0.0
]{113};
\draw (axis cs:8.25,90) node[
  scale=0.8,
  anchor=south,
  text=black,
  rotate=0.0
]{89};
\draw (axis cs:9.25,38) node[
  scale=0.8,
  anchor=south,
  text=black,
  rotate=0.0
]{37};
\end{axis}

\end{tikzpicture}}
 \caption{seq-Phragmén}
        \label{fig:histgram-seqphrag-resampling}
	\end{subfigure}
	\hfill
	\begin{subfigure}[t]{0.32\textwidth}
	\centering
\resizebox{\linewidth}{!}{
\begin{tikzpicture}[scale =0.7]

\definecolor{darkgray176}{RGB}{176,176,176}
\input{tikz_figures/color-def}

\begin{axis}[
legend cell align={left},
legend style={
  fill opacity=0.8,
  draw opacity=1,
  text opacity=1,
  at={(0.03,0.97)},
  anchor=north west,
  draw=lightgray204,
  font=\Large
},
tick align=outside,
tick pos=left,
x grid style={darkgray176},
xlabel={Relative Overlap},
xmin=-0.8625, xmax=9.8625,
xtick style={color=black},
xtick={0,1,2,3,4,5,6,7,8,9},
xticklabel style={rotate=45.0,anchor=east},
xticklabels={0.0–0.1,0.1–0.2,0.2–0.3,0.3–0.4,0.4–0.5,0.5–0.6,0.6–0.7,0.7–0.8,0.8–0.9,0.9–1.0},
y grid style={darkgray176},
ylabel={Frequency},
ymajorgrids,
ymin=0, ymax=220,
ytick style={color=black},
tick label style={color=black, font=\Large},
label style={font =\Large},
]
\draw[draw=black,fill=orange] (axis cs:-0.375,0) rectangle (axis cs:-0.125,0);
\addlegendimage{ybar,ybar legend,draw=black,fill=orange}
\addlegendentry{Approval Score}

\draw[draw=black,fill=orange] (axis cs:0.625,0) rectangle (axis cs:0.875,0);
\draw[draw=black,fill=orange] (axis cs:1.625,0) rectangle (axis cs:1.875,0);
\draw[draw=black,fill=orange] (axis cs:2.625,0) rectangle (axis cs:2.875,0);
\draw[draw=black,fill=orange] (axis cs:3.625,0) rectangle (axis cs:3.875,0);
\draw[draw=black,fill=orange] (axis cs:4.625,0) rectangle (axis cs:4.875,4);
\draw[draw=black,fill=orange] (axis cs:5.625,0) rectangle (axis cs:5.875,27);
\draw[draw=black,fill=orange] (axis cs:6.625,0) rectangle (axis cs:6.875,0);
\draw[draw=black,fill=orange] (axis cs:7.625,0) rectangle (axis cs:7.875,96);
\draw[draw=black,fill=orange] (axis cs:8.625,0) rectangle (axis cs:8.875,205);
\draw[draw=black,fill=blue] (axis cs:-0.125,0) rectangle (axis cs:0.125,5);
\addlegendimage{ybar,ybar legend,draw=black,fill=blue}
\addlegendentry{JR Prevalence}

\draw[draw=black,fill=blue] (axis cs:0.875,0) rectangle (axis cs:1.125,17);
\draw[draw=black,fill=blue] (axis cs:1.875,0) rectangle (axis cs:2.125,64);
\draw[draw=black,fill=blue] (axis cs:2.875,0) rectangle (axis cs:3.125,0);
\draw[draw=black,fill=blue] (axis cs:3.875,0) rectangle (axis cs:4.125,22);
\draw[draw=black,fill=blue] (axis cs:4.875,0) rectangle (axis cs:5.125,104);
\draw[draw=black,fill=blue] (axis cs:5.875,0) rectangle (axis cs:6.125,75);
\draw[draw=black,fill=blue] (axis cs:6.875,0) rectangle (axis cs:7.125,0);
\draw[draw=black,fill=blue] (axis cs:7.875,0) rectangle (axis cs:8.125,38);
\draw[draw=black,fill=blue] (axis cs:8.875,0) rectangle (axis cs:9.125,7);
\draw[draw=black,fill=green] (axis cs:0.125,0) rectangle (axis cs:0.375,0);
\addlegendimage{ybar,ybar legend,draw=black,fill=green}
\addlegendentry{EJR+ Prevalence}

\draw[draw=black,fill=green] (axis cs:1.125,0) rectangle (axis cs:1.375,0);
\draw[draw=black,fill=green] (axis cs:2.125,0) rectangle (axis cs:2.375,0);
\draw[draw=black,fill=green] (axis cs:3.125,0) rectangle (axis cs:3.375,0);
\draw[draw=black,fill=green] (axis cs:4.125,0) rectangle (axis cs:4.375,6);
\draw[draw=black,fill=green] (axis cs:5.125,0) rectangle (axis cs:5.375,126);
\draw[draw=black,fill=green] (axis cs:6.125,0) rectangle (axis cs:6.375,90);
\draw[draw=black,fill=green] (axis cs:7.125,0) rectangle (axis cs:7.375,0);
\draw[draw=black,fill=green] (axis cs:8.125,0) rectangle (axis cs:8.375,76);
\draw[draw=black,fill=green] (axis cs:9.125,0) rectangle (axis cs:9.375,34);
\draw (axis cs:4.75,5) node[
  scale=0.8,
  anchor=south,
  text=black,
  rotate=0.0
]{4};
\draw (axis cs:5.75,28) node[
  scale=0.8,
  anchor=south,
  text=black,
  rotate=0.0
]{27};
\draw (axis cs:7.75,97) node[
  scale=0.8,
  anchor=south,
  text=black,
  rotate=0.0
]{96};
\draw (axis cs:8.75,206) node[
  scale=0.8,
  anchor=south,
  text=black,
  rotate=0.0
]{205};
\draw (axis cs:0,6) node[
  scale=0.8,
  anchor=south,
  text=black,
  rotate=0.0
]{5};
\draw (axis cs:1,18) node[
  scale=0.8,
  anchor=south,
  text=black,
  rotate=0.0
]{17};
\draw (axis cs:2,65) node[
  scale=0.8,
  anchor=south,
  text=black,
  rotate=0.0
]{64};
\draw (axis cs:4,23) node[
  scale=0.8,
  anchor=south,
  text=black,
  rotate=0.0
]{22};
\draw (axis cs:5,105) node[
  scale=0.8,
  anchor=south,
  text=black,
  rotate=0.0
]{104};
\draw (axis cs:6,76) node[
  scale=0.8,
  anchor=south,
  text=black,
  rotate=0.0
]{75};
\draw (axis cs:8,39) node[
  scale=0.8,
  anchor=south,
  text=black,
  rotate=0.0
]{38};
\draw (axis cs:9,8) node[
  scale=0.8,
  anchor=south,
  text=black,
  rotate=0.0
]{7};
\draw (axis cs:4.25,7) node[
  scale=0.8,
  anchor=south,
  text=black,
  rotate=0.0
]{6};
\draw (axis cs:5.25,127) node[
  scale=0.8,
  anchor=south,
  text=black,
  rotate=0.0
]{126};
\draw (axis cs:6.25,91) node[
  scale=0.8,
  anchor=south,
  text=black,
  rotate=0.0
]{90};
\draw (axis cs:8.25,77) node[
  scale=0.8,
  anchor=south,
  text=black,
  rotate=0.0
]{76};
\draw (axis cs:9.25,35) node[
  scale=0.8,
  anchor=south,
  text=black,
  rotate=0.0
]{34};
\end{axis}

\end{tikzpicture}}
 \caption{seq-PAV}
        \label{fig:histogram-seqpav-resampling}
	\end{subfigure}
    \caption{Histograms showing the distribution of relative overlaps between committees selected by different voting rules and the top $k$ candidates according to our three candidates' importance measures for the resampling dataset (332 instances).}
    \label{fig:histogram-vot-resampling}
\end{figure*}

\clearpage

\appendix

\clearpage

\end{document}